\documentclass[twocolumn]{aastex63}

\newcommand{\gr}{g\, -\, r}
\newcommand{\gi}{g\, -\, i}
\newcommand{\gz}{g\, -\, z}
\newcommand{\ug}{u\, -\, g}
\newcommand{\kms}{\rm km~s^{-1}}
\newcommand{\rgc}{R_{\rm GC}}
\newcommand{\vphi}{v_\phi}
\newcommand{\meanvphi}{\langle v_\phi \rangle}
\newcommand{\sigmavphi}{\sigma_{v_\phi}}

\accepted{for publication in the \apj}

\shorttitle{A Blueprint for the Milky Way's Stellar Populations.\ III}
\shortauthors{An \& Beers}

\begin{document}

\title{A Blueprint for the Milky Way's Stellar Populations.\ III. Spatial Distributions and Population Fractions of Local Halo Stars}

\author{Deokkeun An}
\affiliation{Department of Science Education, Ewha Womans University, 52 Ewhayeodae-gil, Seodaemun-gu, Seoul 03760, Republic of Korea; deokkeun@ewha.ac.kr}

\author{Timothy C.\ Beers}
\affiliation{Department of Physics and JINA Center for the Evolution of the Elements, University of Notre Dame, Notre Dame, IN 46556, USA}

\begin{abstract}

We analyze the observed spatial, chemical and dynamical distributions of local metal-poor stars, based on photometrically derived metallicity and distance estimates along with proper motions from the Gaia mission. Along the Galactic prime meridian, we identify stellar populations with distinct properties in the metallicity versus rotational velocity space, including Gaia Sausage/Enceladus (GSE), the metal-weak thick disk (MWTD), and the Splash (sometimes referred to as the ``in situ'' halo). We model the observed phase-space distributions using Gaussian mixtures and refine their positions and fractional contributions as a function of distances from the Galactic plane ($|Z|$) and the Galactic center ($\rgc$), providing a global perspective of the major stellar populations in the local halo. Within the sample volume ($|Z|<6$~kpc), stars associated with GSE exhibit a larger proportion of metal-poor stars at greater $\rgc$ ($\Delta \langle{\rm[Fe/H]}\rangle /\Delta\rgc=-0.05\pm0.02$~dex~kpc$^{-1}$). This observed trend, along with a mild anticorrelation of the mean rotational velocity with metallicity ($\Delta \meanvphi / \Delta\langle{\rm[Fe/H]}\rangle \sim-10\ \kms$~dex$^{-1}$), implies that more metal-rich stars in the inner region of the GSE progenitor were gradually stripped away, while the prograde orbit of the merger at infall became radialized by dynamical friction. The metal-rich GSE stars are causally disconnected from the Splash structure, whose stars are mostly found on prograde orbits ($\ga94\%$) and exhibit a more centrally concentrated distribution than GSE. The MWTD exhibits a similar spatial distribution to the Splash, suggesting earlier dynamical heating of stars in the primordial disk of the Milky Way, possibly before the GSE merger.

\end{abstract}


\section{Introduction}\label{sec:intro}

In a series of papers \citep[][hereafter Papers~I and II, respectively]{an:20,an:21}, we have constructed spatial and chemo-dynamical distributions of metal-poor main-sequence stars in the local volume ($|Z| \leq 6$~kpc) based on large photometric survey databases. These distributions were obtained using photometrically derived metallicities and distances from color-calibrated stellar isochrones, along with proper-motion measurements from Gaia \citep{gaiadr2}. Our approach is essentially based on nearly all-sky imaging surveys with a higher survey filling factor than present spectroscopic surveys, and therefore it can provide a holistic view of the stellar populations in the Milky Way. A ``blueprint" constructed in this way can be used to establish fundamental observational constraints on the formation and evolution of the stellar populations in the Milky Way and for guiding the interpretation of future numerical simulations of galaxy formation.

This powerful approach was demonstrated in Paper~I, which presented phase-space distributions in the rotational velocity ($\vphi$)\footnote{The $\vphi$ parameter is defined as the rotational velocity in the direction of the disk's rotation in the rest frame of the Galactocentric cylindrical coordinate system, assuming $\vphi = 238~{\rm km\ s}^{-1}$ for the circular velocity of the local standard of rest (LSR) and $(U, V, W) = (11.1, 12.24, 7.25)\ {\rm km\ s}^{-1}$ for the Sun's velocity with respect to the LSR \citep{schonrich:12}. Stars with prograde and retrograde orbits have $\vphi > 0$ and $\vphi < 0$, respectively. For the distance to the Galactic center, $8.34$~kpc is adopted \citep{reid:14}.} versus metallicity space. In Paper~II, we improved the accuracy of the maps by utilizing a revised set of stellar isochrones, which incorporated empirical color constraints from the ``double sequences'' in the Gaia color-magnitude diagrams \citep{gaiahrd}. These phase-space maps revealed a number of stellar populations with distinct chemo-kinematical properties, which include previously known structures, such as the thin disk, the canonical thick disk (TD), and the metal-weak TD (MWTD) component having TD-like kinematics with halo-like metallicities \citep{ruchti:11,beers:14,carollo:19}. Stars with halo-like kinematics but TD-like metallicities are also clearly captured in these maps \citep{sheffield:12,bonaca:17,bonaca:20,dimatteo:19,belokurov:20}. In Papers~I and II, we designated them as ``splashed disk'' stars, alluding to the suggested origin of these stars as early scattering of the Milky Way's primordial disk. In the literature, ``in situ'' halo stars are also often used to refer to this component, as they are hypothesized to be a product of star formation in the inner region of the primordial Galaxy \citep[e.g.,][]{gallart:19,naidu:20}. In this paper, we refer to this group of stars as the Splash (SP) following the original name suggested by \citet{belokurov:20}.

In addition to the above populations, we identified two large groups of halo stars with mean metallicities of $\langle {\rm [Fe/H]} \rangle \sim -2.4$ and $-1.6$ on net prograde orbits, which were dubbed as the metal-poor halo (MPH) and the metal-rich halo (MRH), respectively. We also attributed a clump of relatively metal-rich ([Fe/H] $\sim-1.2$) stars on net retrograde orbits to Gaia Sausage/Enceladus \citep[GSE;][]{belokurov:18,helmi:18}, which is often called the Gaia Sausage or the Gaia-Enceladus-Sausage in the literature. Furthermore, the retrograde halo structure (RHS) was defined to capture a group of stars on strong retrograde orbits ($\meanvphi \sim -100\ \kms$), which is less likely to be linked to GSE \citep[e.g.,][]{matsuno:19}. The RHS is probably a collection of a number of substructures, discovered from recent spectroscopic surveys and Gaia \citep[see][and references therein]{koppelman:19,naidu:20,limberg:21a}. Nonetheless, because they were not well separated in our photometric mapping, owing to larger uncertainties in the estimated stellar parameters than from spectroscopic surveys, we treated them as a single component.

Although several of these structures and components described above were inferred from sparsely sampled spectroscopic surveys (e.g., \citealt{chiba:00,carollo:07,beers:14}), others have also been discovered in more recent studies based on a combination of spectroscopic and astrometric data from large spectroscopic surveys and the Gaia mission. However, it is often challenging to obtain a global picture of the Milky Way's populations (even those present in a relatively local volume) from an assemblage of single studies, because each study makes specific selections on the stars included in their samples that are, in most cases, limited by subjective sampling for spectroscopic observations. The registration and cross-validation of stellar populations among different studies can also be confused by complex sampling bias and scale errors in the measurements. As a consequence, the relative importance of each stellar population can be difficult to assess. By obtaining a perspective on the entire set of stellar populations based on large photometric databases, one can combine various inferences from spectroscopic data and build a more consistent picture of the Milky Way's assembly history.

Another powerful advantage of our photometric approach is that the sample sizes can be sufficiently large to be viewed as unbiased in situ probes. In many previous spectroscopic studies, conclusions were drawn from so-called orbital integrations, which assume a specific gravitational potential of the Galaxy to calculate the full three-dimensional orbits and orbital energies of individual stars. For example, spectroscopic data in the local volume ($d < 4$~kpc) were utilized in \citet{carollo:07,carollo:10}  and \citet{beers:12} to explore a systematic change of contributions from two hypothesized populations -- the inner-halo (IH) and the outer-halo (OH) components -- with orbits extending over a larger volume of space ($d < 50$~kpc). In contrast, the current approach only uses observed quantities (metallicities, distances, and proper motions) to construct a global picture, \textit{independent} of the choice of a gravitational potential.

This work presents a detailed analysis of the phase-space distribution of local metal-poor stars, based on parameter estimates in the previous two papers of this series. In the plane of the Galactic prime meridian ($l = 0\arcdeg$ and $180\arcdeg$), $\vphi$ is essentially independent of radial velocities. Therefore, the imaging surveys allow one to take advantage of Gaia's accurate proper-motion measurements and to compute $\vphi$, because, for the time being, only a fraction of the imaging targets have available radial velocity measurements. As shown in Papers~I and II, this generates significant synergy between photometric and astrometric surveys, which helps to split the sample on the basis of $\vphi$ and inspect variations of the observed metallicity distribution function (MDF) in much greater detail than previously possible. The observed fine structures of the MDFs obtained from this approach also make the decomposition into subpopulations substantially more reliable than in previous efforts.

The goal of this work is to identify stellar populations with distinct properties in the rotational velocity versus metallicity space and use Gaussian mixtures to model their phase-space distributions and quantify their relative contributions. We also reorganize the stellar groups identified in Papers~I and II and refine their positions and extents in the context of more recent work in the literature in this fast-moving field.

This paper is organized as follows. A description on the sample is briefly summarized in \S~\ref{sec:sample}, followed by blueprint maps presented in \S~\ref{sec:blueprint}. Decomposition into subpopulations is carried out with a two-step approach, as described in \S~\ref{sec:mdf} and \S~\ref{sec:vphi}, respectively; the observed MDFs in bins of $\vphi$ are first decomposed using four metallicity groups, and then the $\vphi$ distribution of each group is decomposed into individual stellar populations. In \S~\ref{sec:spatial}, the fractional contribution of each population is presented as a function of the distance from the Galactic plane in the Galactocentric cylindrical coordinate system ($|Z|$) and distance from the Galactic center in the Galactocentric spherical system ($R_{\rm GC}$). We discuss the most important results from our mapping in Section~\ref{sec:discussion}.

\section{Sample and Stellar Parameters}\label{sec:sample}

Metallicity and distance estimates reported in Papers~I and II are mainly based on the application of stellar isochrones to photometry in Data Release 14 (DR14) of the Sloan Digital Sky Survey \citep[SDSS;][]{dr14}. In some survey areas, the SDSS Legacy Survey database was supplemented by deep $u$-band photometry from the Southern Cap $u$-band Sky Survey \citep[SCUSS;][]{zou:16} or more precise $griz$ measurements from Pan-STARRS~1 (PS1) photometry \citep{chambers:16}. However, we focus on distances and metallicities derived from the SDSS data in this work, because the SCUSS survey area in the Southern Galactic Hemisphere overlaps little with the $\pm30\arcdeg$ band around the Galactic prime meridian (see Figure~1 in Paper~I), and metallicity estimates based on PS1 are less accurate owing to the limited accuracy of stellar isochrones in the PS1 passbands. The gain from the adoption of PS1 is also not significantly different from the case with SDSS, as discussed in Paper~II. The total area covered in this work is approximately $16,800\ {\rm deg}^2$ ($12,500\ {\rm deg}^2$ in the northern Galactic hemisphere); see Paper~I for more information.

The distance and metallicity estimates of stars analyzed in this study are those obtained in Paper~II, based on a revised set of empirically calibrated stellar isochrones using Gaia's double sequences \citep{gaiahrd}. These sequences are composed of two distinct, dominant stellar populations of stars with large transverse motions in the local volume and are attributed to the SP (as well as from the TD) and GSE, respectively. Gaia's double sequences were used to register stellar models at [Fe/H] $= -1.3$ and $-0.4$, resulting in a mild difference in the photometrically derived metallicities and distances between Paper~I and Paper~II (see Figure~8 in Paper~II). These sequences are slightly redder than those based on parallax measurements in Gaia's Early DR3 \citep[EDR3;][]{edr3} with zero-point corrections as suggested by the Gaia team \citep{lindegren:21}: $0.03$-$0.04$~mag in $\ug$, $\sim0.02$~mag in $\gr$ and $\gi$, and $0.02$--$0.03$~mag in $\gz$. Because these differences are relatively small, we adopt photometric estimates (distance and metallicity) from the original calibration in Paper~II for this work.

For the main stellar sample, the $\vphi$ of individual stars are computed using photometric distances and proper-motion measurements from Gaia EDR3. We restrict our analysis to stars within $\pm30\arcdeg$ from the Galactic prime meridian as in the previous papers of this series. In some of the analyses below, additional samples are constructed by combining parallaxes and proper motions from Gaia EDR3 with photometric metallicities, for which Gaia parallaxes are used as a prior.

\section{A Blueprint for the Local Stellar Halo}\label{sec:blueprint}

\begin{figure*}
\centering
\includegraphics[scale=0.36]{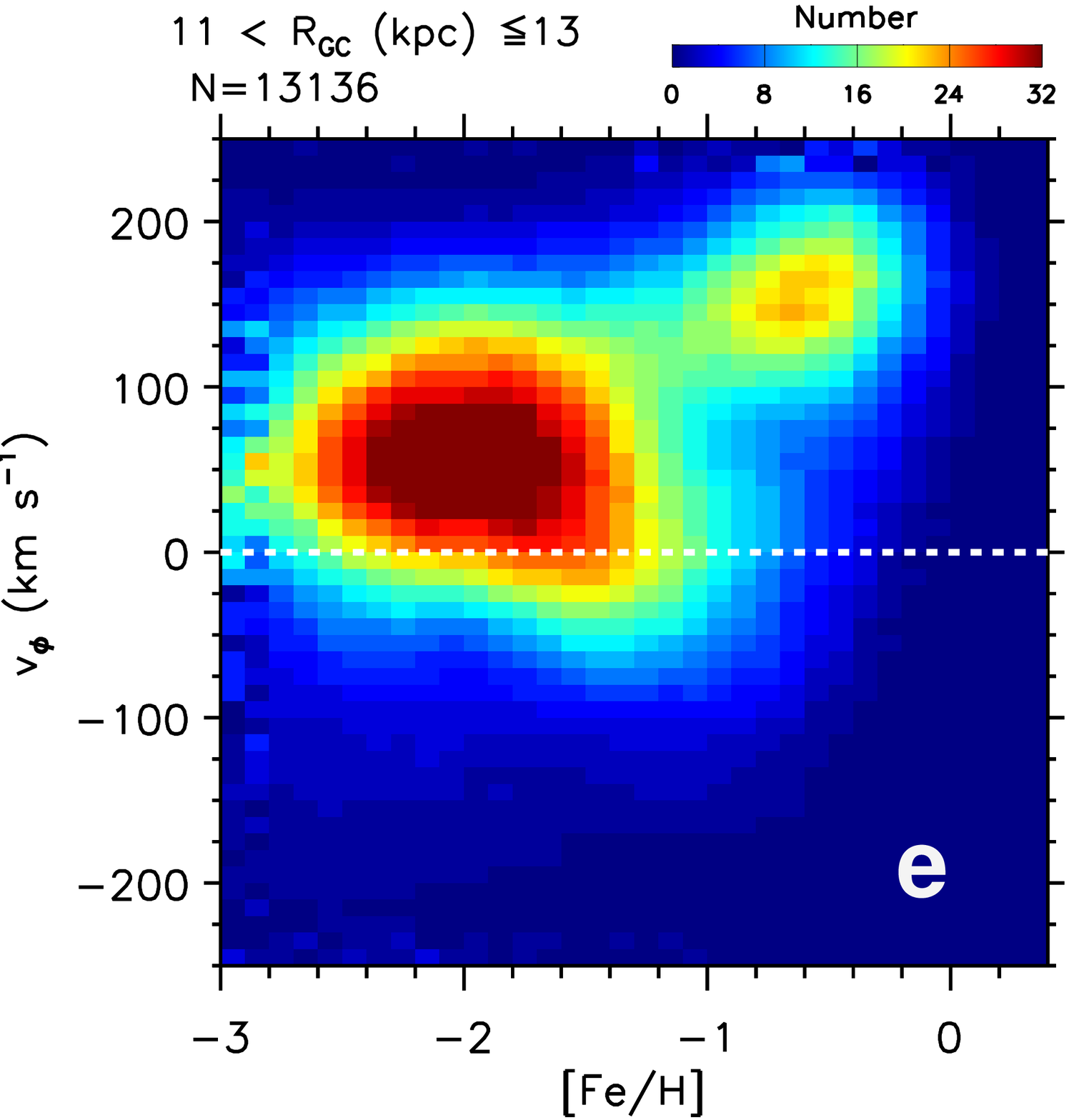}
\includegraphics[scale=0.36]{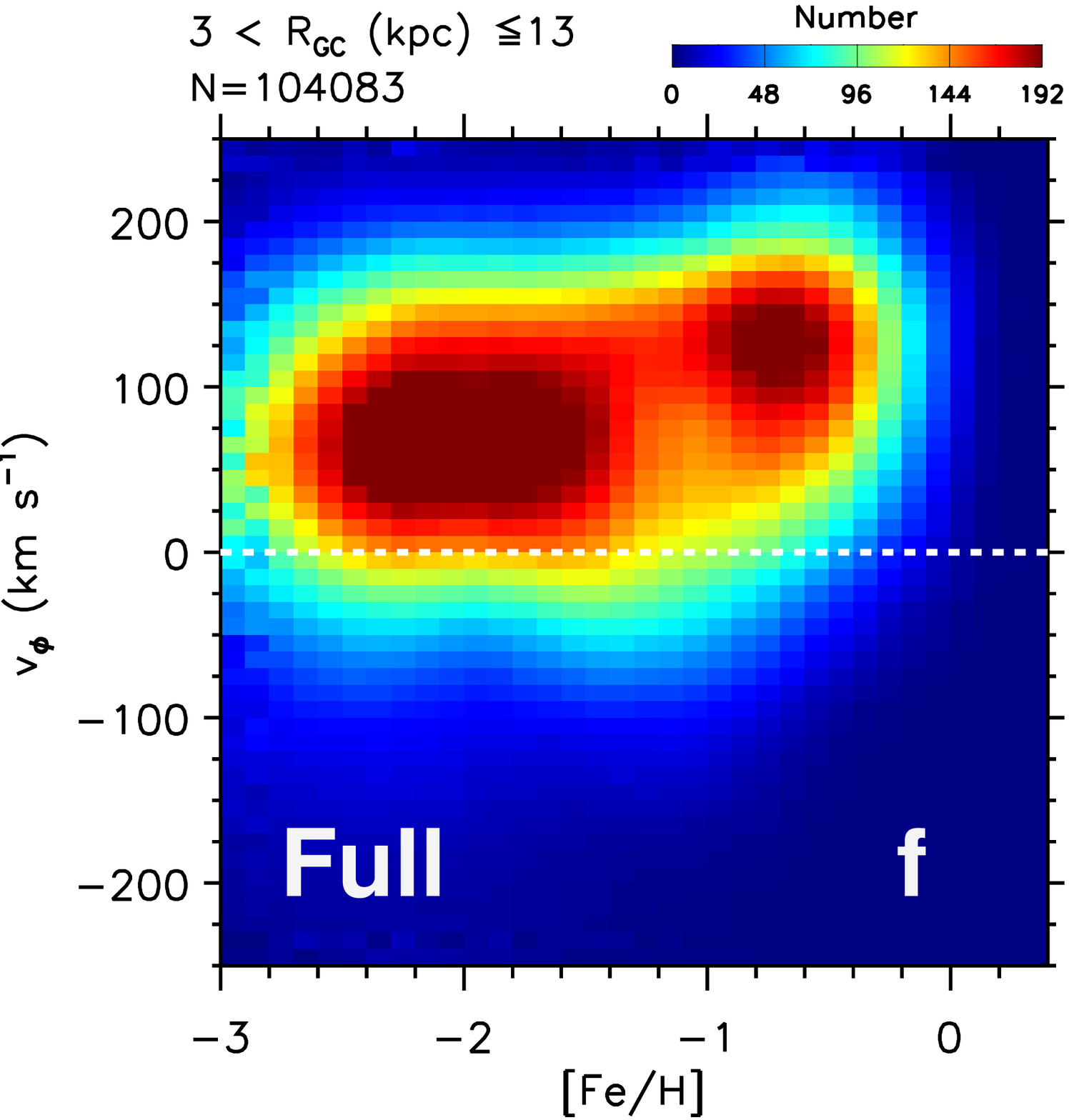}
\includegraphics[scale=0.36]{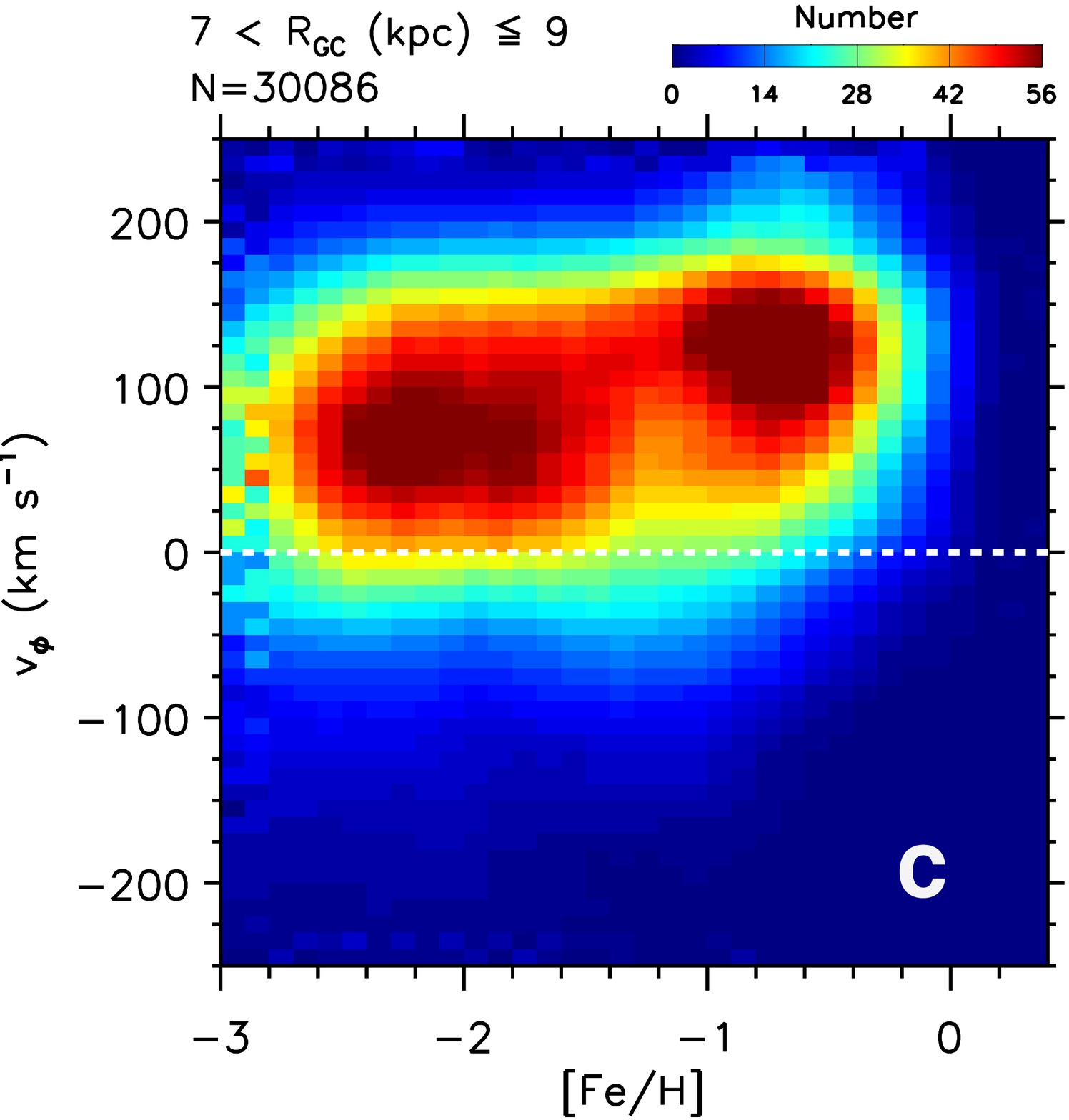}
\includegraphics[scale=0.36]{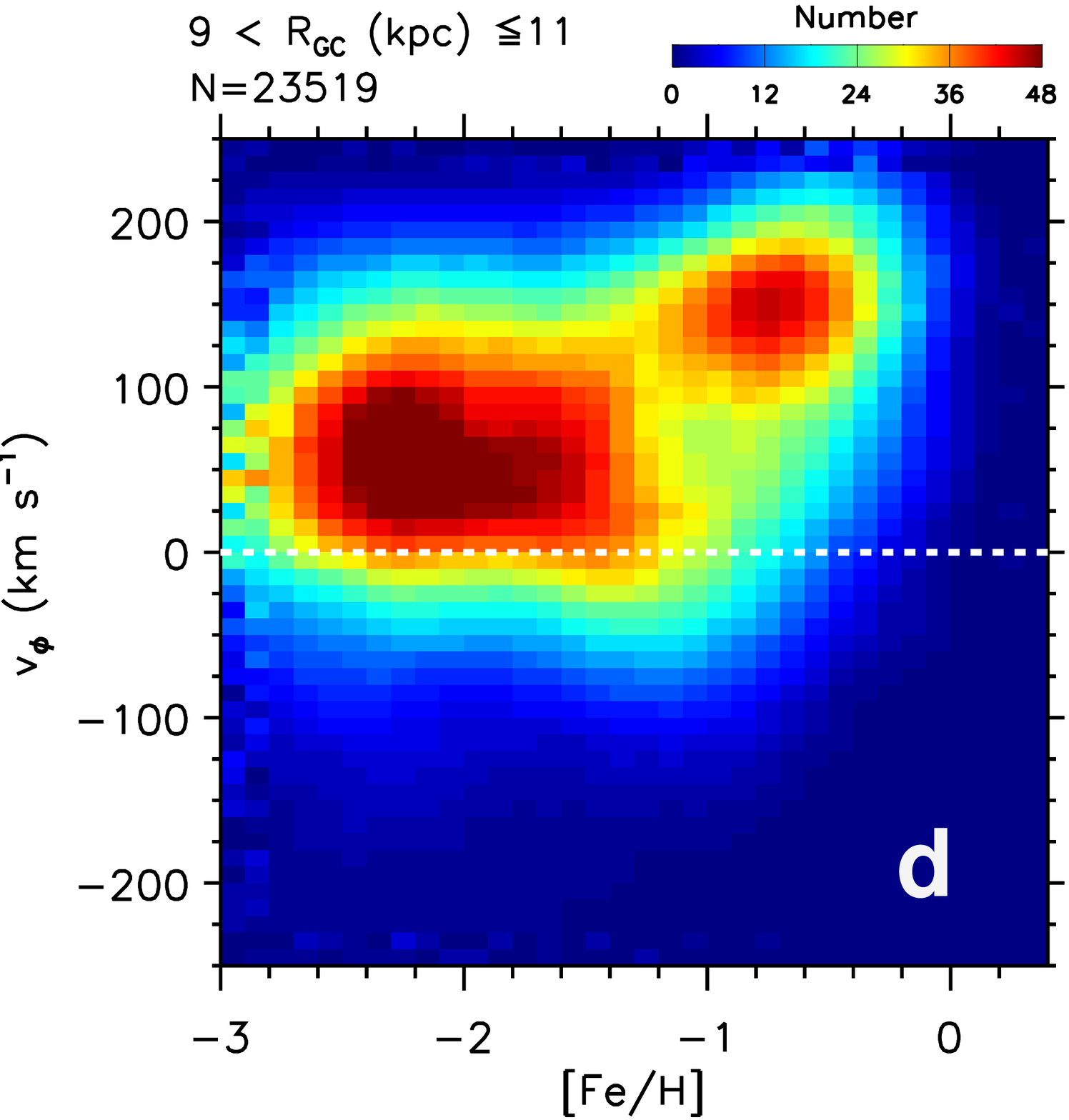}
\includegraphics[scale=0.36]{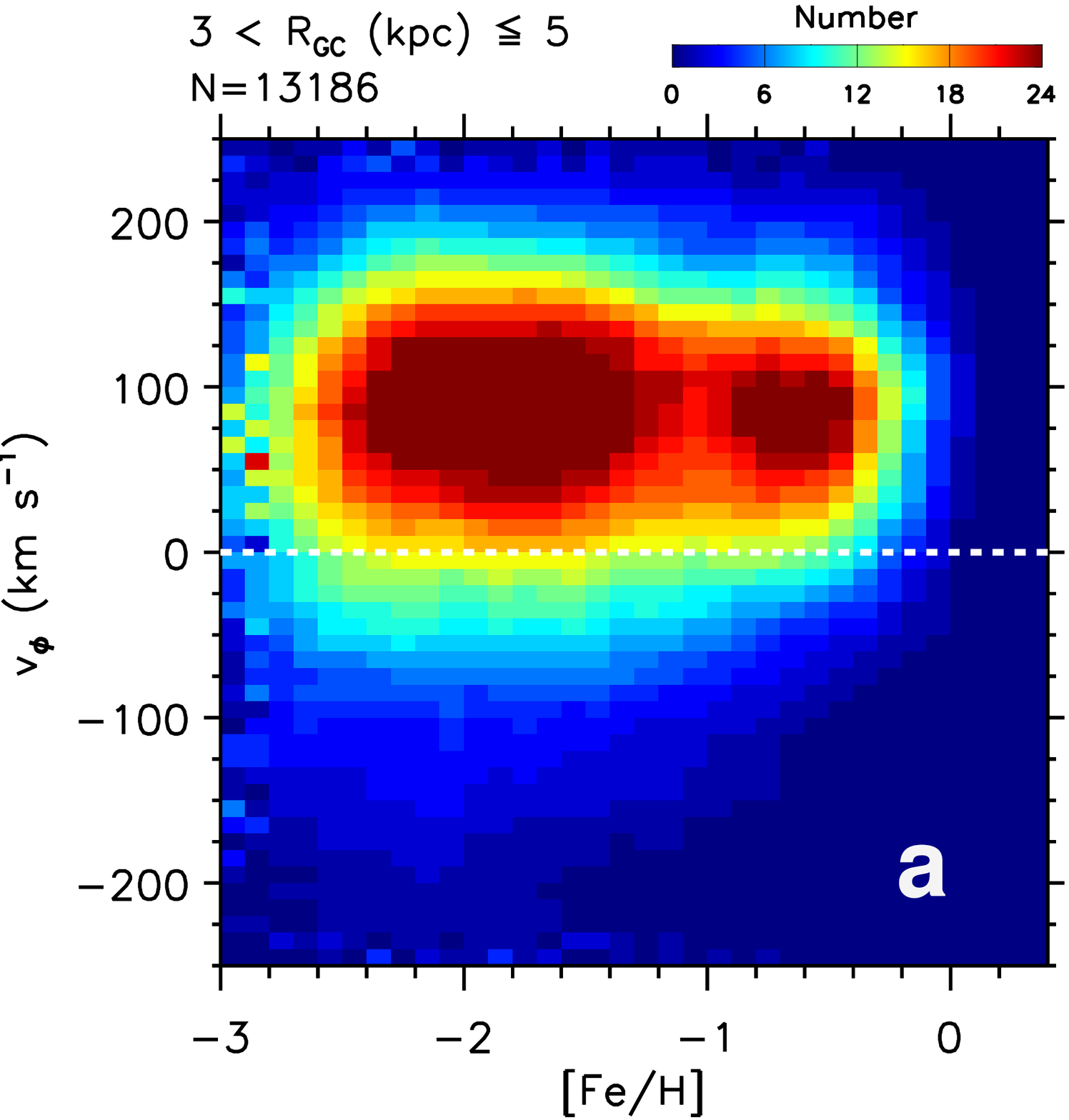}
\includegraphics[scale=0.36]{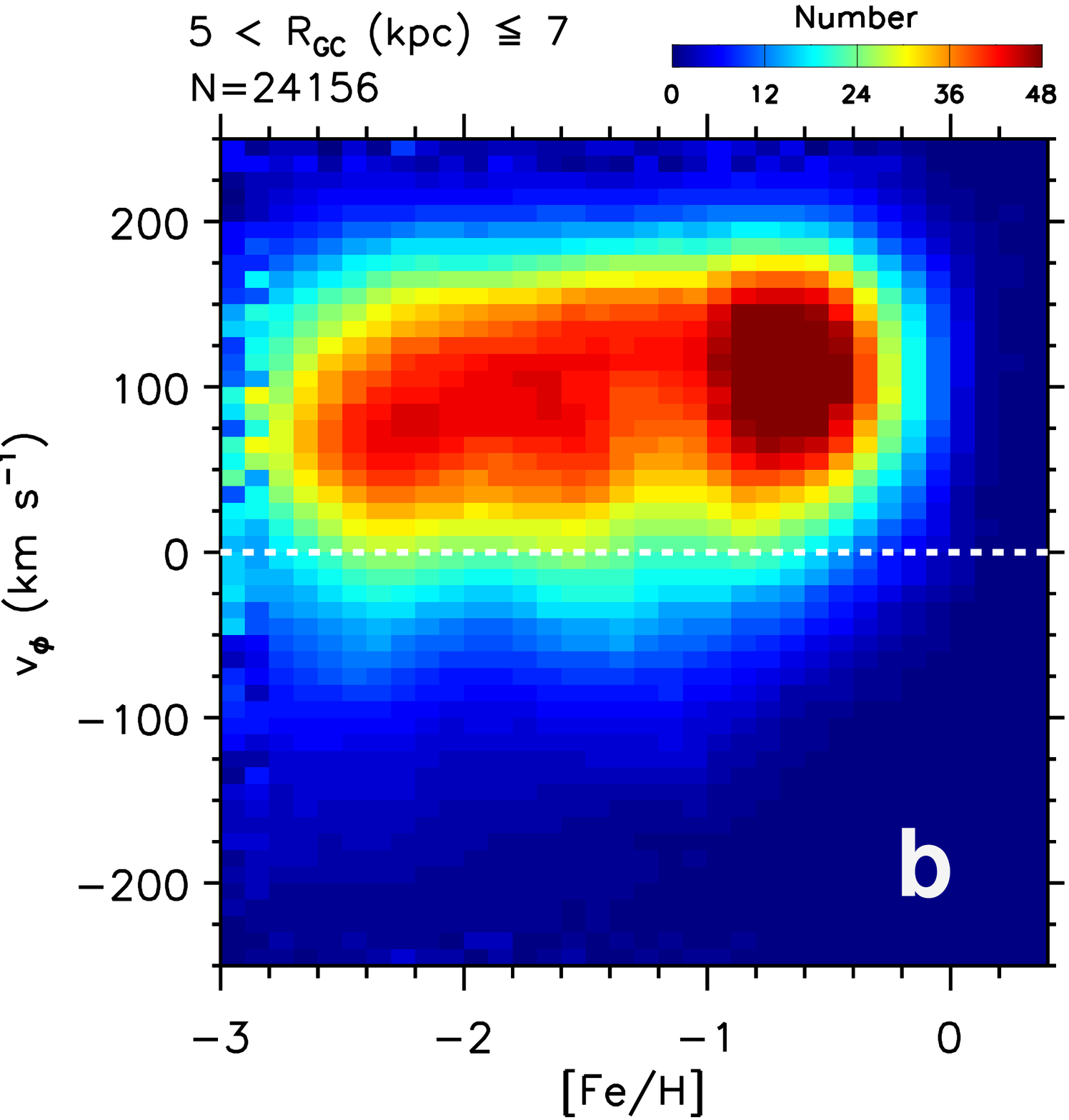}
\caption{The $\vphi$ vs.\ [Fe/H] distributions of the main photometric sample at $3.5 < |Z| \leq 6$~kpc, for different cuts in the distance from the Galactic center ($\rgc$) in the spherical coordinate system. Note that the panel on the upper right includes the full sample over $3 < \rgc \leq 13$ kpc.}
\label{fig:blueprint}
\end{figure*}

Figure~\ref{fig:blueprint} shows $\vphi$-[Fe/H] distributions of the main photometric sample as a function of $\rgc$. Only stars at large distances from the Galactic plane ($3.5 < |Z| \leq 6$~kpc) are displayed to minimize contributions from stars in the thin disk and TD. In the top right panel, where all stars from the other panels are combined, the metal-rich component ($\langle {\rm [Fe/H]} \rangle \sim-0.7$) with large rotation ($\meanvphi \sim 120\ \kms$) is well separated from the metal-poor component with slower rotation ($\meanvphi \sim 70\ \kms$). If we focus on the metal-rich clump ([Fe/H] $\sim -0.7$), Figure~\ref{fig:blueprint} reveals a rapid change of its mean rotation ($\meanvphi$) with $\rgc$, in which $\meanvphi$ increases from $\sim80\ \kms$ in the innermost $\rgc$ bin to $\meanvphi \sim150\ \kms$ at $11 < \rgc \leq 13$~kpc. However, a shift in $\meanvphi$ of a single population is unlikely to be the cause of this change. Instead, the large dispersion in $\vphi$ at intermediate $\rgc$ bins and a switch from $\meanvphi\sim80\ \kms$ to $\sim150\ \kms$ across the bins indicate that there are two overlapping populations with similar metallicities, but with distinct $\meanvphi$. This indicates that the clump is likely a superposition of the SP and TD, according to which the SP is more pronounced at smaller $\rgc$.

In Paper II \citep[see also][]{an:13,an:15}, we tentatively assigned metal-poor stars on prograde orbits centered at [Fe/H] $\approx-2.4$ to the MPH, in contrast to the more metal-rich component at [Fe/H] $\approx-1.6$ (MRH), {\`a} la a dichotomy between the IH and the OH \citep{carollo:07,carollo:10}, based on a two-dimensional Gaussian fit in the $\vphi$-[Fe/H] plane. While there is a significant overlap between the two components in Figure~\ref{fig:blueprint}, these metal-poor components also exhibit a dramatic change with $\rgc$. At small $\rgc$, it shows almost a uniform $\vphi$, independent of metallicity. However, at $11 < \rgc \leq 13$~kpc, the $\vphi$-[Fe/H] distribution reveals an extended metal-rich branch at $\vphi \sim 0\ \kms$ from the MRH. The same structure may exist in other $\rgc$ bins, but it is not as clearly seen as for $\rgc > 11$~kpc, because the region is dominated by stars from the SP. Figure~\ref{fig:blueprint} shows that the parameter space, delineated by $-1.0 \la {\rm [Fe/H]} \la -0.4$ and $-40 \la \vphi \la 60\ \kms$, is occupied not only by stars in the low-$\vphi$ tail of the SP but also by stars in the metal-rich extension of the MRH.

However, a question remains as to whether the GSE structure identified in our previous work represents the same entity as those found in the original analyses by \citet{belokurov:18} and \citet{helmi:18}. The Gaia Sausage refers to an oblate shape of a velocity ellipsoid of halo stars in a net prograde rotation \citep{belokurov:18}, while Gaia Enceladus was discovered as a group of stars with coherent kinematics on slightly retrograde and elongated orbits \citep{helmi:18}. The two studies reported somewhat discrepant $\meanvphi$ -- in more recent investigations, \citet{belokurov:20} found $\meanvphi=0\ \kms$ for the Gaia Sausage based on Gaia DR2 \citep[see also][]{naidu:20}, while \citet{helmi:20} refined the mean rotation of Gaia Enceladus to $\meanvphi \sim -20\ \kms$, based on a relatively small number ($N=29$) of stars that are on the [$\alpha$/Fe]-poor sequence at [Fe/H] $>-1.3$. On the other hand, in Papers~I and II, the centroid of GSE was measured as $-1.3 \la \langle {\rm [Fe/H]} \rangle \la -1.0$ and $\meanvphi \sim -60\ \kms$ at $4 < |Z| \leq 6$~kpc, where the contributions from the thin disk and TD are not as important as for lower $|Z|$ bins. Although it remains unclear whether the Gaia Sausage and Gaia Enceladus should be regarded as exactly the same stellar population, the mean velocities of GSE in Papers~I and II are too small (retrograde) compared to those derived in \citet{belokurov:18} and \citet{helmi:18}. It may be that modeling the MRH using a two-dimensional Gaussian was imperfect in Paper~II, leaving behind residual counts, or that the way we normalized the $\vphi$-[Fe/H] distribution in bins of $\vphi$ may have introduced a false trigger in Paper~I.

\begin{figure*}
\centering
\includegraphics[scale=0.40]{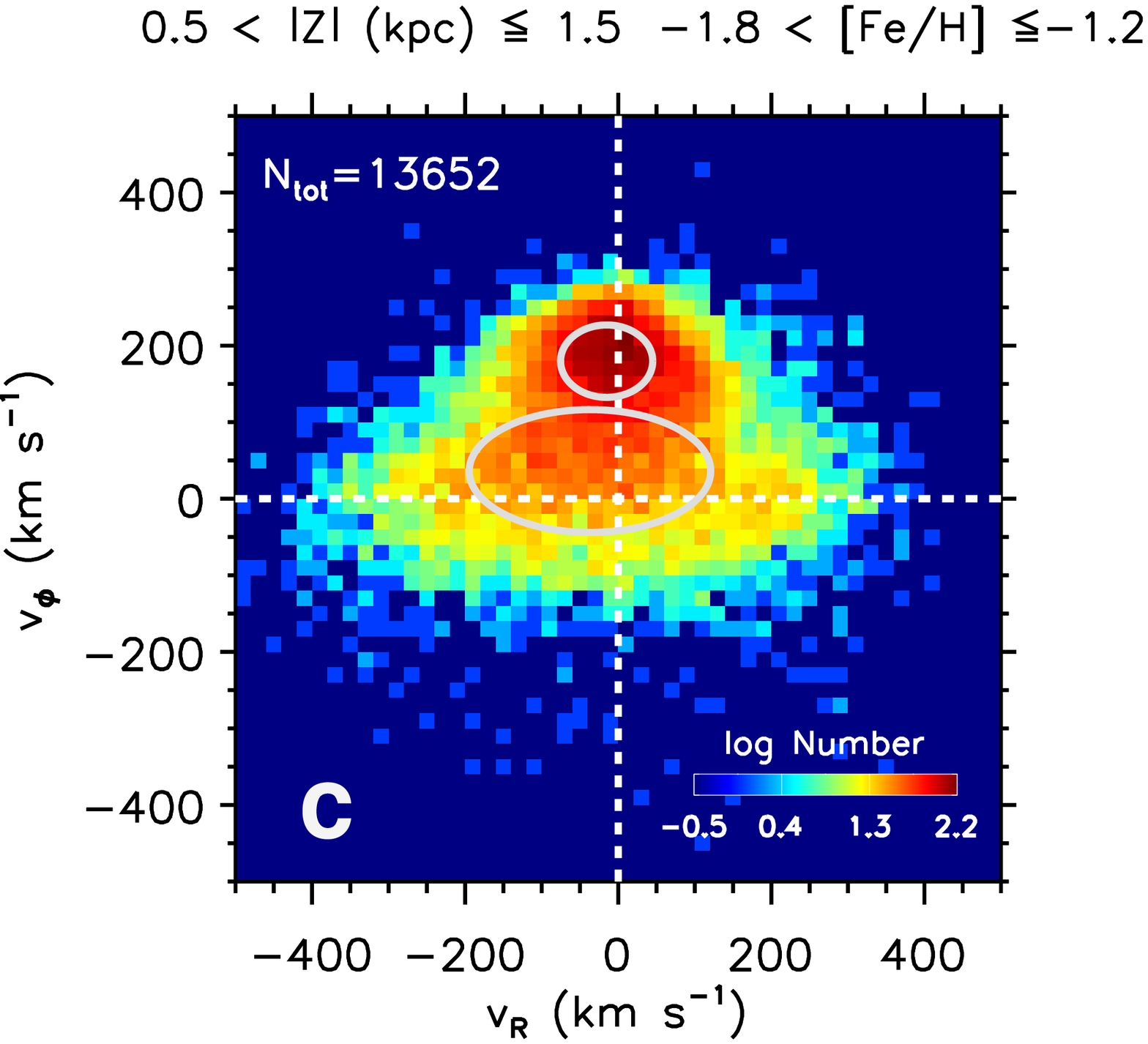}
\includegraphics[scale=0.40]{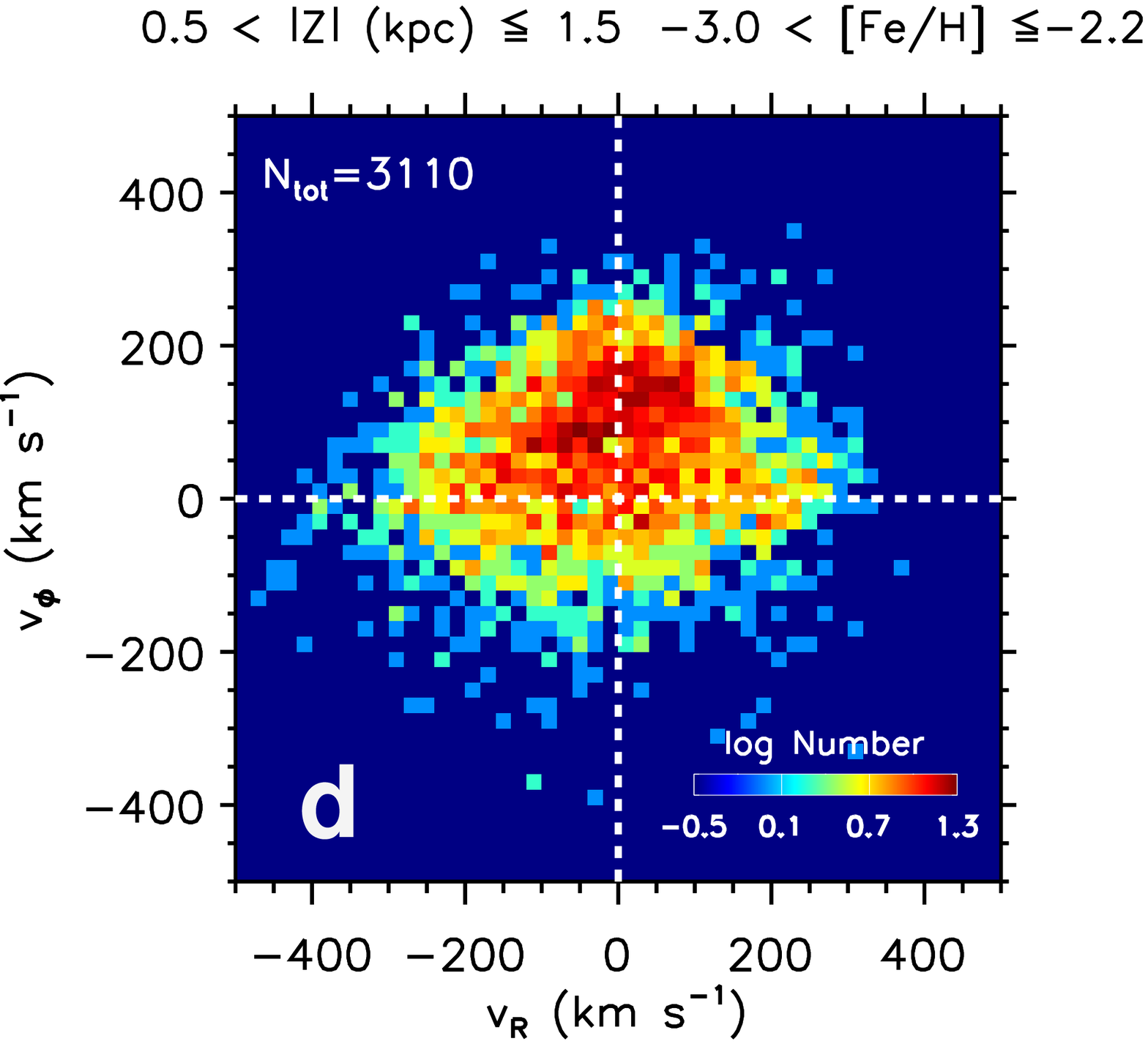}
\includegraphics[scale=0.40]{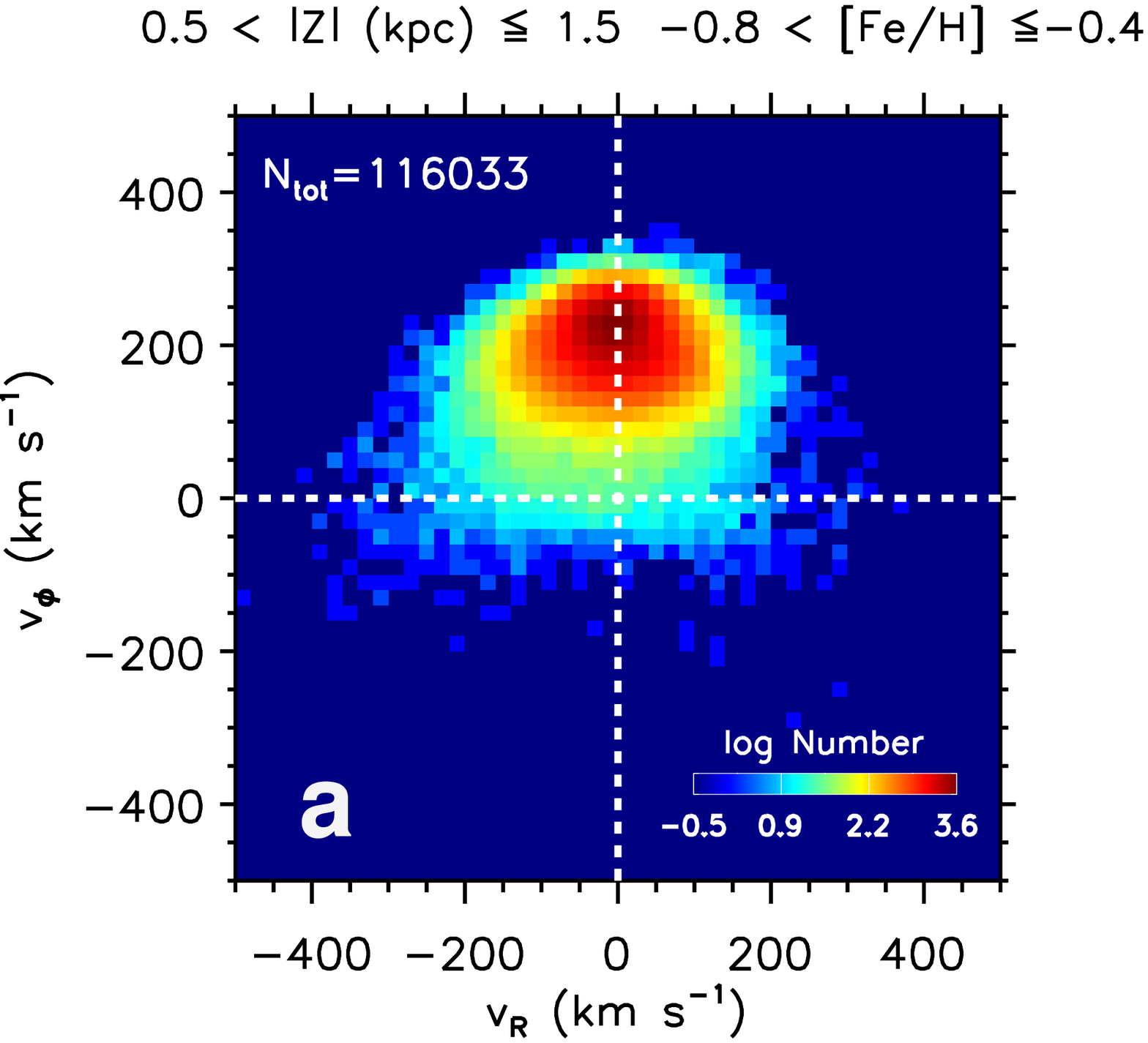}
\includegraphics[scale=0.40]{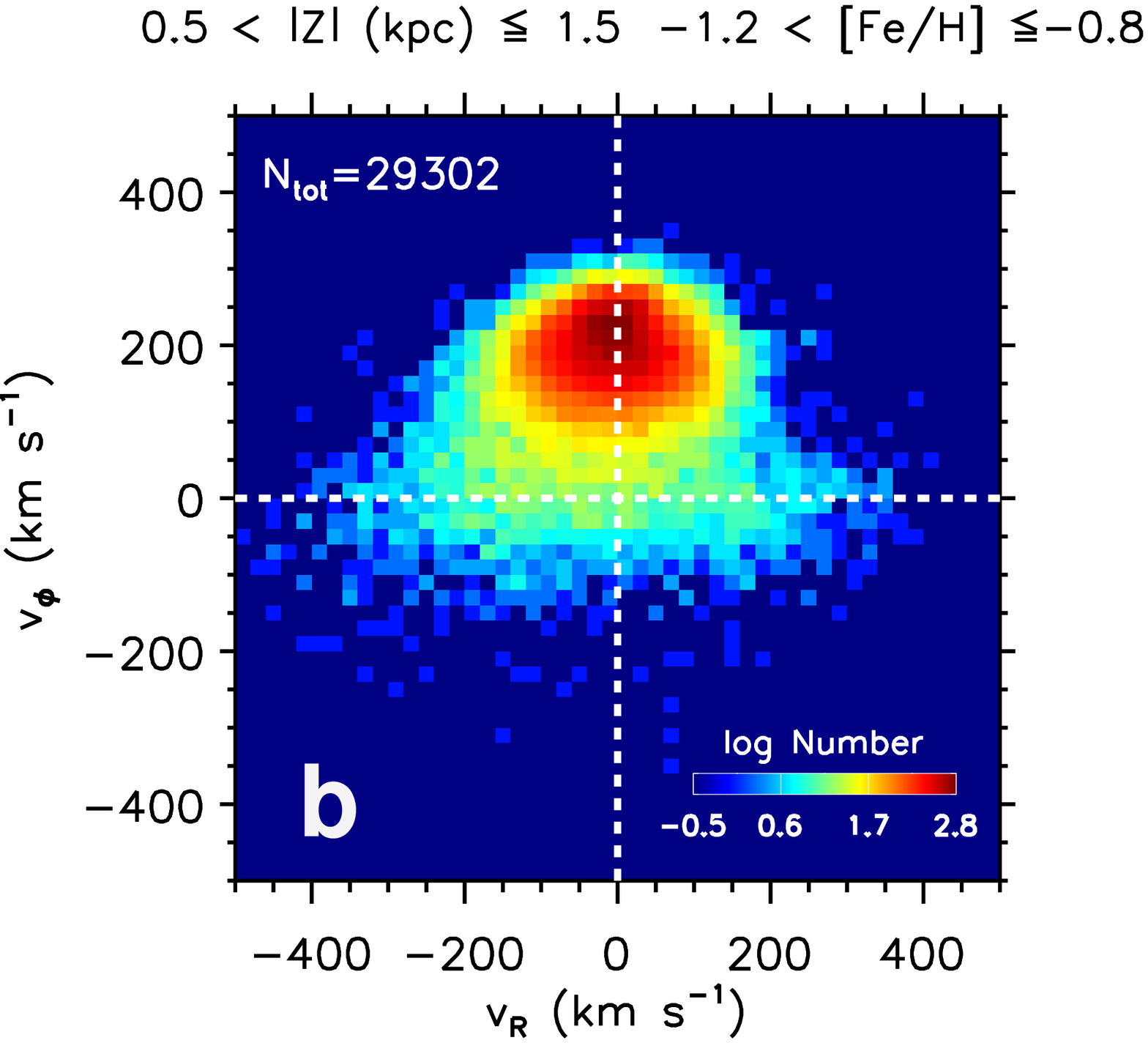}
\caption{The $\vphi$ vs.\ $v_R$ distributions (in Galactocentric cylindrical coordinates) of stars at $0.5 < Z \leq 1.5$~kpc in four different metallicity bins. Both velocities are computed from proper motions and parallaxes ($\sigma_\pi/\pi < 0.3$) from Gaia EDR3 at $|b| > 60\arcdeg$. Photometric metallicities derived with Gaia priors are used to divide stars into the above four subsamples. The gray ellipses in $-1.8 < {\rm [Fe/H]} \leq -1.2$ (top left) indicate the best-fitting Gaussians to the observed $v_\phi$--$v_R$ distribution.}
\label{fig:vrad}
\end{figure*}

On the contrary, the chemical and kinematical properties of GSE in \citet{belokurov:18,belokurov:20} are more similar to those of the MRH in our mapping. To investigate this further, we construct the velocity distribution of stars in $\vphi$ and $v_R$ in Figure~\ref{fig:vrad}, where $v_R$ is the radial velocity in the Galactocentric cylindrical coordinate system. We limit the sample to $|b| > 60\arcdeg$ in order to make velocity vectors computed from Gaia's proper motions parallel to the Galactic plane. We also use Gaia parallaxes ($\sigma_\pi / \pi < 0.3$) by restricting distances to $0.5 < |Z| \leq 1.5$~kpc, where $v_R$ asymptotically approaches the radial component of the velocity vector in spherical coordinates.

As found in \citet{belokurov:18,belokurov:20}, the metal-poor stars ($-1.8 < {\rm [Fe/H]} \leq -1.2$) with small net $\vphi$ exhibit a sausage-like, extended distribution in Figure~\ref{fig:vrad}, in addition to a group of stars in disk-like rotation ($\meanvphi \sim 180\ \kms$). Our $\vphi$-$v_R$ distribution is modeled using two-dimensional Gaussians (gray ellipses), from which we find $\sigmavphi = 80\ \kms$ and $\sigma_{v_R} = 158\ \kms$, which  are comparable to those of \citet{belokurov:18,belokurov:20}. However, we find a net rotation $\meanvphi = 36\ \kms$, which is higher than the $0\ \kms$ reported in \citet{belokurov:20}, who revised their original estimate, $20$--$30\ \kms$ \citep{belokurov:18}, by replacing photometric distances with Gaia DR2 parallaxes in their sample.

In Appendix~\ref{sec:scale}, we use stars with full three-dimensional kinematics in Gaia EDR3 to test the accuracy of our conversion from proper motions to $\vphi$. The comparison yields a scale difference of $\sim20\ \kms$ in $\vphi$ for stars with large velocity lags ($\vphi < 100\ \kms$), when derived from proper-motion data alone. The sense of the difference is that these velocities are systematically higher than true values. As also shown in Appendix~\ref{sec:scale}, our photometric distance scale is in overall good agreement with Gaia EDR3, although it is on average $2.5\%\pm0.5\%$ longer than Gaia distances for stars at $-2 < {\rm [Fe/H]} < -1$. The above two systematic uncertainties change $\vphi$ in opposite directions, so our $\vphi$ estimates are expected to be larger by $\sim15\ \kms$ for stars with large transverse motions, leading to a true mean velocity $\meanvphi\sim20\ \kms$ of the MRH (the sausage-like structure in the top left panel of Figure~\ref{fig:vrad}).

Our peak $\vphi$ for the MRH is comparable to the net rotation of the halo in the local volume ($< 4$~kpc) as traced by K giants ([Fe/H] $<-1$), $\meanvphi = 33\pm5\ \kms$ (\citealt{tian:19}, after correcting for the LSR difference; see also \citealt{deason:17,kafle:17}). Furthermore, \citet{iorio:21} identified field RR Lyrae stars that are likely associated with the Gaia Sausage and found that they exhibit a small net prograde rotation ($\approx15\ \kms$), which is most clearly seen inside of $\rgc \sim 10$~kpc. We conclude, based on similarities in the chemical and kinematical properties, that the MRH in our mapping represents the main body of GSE.

Having established a close link between the MRH and GSE, it is conceivable that the clump with a negative spin, which was previously attributed to GSE in Papers~I and II, partly forms the metal-rich extension of GSE. The existence of metal-rich ([Fe/H] $>-1$) members in GSE can also be seen in spectroscopic and photometric MDFs. \citet{helmi:18} presented an MDF of potential GSE members based on the Apache Point Observatory Galactic Evolution Experiment \citep[APOGEE;][]{majewski:17}, which shows a peak at [Fe/H] $\approx-1.4$ and includes a sizable fraction of metal-rich stars\footnote{A subset of their $\alpha$-rich stars from the same kinematic selection forms a secondary peak at [Fe/H] $\approx-0.6$, if they are true GSE members.} \citep[see also][]{mackereth:19}. Furthermore, \citet{naidu:20} presented their spectroscopic MDF of stars on high eccentric orbits, which qualitatively agrees with the result in the former study. These MDFs also agree with the photometrically derived MDF in \citet{feuillet:20}. Perhaps the scale difference of $\Delta \meanvphi \sim20\ \kms$ between the Gaia Sausage in \citet{belokurov:18,belokurov:20} and Gaia Enceladus in \citet{helmi:18} is caused by the fact that the sample in \citet{helmi:18} included a sizable fraction of metal-rich stars with systematically slower (retrograde) orbits than those of the metal-poor counterpart.

The spectroscopic MDFs in the above studies also revealed a long tail of metal-poor stars below [Fe/H] $\approx-2$. In this work, we mostly inherited the same identification of stellar populations as in Papers~I and II, but it has become increasingly clear that some fraction of stars in the MPH constitute the metal-poor tail of GSE. Below, we refer to GSE as a combination of (or a part of) the MPH and the MRH in Papers~I and II, as well as its metal-rich ([Fe/H] $> -1$) extension.

\section{Decomposition of the $\vphi$-[Fe/H] Distribution}\label{sec:decomp}

In the following analysis, we extract information on individual stellar populations from the MDFs and $\vphi$ distributions. The observed $\vphi$-[Fe/H] distributions are decomposed into subpopulations by fitting multiple Gaussian functions. Because the stellar populations significantly overlap with each other in metallicity space, the decomposition is performed in a two-stage process. First, the observed MDF for each $\vphi$ bin is fit using four principle components of metallicity groups. Then, for each metallicity group, the $\vphi$ distribution is decomposed into individual stellar populations using Gaussian mixtures. A decomposition using multiple Gaussians requires a sizable number of stars, so we inspect the distribution of each population in bins of either $|Z|$ or $\rgc$.

\subsection{Metallicity Distribution Functions}\label{sec:mdf}

\begin{figure*}
\centering
\includegraphics[scale=0.36]{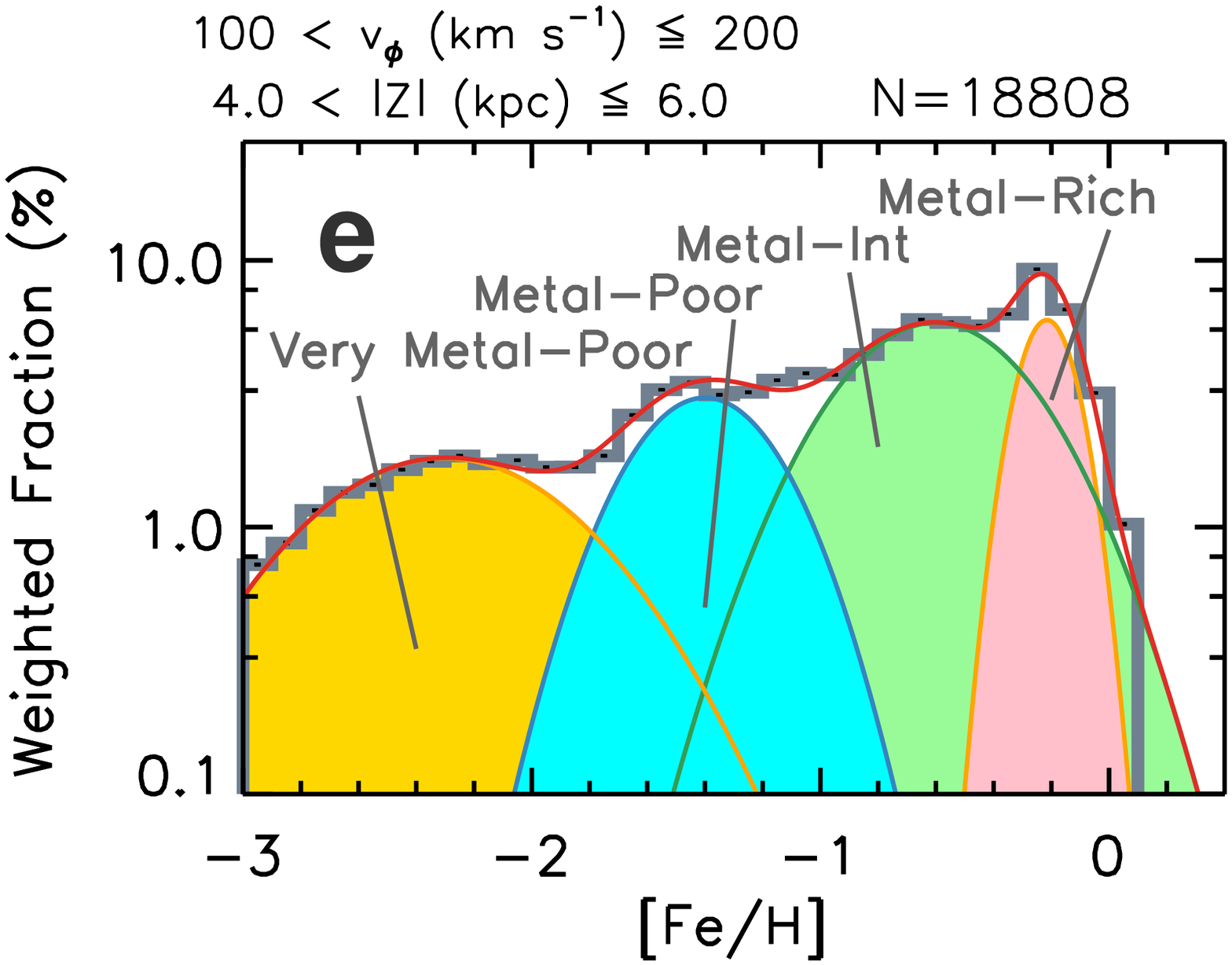}
\includegraphics[scale=0.36]{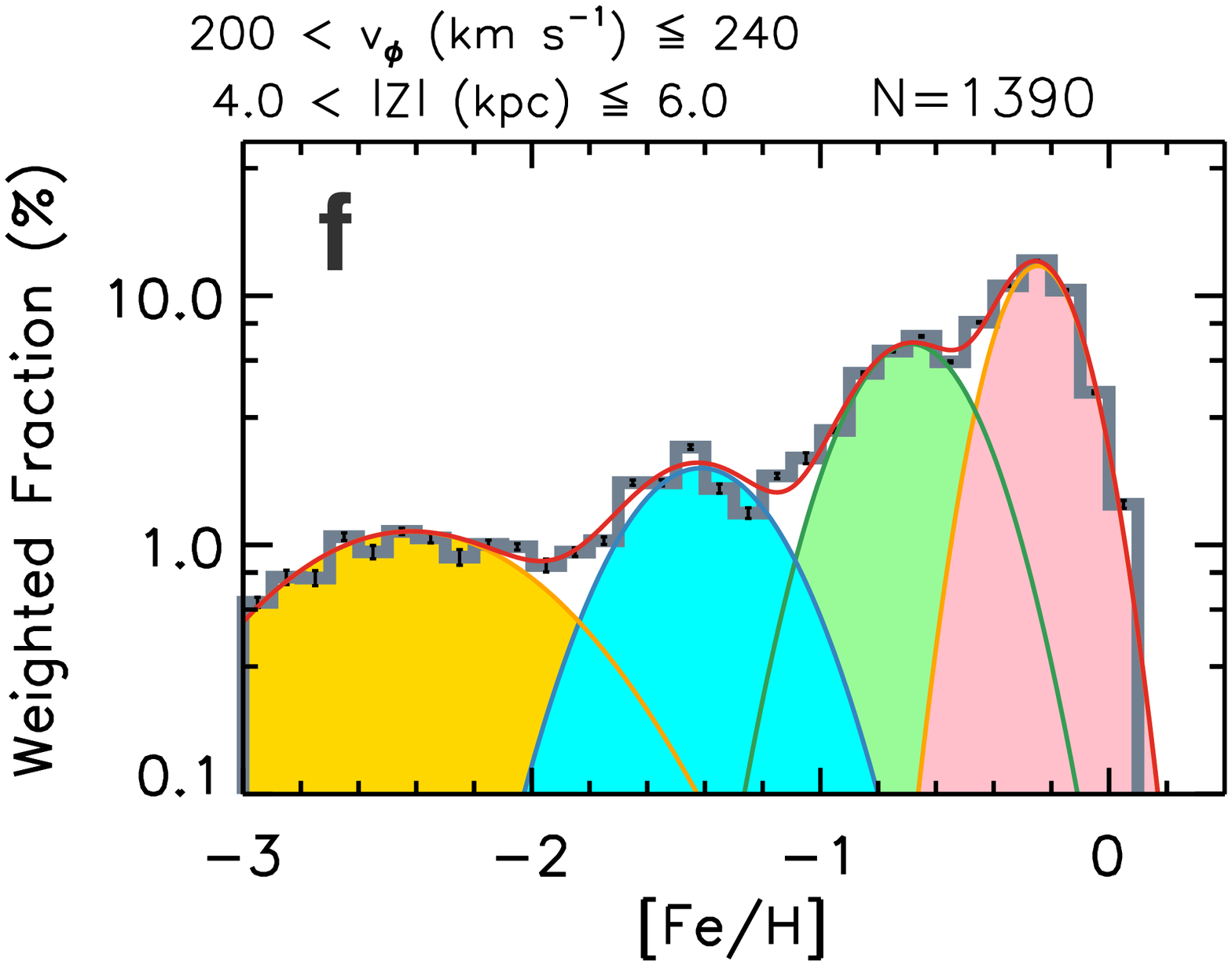}
\includegraphics[scale=0.36]{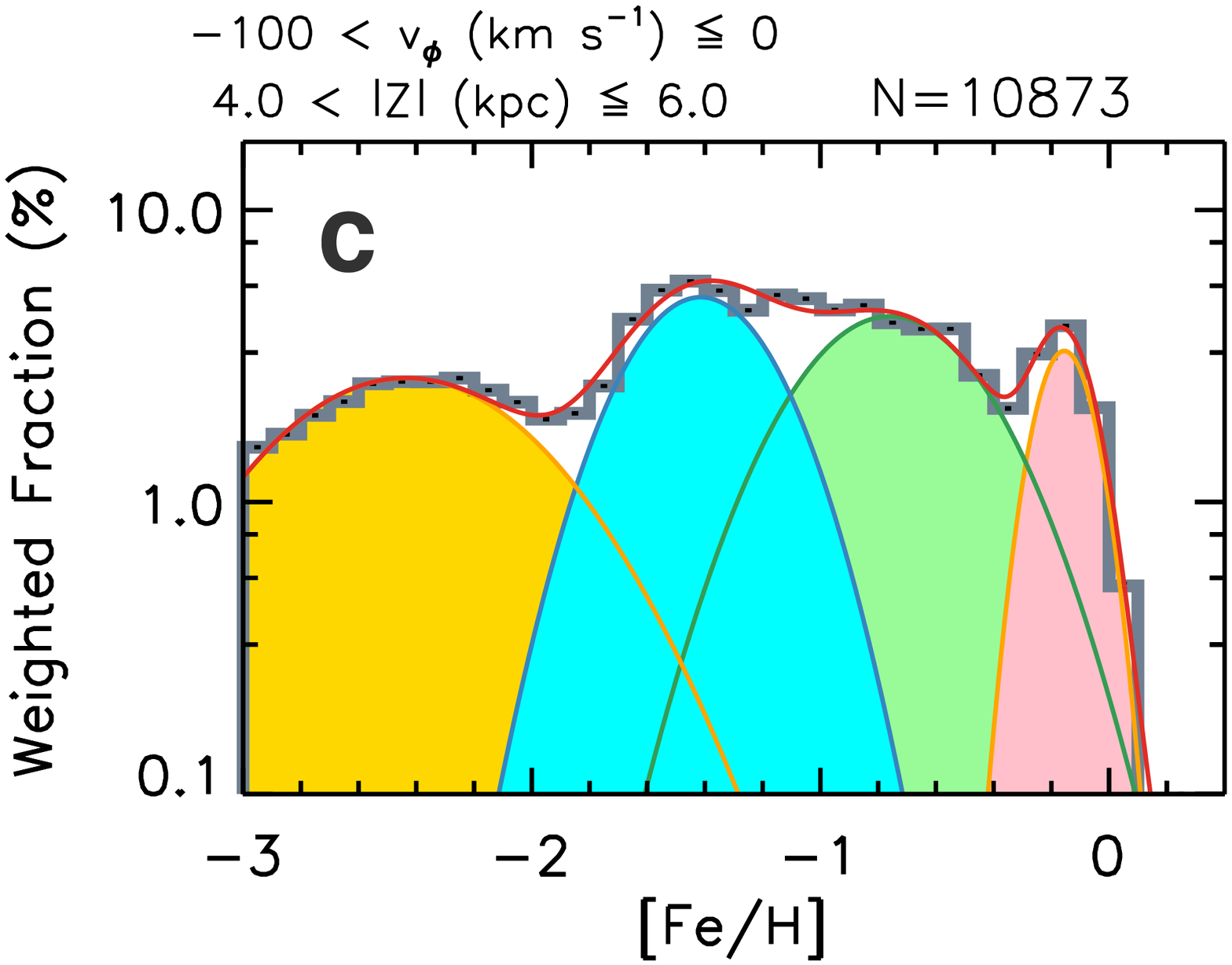}
\includegraphics[scale=0.36]{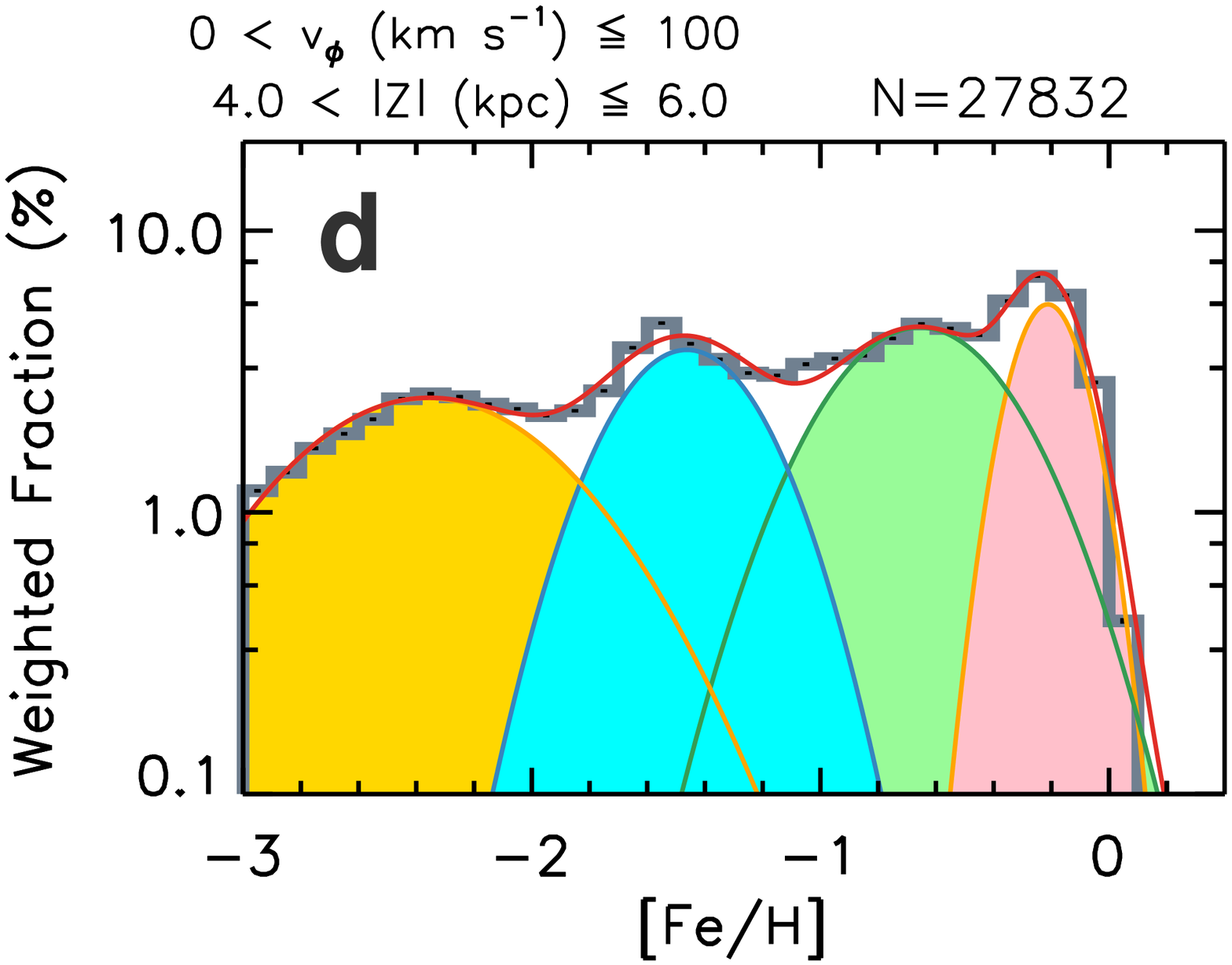}
\includegraphics[scale=0.36]{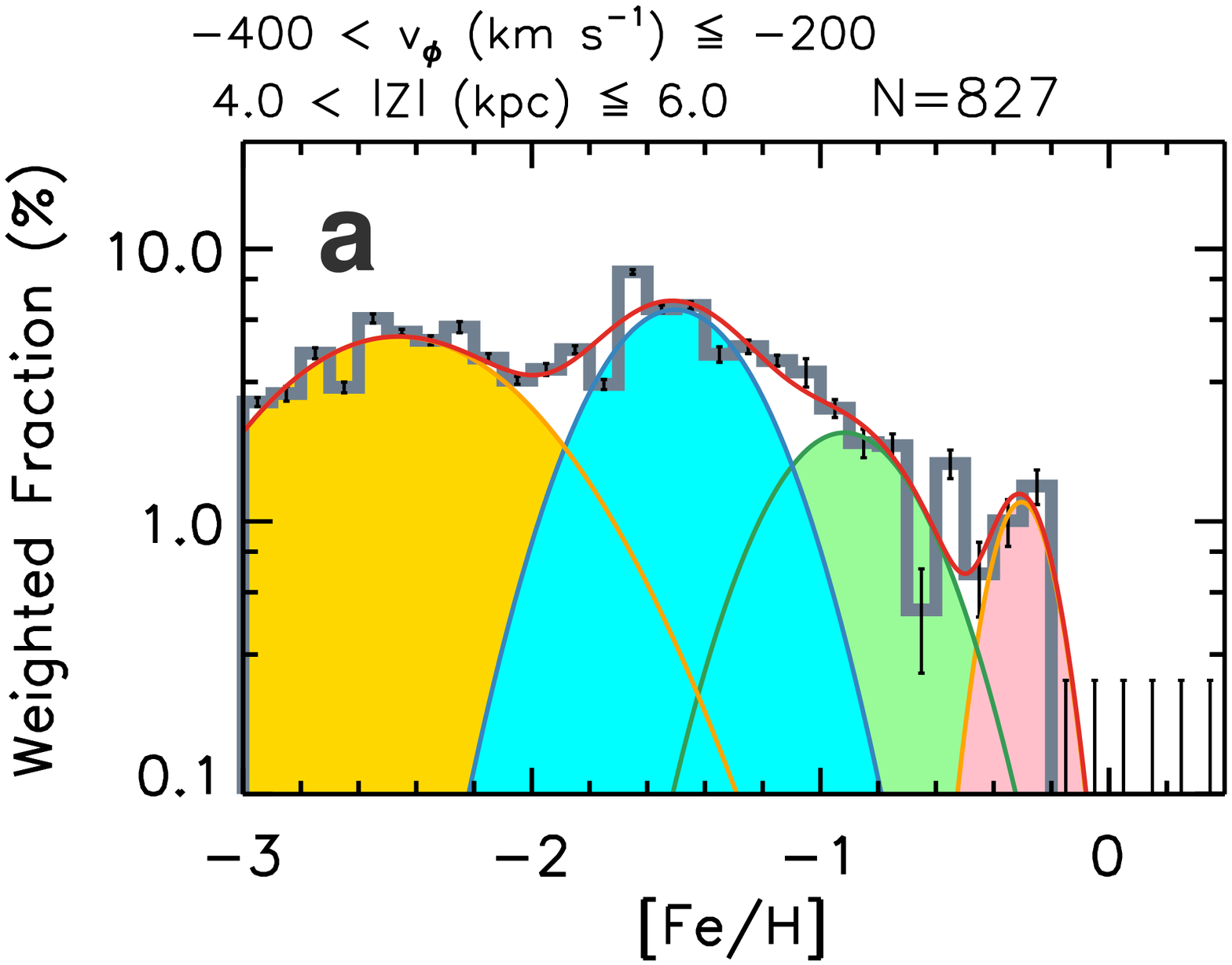}
\includegraphics[scale=0.36]{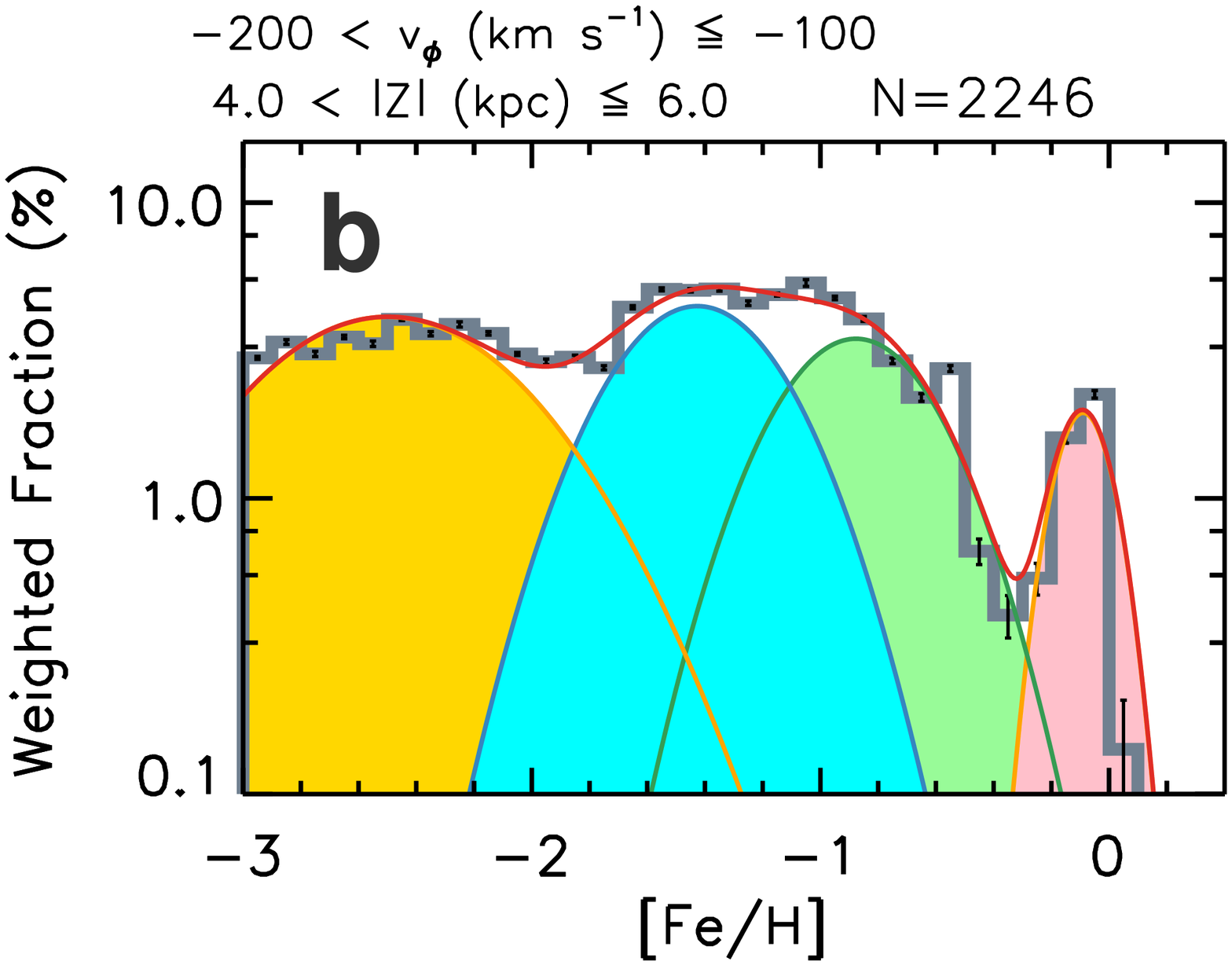}
\caption{Examples of the decomposition of MDFs. Each MDF is taken from slices of $|Z|$ ($4$--$6$~kpc in the above examples) and $\vphi$ and is fit using a principal set of four Gaussians that are used throughout this study. Error bars are shown from bootstrap sampling, although most bins have very small uncertainties owing to the broad $\vphi$ bins in the above examples.}
\label{fig:mdf}
\end{figure*}

\begin{deluxetable*}{lcccc}
\tablecaption{Nomenclature Used in This Work\label{tab:tab1}}
\tabletypesize{\scriptsize}
\tablehead{
   \colhead{} &
   \multicolumn{4}{c}{Metallicity Groups} \\
   \cline{2-5}
   \colhead{} &
   \colhead{Very Metal-Poor} &
   \colhead{Metal-Poor} &
   \colhead{Metal-Intermediate} &
   \colhead{Metal-Rich} \\
   \colhead{Stellar Population} &
   \colhead{$\langle {\rm [Fe/H]} \rangle \approx-2.4$} &
   \colhead{$\langle {\rm [Fe/H]} \rangle \approx-1.5$} &
   \colhead{$-1.0\la \langle {\rm [Fe/H]} \rangle \la-0.5$} &
   \colhead{$-0.4\la \langle {\rm [Fe/H]} \rangle \la-0.2$}
}
\startdata
Thin disk & \nodata & \nodata & maybe & \checkmark \\
Thick disk (TD) & \nodata & \nodata & \checkmark & \checkmark \\
Splash (SP) & \nodata & \nodata & \checkmark & \checkmark \\
Metal-weak thick disk (MWTD) & \checkmark & \checkmark & \nodata & \nodata \\
Gaia Sausage/Enceladus (GSE) & \checkmark\tablenotemark{\scriptsize a} & \checkmark\tablenotemark{\scriptsize b} & \checkmark & \checkmark \\
Retrograde halo structure (RHS) & \checkmark & \checkmark & \checkmark & \checkmark \\
\enddata
\tablenotetext{a}{Designated as the metal-poor halo (MPH) in Papers~I and II.}
\tablenotetext{b}{Designated as the metal-rich halo (MRH) in Papers~I and II.}
\end{deluxetable*}

Figure~\ref{fig:mdf} shows MDFs at $4 < |Z| \leq 6$~kpc in slices of $\vphi$. Given the observed change of the MDFs due to varying contributions from the underlying stellar populations, we find that at least four Gaussian functions are needed to fit the observed MDFs in all $\vphi$ bins (see also Paper~I). In Figure~\ref{fig:mdf}, these metallicity groups are indicated by different colors; we refer to these fitting components as (i) very metal-poor (${\rm \langle {\rm [Fe/H]} \rangle} \approx-2.4$; orange shaded), (ii) metal-poor (${\rm \langle {\rm [Fe/H]} \rangle} \approx-1.5$; blue shaded), (iii) metal-intermediate ($-1.0 \la {\rm \langle {\rm [Fe/H]} \rangle} \la -0.5$; green shaded), and (iv) metal-rich ($-0.4 \la {\rm \langle {\rm [Fe/H]} \rangle} \la -0.2$; red shaded) groups, according to their relative metal abundances. The existence of the metal-poor tail portrayed by the very metal-poor group is likely real and is supported by a plateau or a distinct peak in a spectroscopic MDF at a similar metallicity range \citep[e.g.,][]{fernandezalvar:17}; see below for additional evidence.

Table~\ref{tab:tab1} summarizes the information on the metallicity groups considered in the MDF fitting and their association with previously known individual stellar populations. As we show in the following $\vphi$ decomposition, each metallicity group contains multiple stellar populations, having different $\vphi$ and spatial distributions from one another. Most stellar populations can be found across different metallicity groups, as they cover a relatively wide metallicity range. The RHS is likely composed of multiple populations (see \S~\ref{sec:intro}) but simply indicates a fitted component in our Gaussian decomposition, which includes a large fraction of stars on retrograde orbits.

In each $\vphi$ bin, four Gaussians are fit simultaneously over the entire metallicity range, and the derived parameters are used as an initial guess on the next iteration. Discrepant Gaussian parameters are replaced by values obtained from nearby $\vphi$ bins. For the data used in the main analysis, finer steps in $\vphi$ of $10$~km~s$^{-1}$ are used with a window size of $10$~km~s$^{-1}$. Because each bin contains a lower number of stars than in Figure~\ref{fig:mdf}, some of the $\vphi$ bins have weaker constraints on Gaussian parameters. Based on experiments of varying ranges of $\vphi$ and $|Z|$, we find that the centroid and dispersion of Gaussians that represent the very metal-poor or the metal-poor metallicity group change little. For this reason, we limit the Gaussian parameters for these groups as follows in the whole survey volume: $-1.6 \leq \langle {\rm [Fe/H]} \rangle \leq-1.4$ and $0.25 \leq \sigma({\rm [Fe/H]}) \leq 0.35$ dex for the metal-poor group, and $-2.5 \leq \langle {\rm [Fe/H]} \rangle \leq -2.3$ and $0.35 \leq \sigma({\rm [Fe/H]}) \leq 0.45$ dex for the very metal-poor group.

The MDF decomposition using multiple Gaussians is weighted by uncertainties, which are computed by bootstrapping the observed sample. In each $|Z|$ or $\rgc$ bin, a sample of stars with the same size as the original data set is drawn at random with replacement, and the MDFs are constructed from $10^3$ simulated data sets. A standard deviation from the simulated MDFs is taken as the uncertainty in the observed MDF.

\begin{figure}
\centering
\includegraphics[scale=0.42]{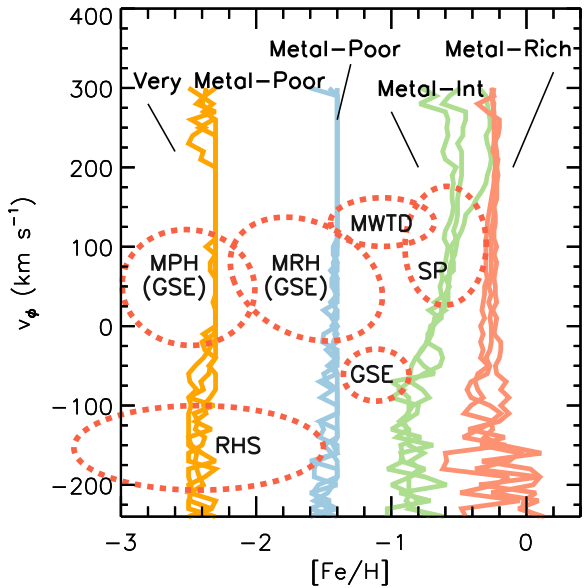}
\caption{Central [Fe/H] values of the four Gaussian components used in the decomposition of MDFs (see Figure~\ref{fig:mdf}). Four different sets of traces are shown from $1 < |Z| \leq 2$~kpc, $2 < |Z| \leq 3$~kpc, $3 < |Z| \leq 4$~kpc, and $4 < |Z| \leq 6$~kpc. Overlaid are {\it provisional} locations and extents of various stellar populations at $4 < |Z| \leq 6$~kpc (dotted ellipses), as presented in Paper~II (see its Figure~8; see also Figure~5 in Paper~I). Note that we reassign the MPH and MRH to GSE in the current work.}
\label{fig:groups}
\end{figure}

Figure~\ref{fig:groups} shows the central metallicity values for the four metallicity groups in the MDF fitting over the full $\vphi$ range. Contrary to other metallicity groups, the central metallicity of the metal-intermediate group shifts mildly with $\vphi$, from [Fe/H] $\sim-0.9$ at $\vphi \la -100\ \kms$ to [Fe/H] $\sim-0.5$ at $\vphi \ga 100\ \kms$. This is mainly due to a gradual shift in stellar populations from the RHS and GSE in the lower-$\vphi$ region to the canonical TD and SP in the higher-$\vphi$ region. For highly retrograde stars, the central [Fe/H] values of the metal-rich and metal-intermediate groups are rather uncertain, because such stars are rare in this portion of the phase space.

We note that our MDF decomposition is not intended to separate thin-disk and TD stars using Gaussians for the metal-intermediate and metal-rich groups. We simply employ them to fully describe the observed metal-rich portion of the MDF with a minimum set of Gaussians. In Appendix~\ref{sec:disk}, we provide a detailed look into the observed properties of disk stars based on our derived parameters, but we defer detailed modeling of their observed $\vphi$-[Fe/H] distributions to future studies, since the goal of this paper is to study the relative fractions of stellar populations in the Galactic halo. For the following analysis, we sum the fractions from the two metal-rich groups to construct a combined $\vphi$ distribution.

\subsection{$\vphi$ Distributions}\label{sec:vphi}

Figures~\ref{fig:vphi}--\ref{fig:vphi3} show the distributions of $\vphi$, as a function of $|Z|$ (left columns) and $\rgc$ (right columns), for each set of metallicity groups. The samples are divided into five distance bins in $\Delta |Z|=0.5$--$2$~kpc from $0.5$ to $6$~kpc from the Galactic plane, and in bins of $\Delta \rgc = 2$~kpc from $\rgc=3$~kpc to $13$~kpc. Stars within $|Z|<0.5$~kpc are not included in this study, because main-sequence stars in this volume are often brighter than the bright limit of the SDSS imaging survey, which can lead to a bias in the metallicity and $\vphi$ distributions. Similarly, although there are more distant stars ($|Z| > 6$~kpc) in the original sample, the upper distance limit is set to make the sample less influenced by a metallicity-dependent bias. The most distant bin ($4 < |Z| \leq 6$~kpc) has a wider $|Z|$ range ($\Delta |Z| = 2$~kpc) to keep as many stars as possible. In the right panels, the samples are limited to $|Z| > 3$~kpc, in order to reduce the contributions by disk stars and to better inspect halo populations. The $\vphi$ distributions are smoothed using a five-point boxcar average and are shown in terms of the fraction of stars in each $|Z|$ or $\rgc$ bin.

Below, each observed $\vphi$ distribution is modeled using Gaussian mixtures. The number of Gaussians in our decomposition is set primarily (but not exclusively) based on previous identification of stellar populations on the $\vphi$-[Fe/H]-$|Z|$ space (see Figure~5 in Paper~I and Figure~8 in Paper~II). This requires a minimum set of three to four stellar populations for each metallicity group (see also Table~\ref{tab:tab1}). Our approach follows the general practice in the literature but can be regarded as an improvement over the previous studies for the following reasons. First, the $\vphi$ distributions are extracted from individual metallicity groups, and therefore the assignment of a fitted component to each stellar population is more straightforward. Compared to using a simple metallicity cut to construct a $\vphi$ distribution, this approach can significantly lower the amount of contamination by other populations. Second, since the observed $\vphi$ distributions based on highly precise proper-motion measurements show a number of fine structures, their changes can be followed as one explores different $|Z|$ or $\rgc$ bins. This not only helps to identify individual stellar populations but also provides important constraints on the Gaussian parameters of each fitted component.

A detailed explanation is provided below for the decomposition of the $\vphi$ distributions for each metallicity group. We provide the cases for the very metal-poor and metal-poor groups together, because of their similar $\vphi$ distributions. For the metal-intermediate and metal-rich groups, we combine their $\vphi$ distributions, for the reason mentioned above.

\subsubsection{The Very Metal-poor and Metal-poor Groups}

\begin{figure*}
  \centering
  \textbf{Very Metal-Poor Group ($-2.5 < \langle {\rm [Fe/H]} \rangle < -2.3$)}
  \gridline{\fig{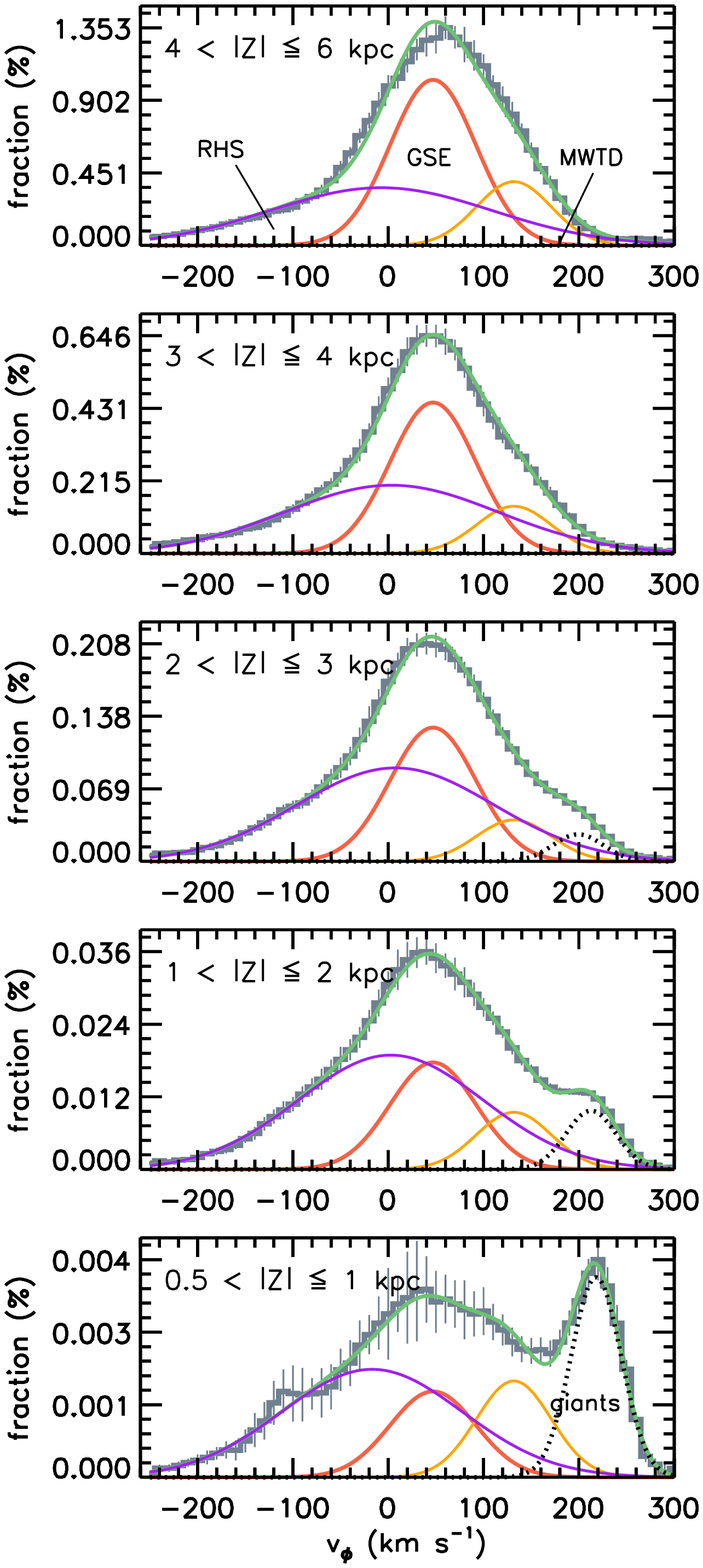}{0.5\textwidth}{\textbf{(a) Decomposition in bins of $|Z|$}}
  \fig{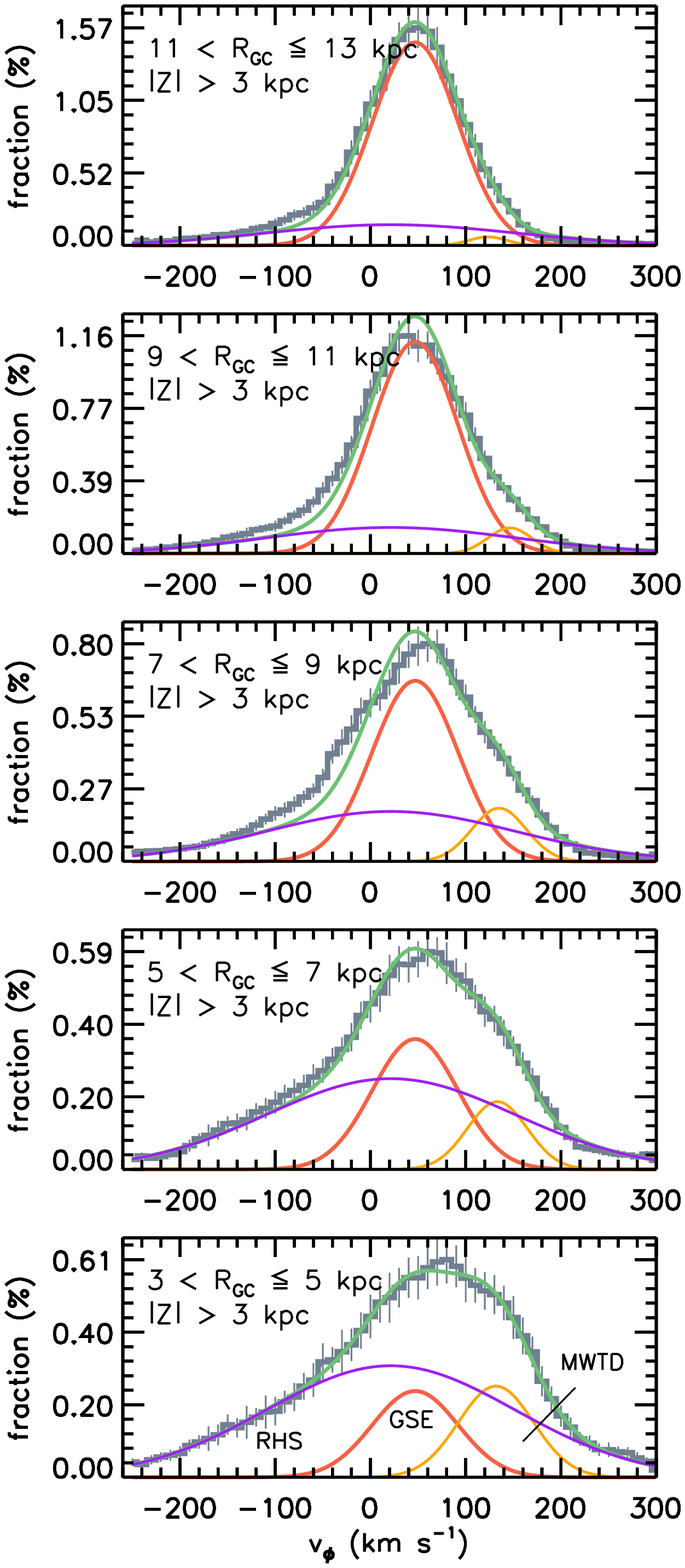}{0.5\textwidth}{\textbf{(b) Decomposition in bins of $\rgc$}}
  }
  \caption{The observed $\vphi$ distributions of stars in the very metal-poor group, as a function of $|Z|$ (left column) and $\rgc$ (right column). The ordinate indicates the fraction of stars in each $\vphi$ bin relative to the entire sample in a given $|Z|$ or $\rgc$ slice. Only stars at $|Z| > 3$~kpc are included in the right column of panels. Each distribution is decomposed using Gaussian mixtures as shown by various lines (see text); the green line shows the sum of all of these individual contributions. The black dotted line at small $|Z|$ indicates distant giants that are misidentified as main-sequence stars but are revealed as an agglomerate at $\meanvphi\sim220\ \kms$.}
\label{fig:vphi}
\end{figure*}

\begin{figure*}
  \centering
  \textbf{Metal-Poor Group ($-1.6 < \langle {\rm [Fe/H]} \rangle < -1.4$)}
  \gridline{\fig{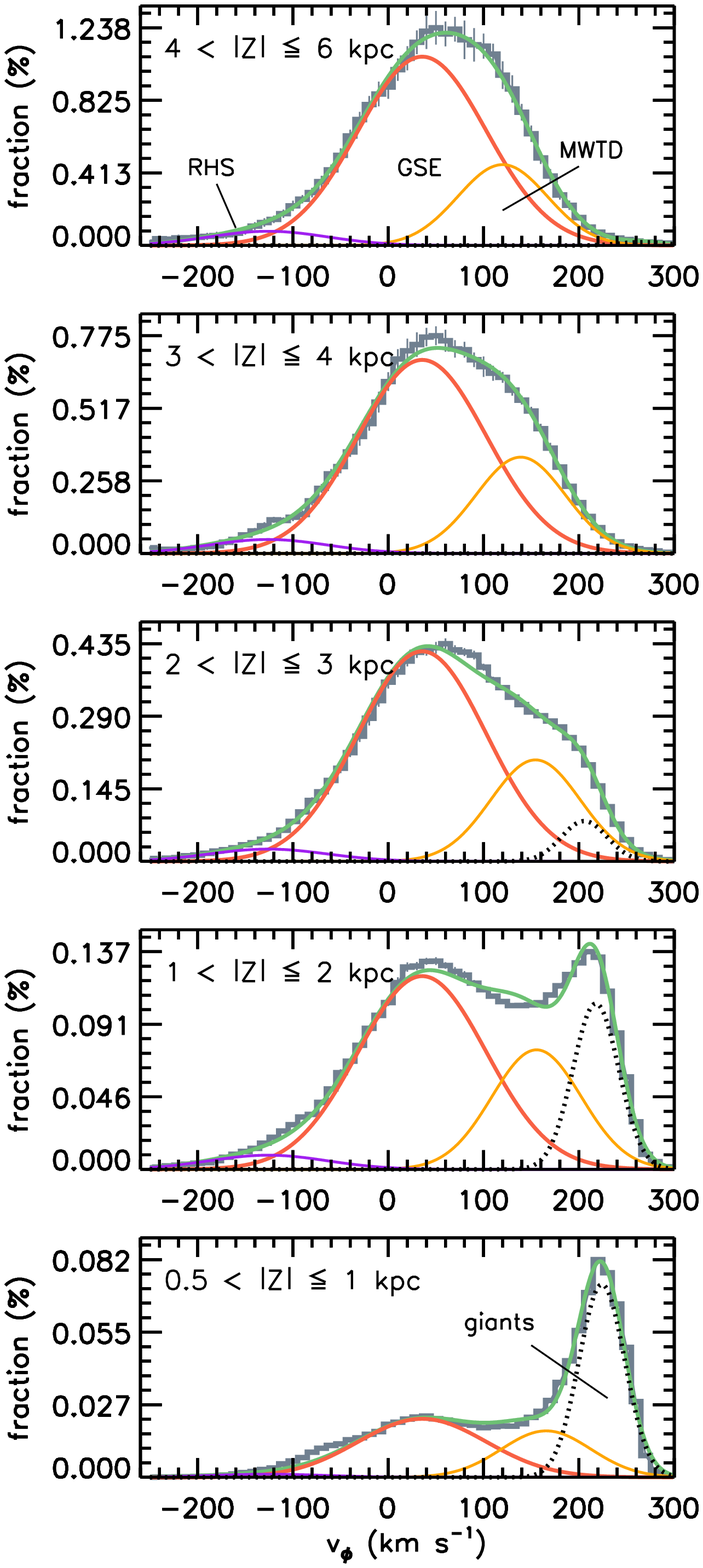}{0.5\textwidth}{\textbf{(a) Decomposition in bins of $|Z|$}}
  \fig{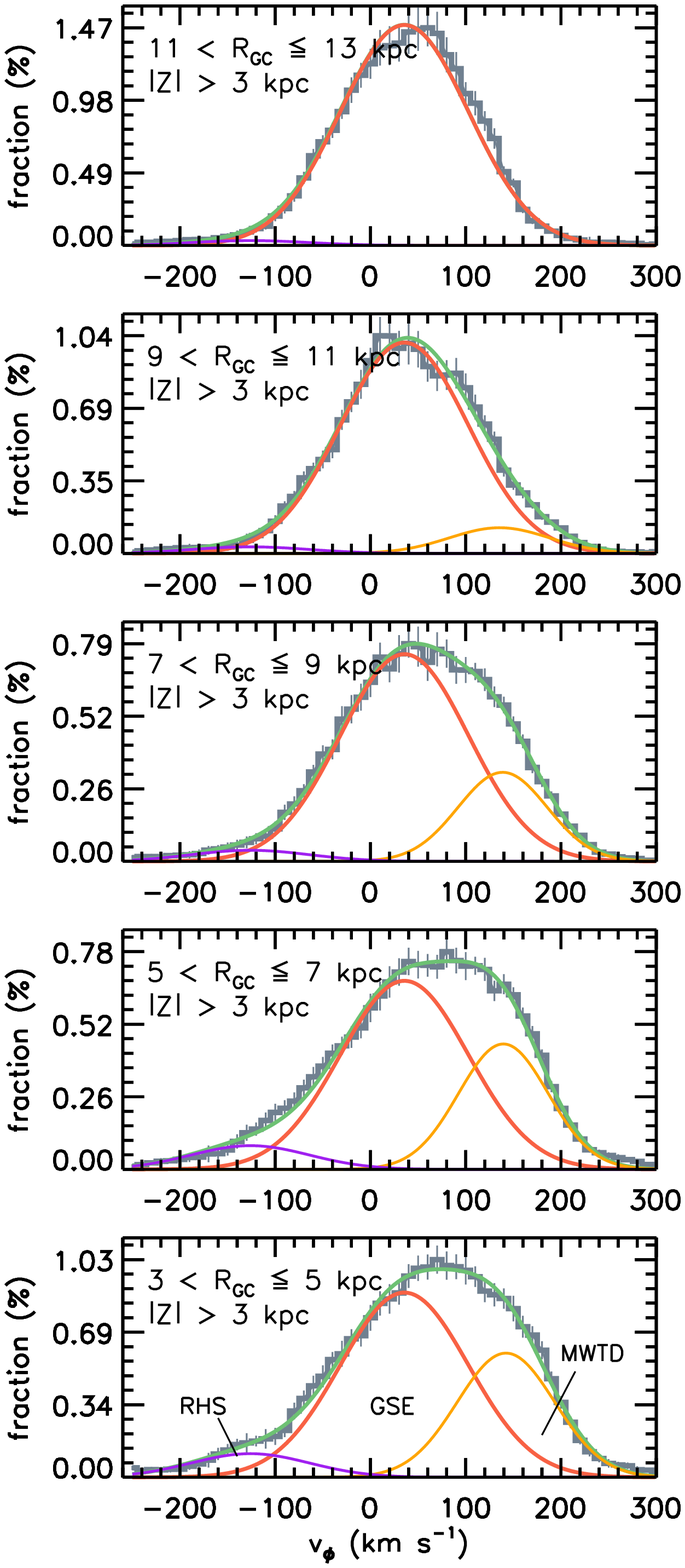}{0.5\textwidth}{\textbf{(b) Decomposition in bins of $\rgc$}}
  }
  \caption{Same as in Figure~\ref{fig:vphi}, but for the metal-poor group.}
  \label{fig:vphi2}
\end{figure*}

As shown in Figures~\ref{fig:vphi} and \ref{fig:vphi2}, the overall distribution of the metal-poor group resembles that of the very metal-poor group. Both $\vphi$ distributions in bins of $|Z|$ are apparently monotonic at $|Z| > 2$~kpc, with a small net prograde rotation ($\langle \vphi \rangle \sim 50\ \kms$) and smoothly decreasing fractions on each side of the peak. At $0.5 < |Z| \leq1$~kpc, an additional strong peak at $\sim220\ \kms$ is seen; however, these are mostly misidentified giants in our sample. Based on broadband photometry alone, distant ($> 10$~kpc) giants in the Galactic halo are difficult to discriminate from the far more numerous nearby main-sequence stars in our sample. Since $\ug$ colors of giants are systematically redder than main-sequence stars at a given $\gr$, giants that are misidentified as main-sequence stars may appear more metal-rich by up to $\Delta {\rm [Fe/H]} \sim 1$ dex \citep{an:13}. Furthermore, because of their intrinsically higher luminosities, distances to giants are greatly underestimated. Therefore, misidentified giants have a strong impact at $|Z| < 2$~kpc, but their strength diminishes rapidly at larger $|Z|$. Furthermore, because of the large  underestimation of their distances, their (true) large velocity lags are transformed into $\vphi \ga 200\ \kms$ in our mapping.

In comparison, a notable change in the $\vphi$ distribution can be seen along $\rgc$. At large $\rgc$, the distribution is narrow, with a standard deviation of $\sigmavphi \sim 45$--$70\ \kms$. However, it becomes more skewed toward higher $\vphi$ at smaller $\rgc$ with a progressively larger dispersion. The fraction of retrograde stars ($\vphi < 0\ \kms$) also increases rapidly with decreasing $\rgc$. Asymmetric photometric errors across the sky are very unlikely to produce such spreads, as these stars are located at nearly the same distances from the Sun. The observed differentiation not only proves the presence of multiple stellar populations in these metallicity groups but also provides strong evidence on their varying contributions with $\rgc$.

As seen from the intercomparison of the $\vphi$ distribution at $3 < \rgc \leq 5$~kpc and $11 < \rgc \leq 13$~kpc, at least three components emerge from the observed $\vphi$-$\rgc$ distribution -- the central component at a mean rotation $\meanvphi \sim 40\ \kms$, the wing component at $\meanvphi \sim 140\ \kms$, and the one containing a large reservoir of stars on retrograde orbits. In Papers~I and II, we also identified two components in the metallicity range covered by the very metal-poor and metal-poor groups -- a major component with prograde orbits and the much weaker group(s) of stars on retrograde orbits, which are dubbed as the MPH/MRH and RHS, respectively (see dashed curves in Figure~\ref{fig:groups}). On the other hand, the third component at $\meanvphi \sim140\ \kms$ was not clearly visible on the phase-space maps in Papers~I and II, which can be understood as a result of its relatively small contribution at $4 < |Z| \leq 6$~kpc.

We assign the central component to GSE and the retrograde component to the RHS, although the latter covers a wide range of $\vphi$, including a sizable fraction of stars on prograde orbits, as revealed in the following decomposition. The rapid evolution of the wing component at $\meanvphi \sim 140\ \kms$ can be interpreted as a varying contribution from the MWTD with $\rgc$, based on the similar $\meanvphi$ and metallicity to its original identification \citep[$100 \la \vphi \la 150\ \kms$ and $-1.8 \la {\rm [Fe/H]} \la -0.8$;][and references therein]{carollo:10,ruchti:11,beers:14,carollo:19}. Large errors in the measured quantities could be responsible for the appearance of this component at [Fe/H] $<-2$. However, the presence of such very metal-poor stars in the spectroscopic MWTD samples \citep{ruchti:11,beers:14} suggests that the above component from the very metal-poor group represents the metal-poor tail of the MWTD population. The MWTD may also include some stars from the Helmi stream \citep{helmi:99,dodd:21}, because of their similar metallicity and $\vphi$ distributions. However, its contribution is likely negligible compared to that of the MWTD \citep{naidu:20}.

The necessity for more than a single stellar population can also be seen from the kinematical properties of spectroscopically confirmed, very metal-poor ([Fe/H] $\la -2$) stars. According to recent studies \citep{dimatteo:20,sestito:20,venn:20,cordoni:21}, a large number of very metal-poor stars exhibit eccentric ($e \ga 0.8$) orbits, in contrast to significantly smaller eccentricities of stars at large positive or negative $\vphi$. Given the small $\vphi$ of these stars, they probably represent a metal-poor tail of GSE (see also \S~\ref{sec:blueprint}). This is also consistent with \citet{limberg:21a}, who found a number of dynamically tagged groups of very metal-poor stars with similar dynamical properties to GSE. Meanwhile, about $20\%$ of very metal-poor stars in the above studies are confined to $|Z| < 3$~kpc and show disk-like kinematics, many of which may have a close link to the MWTD.

The other important observational constraint from large spectroscopic surveys \citep{sestito:20,carter:21,cordoni:21,limberg:21b} is that the $\vphi$ distribution of these very metal-poor stars is skewed toward prograde rotation. Such a ``prograde bias'' seems to be a generic feature of galaxy formation simulations of the Milky Way analogs \citep{santistevan:20,sestito:21}. This is also clearly demonstrated in our data as well, in which stars on prograde orbits always dominate the sample in all $|Z|$ or $\rgc$ slices. For the very metal-poor group (Figure~\ref{fig:vphi}), the observed fraction of prograde stars is $\sim70\%$  at $2 < |Z| \leq 3$~kpc, where the contamination from giants can be neglected. This number is comparable to the $\sim65\%$ value reported by \citet{sestito:20}, whose sample is mostly confined to $|Z| \leq 3$~kpc. Our estimated fraction is also in good agreement with a finding in \citet{carter:21} that very metal-poor stars in their spectroscopic sample comprise $>70\%$ prograde stars at $1 \la |Z| \la 3$~kpc.

To estimate the fractional contributions of individual components, we decompose the $\vphi$ distributions using a set of three Gaussians (plus one extra component adopted for the background giants; see below), each of which is assigned to GSE, RHS, and MWTD, respectively. Individual components extracted from the following decomposition are shown by various lines in Figures~\ref{fig:vphi} and \ref{fig:vphi2}; the green line indicates the sum of all contributions. At $11 < \rgc \leq 13$~kpc, we assume that the distribution is dominated by GSE, as is evident from its single Gaussian-like appearance. The mean rotational velocity ($\meanvphi$) and the standard deviation ($\sigmavphi$) of the GSE are derived from this bin and are fixed in the other $\rgc$ bins (right columns in Figures~\ref{fig:vphi} and \ref{fig:vphi2}), while searching for the best-fitting set of other Gaussians. Reassuringly, $\meanvphi$ and $\sigmavphi$ of the ``Sausage'' in the $\vphi$-$v_R$ distribution of Figure~\ref{fig:vrad} are approximately the same as those found here. In the modeling for $|Z|$ bins (left columns), the $\meanvphi$ and $\sigmavphi$ for GSE (and the MWTD for the very metal-poor group) are taken from the best-fitting models in the right columns and are held fixed in the final derivation of Gaussian parameters.

The RHS is possibly related to a collection of several, recently claimed structures with large retrograde motions. For the very metal-poor group, the resulting fit requires a very wide spread of the RHS centered at nearly zero rotation, contrary to what its name originally meant. It is likely that the RHS centered at $\langle {\rm [Fe/H]} \rangle \sim -2.4$ includes Thamnos \citep{koppelman:19} and I'itoi \citep{naidu:20}. Our decomposition suggests that the RHS also comprises a significant number of very metal-poor stars on prograde orbits, some of which may be associated with the polar stream LMS-1 \citep{yuan:20}, also known as Wukong \citep{naidu:20}. For the very metal-poor group, the $\meanvphi$ and $\sigmavphi$ of the RHS are taken from $3 < \rgc \leq 5$~kpc and are used for the modeling in other $\rgc$ bins.

The metal-poor group has a significantly smaller fraction of stars in the RHS than does the very metal-poor group. However, there are still many stars seen with large retrograde orbits ($\vphi < -100\ \kms$), which cannot be modeled using the GSE component alone. Given the metallicity range of this group, such highly retrograde stars can tentatively be associated with Sequoia \citep{myeong:19} and Arjuna \citep{naidu:20}. When GSE and the MWTD components are constrained as described above, the RHS has $\meanvphi = -126\ \kms$ and $\sigmavphi = 62\ \kms$ at $2 < |Z| \leq 6$~kpc. To minimize the degree of freedom in the decomposition, these $\meanvphi$ and $\sigmavphi$ are used in all $|Z|$ and $\rgc$ bins.

At $0.5 < |Z| \leq 1$~kpc, the narrow peak from the misidentified giants is modeled using an additional Gaussian component, while holding $\meanvphi$ and $\sigmavphi$ of the other components fixed from $3 < \rgc \leq 5$~kpc. To avoid overfitting in the modeling, the $\meanvphi$ or $\sigmavphi$ of the background giants are fixed in the other $|Z|$ bins in the following iteration. The resulting fit (green line) shows small deviations from the observed data, which could be improved by relaxing the above conditions. Although our modeling is not as satisfactory as those obtained from fully unconstrained fits, we believe that limiting the degree of freedom as delineated above provides more physically meaningful results.

\subsubsection{The Metal-intermediate and Metal-rich Groups}

\begin{figure*}
  \centering
  \textbf{Metal-Int + Metal-Rich Groups ($-1.0 < \langle {\rm [Fe/H]} \rangle < -0.2$)}
  \gridline{\fig{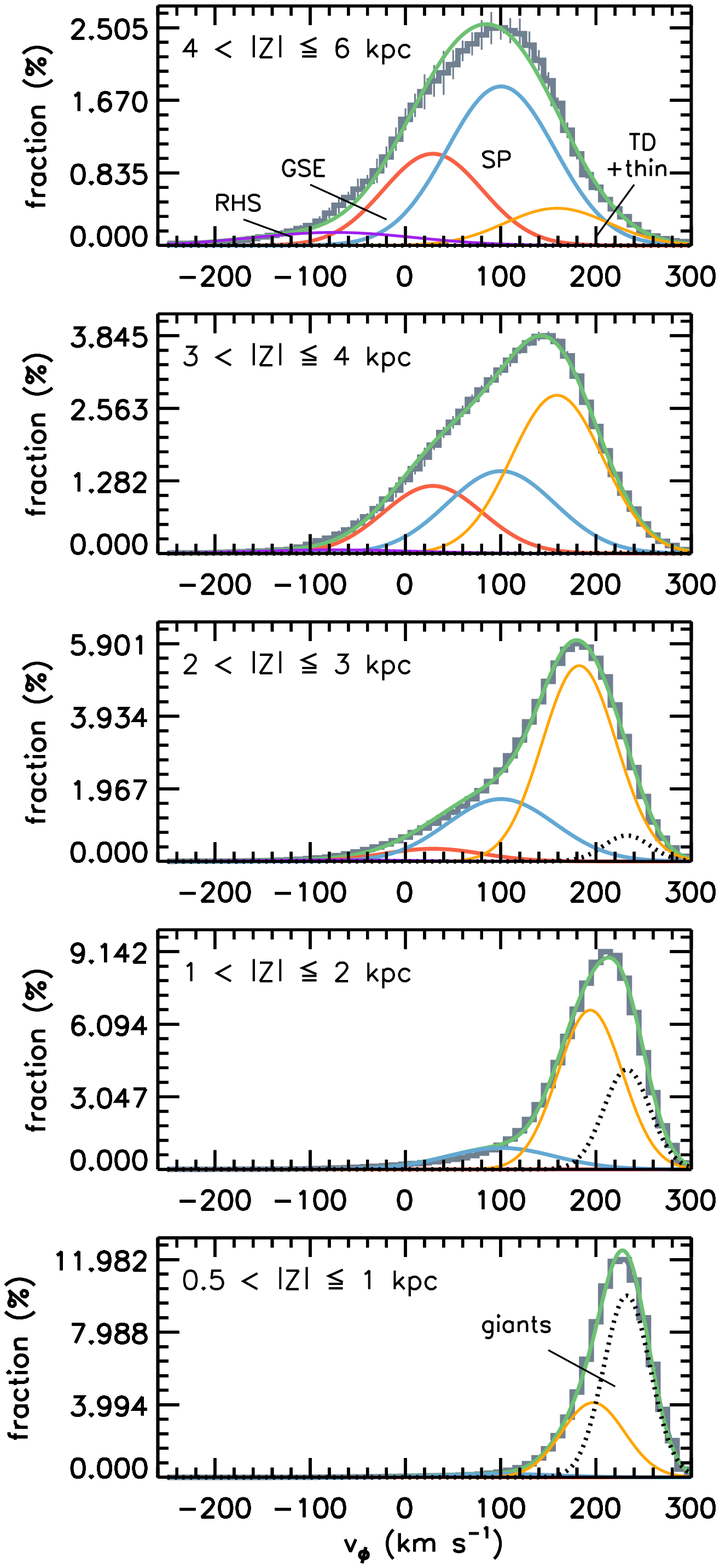}{0.5\textwidth}{\textbf{(a) Decomposition in bins of $|Z|$}}
  \fig{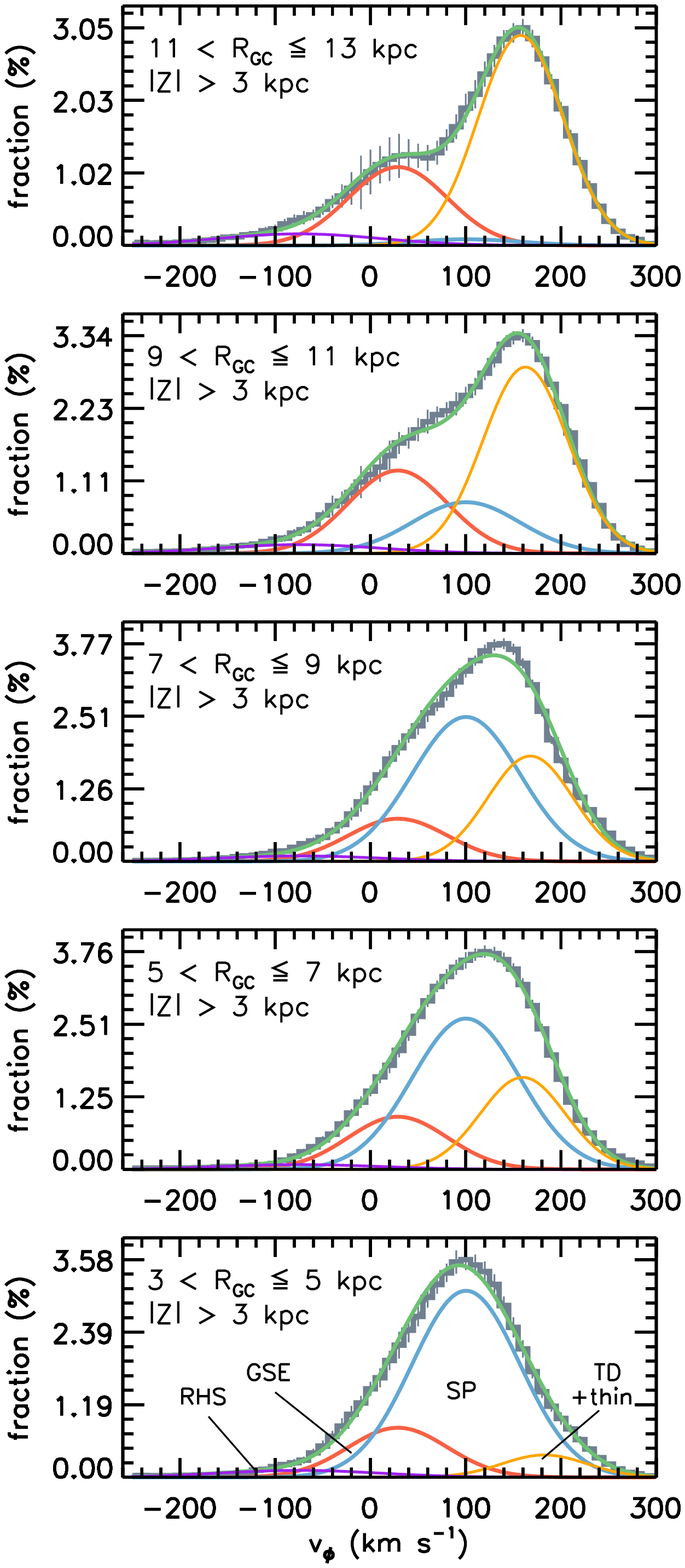}{0.5\textwidth}{\textbf{(b) Decomposition in bins of $\rgc$}}
  }
  \caption{Same as in Figure~\ref{fig:vphi}, but for the combined data set from the metal-intermediate and metal-rich groups.}
  \label{fig:vphi3}
\end{figure*}

As shown in Figure~\ref{fig:vphi3}, the sum of the metal-intermediate and metal-rich groups exhibits a dramatic change in the $\vphi$ distribution with $\rgc$ (right column). Most notably, it shows clear differentiation of a single-peak distribution at $3 < \rgc \leq 5$~kpc into a double-peak structure at $11 < \rgc \leq 13$~kpc, having peak values of $\vphi \sim 30\ \kms$ and $\sim160\ \kms$, respectively. As discussed in \S~\ref{sec:blueprint}, the high-$\vphi$ clump belongs to the canonical TD and/or thin disk, while we attribute the lower-$\vphi$ clump to GSE. The striking appearance of the double clumps is induced by the rapid reduction of a third stellar component (the SP) at large $\rgc$, which is distinct from the TD+thin disk and GSE (see \S~\ref{sec:blueprint}). Its existence can also be hinted at by a mild change of curvature at $\vphi \sim 100\ \kms$ as a function of $\rgc$. The spatial variation of each component with $\rgc$ suggests that the SP component is disconnected from GSE and/or TD+thin disk (see also Figure~\ref{fig:blueprint}).

We use the observed $\vphi$ distributions at $3 < \rgc \leq 5$~kpc and $11 < \rgc \leq 13$~kpc as a basis for constraining Gaussian parameters in the modeling. In order to constrain $\meanvphi$ and $\sigmavphi$ of the SP, we take advantage of the narrow $\vphi$ distribution in $3 < \rgc \leq 5$~kpc, which are subsequently adopted in the modeling throughout all volumes. The $\sigmavphi$ of the TD is constrained at $11 < \rgc \leq 13$~kpc, assuming that the large $\vphi$ peak is entirely contributed by this component. While holding a subset of Gaussian parameters fixed in this way, we derive the $\meanvphi$ and $\sigmavphi$ of GSE at $11 < \rgc \leq 13$~kpc, where the GSE component can be most clearly isolated. The resulting Gaussian center of the GSE component is $\meanvphi = 29\ \kms$ and is smaller than $\meanvphi \approx 36\ \kms$ and $47\ \kms$ from the metal-poor and very metal-poor groups, respectively. In addition, the RHS is required to describe the low-$\vphi$ tail, although it probably does not represent the same populations as those found in the two metal-poor groups. The $\meanvphi$ and $\sigmavphi$ of the RHS from $3 < \rgc \leq 5$~kpc are adopted in all bins. The Gaussian center and width of the misidentified giants are taken from $0.5 < |Z| \leq 1$~kpc.

\citet{fernandezalvar:19} found a group of metal-rich ([Fe/H] $\sim-0.5$) stars with intermediate [Mg/Fe] $\sim +0.1$, located at a large $|Z|$ up to $\sim10$~kpc. These stars exhibit nearly circular, large prograde orbits ($\meanvphi \ga 200\ \kms$). In all aspects, their properties are similar to those associated with Aleph in \citet{naidu:20}. Our $\vphi$ distributions do not show any clear signature of this structure, but such stars may constitute a part of the high-$\vphi$ tail at large $|Z|$, if they exist in large numbers in the local volume.

A number of studies noted a small fraction of stars with TD-like metallicities on retrograde orbits, unlike the majority of stars showing significant rotation \citep{nissen:10,hayes:18,haywood:18,koppelman:19,amarante:20}. They were interpreted as a low-velocity tail of the SP, which was possibly formed by dynamical heating of old disk stars by satellite accretion \citep[e.g.,][]{purcell:10,haywood:18,belokurov:20}. On the other hand, our decomposition indicates that the SP population is mostly confined to prograde orbits, while these counterrotating stars are more likely associated with the metal-rich extension of GSE (see also \S~\ref{sec:blueprint}). For the SP, our measured $\meanvphi \approx 100\ \kms$ is significantly larger than the estimate in \citet{belokurov:20}, who obtained $25\ \kms$ from stars with $-0.7 < {\rm [Fe/H]} < -0.2$ at $2 < |Z| \leq 3$~kpc. We note that they modeled a $\vphi$ distribution using three Gaussian components (thin disk, TD, and SP), but without GSE.

In the above decomposition, we denote ``TD+thin'' to account for a (significant) contribution from the thin disk, especially in lower $|Z|$ bins. Besides, \citet{carollo:19} found a large group of stars with elevated $\alpha$-element abundances with respect to those of the canonical TD at  $-1 \la {\rm [Fe/H]} \la -0.5$, which they attributed to the MWTD (possibly representing the metal-rich subgroup of this population). This implies that some fraction of the TD at large $|Z|$ in these metallicity groups is made up of MWTD stars (and probably thin-disk stars to a lesser degree). Nevertheless, the contamination fraction is not likely significant, given the observed $\meanvphi$ and $|Z|$ gradient of $\vphi$ that are similar to those found for the canonical TD from spectroscopic studies---at $1 < |Z| \leq 4$~kpc, the $\meanvphi$ of the TD+thin disk decreases at a rate of $\Delta \vphi / \Delta |Z| \approx -19\pm6\ \kms\ {\rm kpc}^{-1}$, which is not greatly different from spectroscopic estimates ($-9\ \kms\ {\rm kpc}^{-1}$ in \citealt{lee:11}; see also \citealt{yan:19} and references therein). In Appendix~\ref{sec:disk}, we examine the behaviors of the metal-rich stars in our photometric sample in more detail.

\begin{deluxetable*}{lccc}
\tablecaption{Rotational Velocities of Individual Components in bins of $|Z|$
\label{tab:tab2}}
\tabletypesize{\scriptsize}
\tablehead{
   \colhead{} &
   \multicolumn{3}{c}{Mean $\vphi$ ($\kms$), Dispersion ($\kms$), Fraction (\%)\tablenotemark{\scriptsize a}} \\
   \cline{2-4}
   \colhead{Stellar} &
   \colhead{Very Metal-poor Group} &
   \colhead{Metal-poor Group} &
   \colhead{Metal-int + Metal-rich Groups} \\
   \colhead{Population} &
   \colhead{$\langle {\rm [Fe/H]} \rangle \approx-2.4$} &
   \colhead{$\langle {\rm [Fe/H]} \rangle \approx-1.5$} &
   \colhead{$-1.0\la \langle {\rm [Fe/H]} \rangle \la-0.2$}
}
\startdata
\cline{1-4}
\multicolumn{4}{c}{$4 < |Z| (\rm kpc) \leq 6$} \\
\cline{1-4}
TD+thin disk & \nodata & \nodata & $[159.0\pm3.0],54.4\pm9.7,{\bf5.9\pm2.2}$ \\
SP & \nodata & \nodata & $[100.3\pm2.6],[56.6\pm5.5],{\bf26.1\pm5.9}$ \\
MWTD & $[131.7\pm7.0],[39.3\pm6.1],{\bf3.9\pm1.0}$ & $120.0\pm6.1,[47.0\pm4.4],{\bf5.4\pm2.7}$ & \nodata \\
GSE & $[46.9\pm2.7],[45.1\pm1.9],{\bf11.7\pm2.0}$ & $[35.6\pm2.0],[67.4\pm1.5],{\bf18.2\pm4.4}$ & $[28.7\pm5.9],[52.0\pm4.8],{\bf13.8\pm3.7}$ \\
RHS & $-7.6\pm9.2,116.3\pm5.5,{\bf10.5\pm2.1}$ & $[-126.0\pm40.1],[62.1\pm16.0],{\bf1.3\pm2.4}$ & $[-74.8\pm32.4],[84.5\pm10.8],{\bf3.2\pm1.6}$ \\
\cline{1-4}
\multicolumn{4}{c}{$3 < |Z| (\rm kpc) \leq 4$} \\
\cline{1-4}
TD+thin disk & \nodata & \nodata & $[159.0\pm3.0],48.1\pm2.1,{\bf33.7\pm6.7}$ \\
SP & \nodata & \nodata & $[100.3\pm2.6],[56.6\pm5.5],{\bf20.7\pm5.8}$ \\
MWTD & $[131.7\pm7.0],[39.3\pm6.1],{\bf1.4\pm0.6}$ & $138.9\pm7.5,[47.0\pm4.4],{\bf4.0\pm2.2}$ & \nodata \\
GSE & $[46.9\pm2.7],[45.1\pm1.9],{\bf5.1\pm1.3}$ & $[35.6\pm2.0],[67.4\pm1.5],{\bf11.7\pm3.9}$ & $[28.7\pm5.9],[52.0\pm4.8],{\bf15.6\pm3.3}$ \\
RHS & $3.2\pm8.6,112.1\pm4.7,{\bf5.7\pm1.7}$ & $[-126.0\pm40.1],[62.1\pm16.0],{\bf0.8\pm2.0}$ & $[-74.8\pm32.4],[84.5\pm10.8],{\bf1.4\pm0.9}$ \\
\cline{1-4}
\multicolumn{4}{c}{$2 < |Z| (\rm kpc) \leq 3$} \\
\cline{1-4}
TD+thin disk & \nodata & \nodata & $[182.4\pm9.6],39.0\pm2.2,{\bf54.5\pm2.1}$ \\
SP & \nodata & \nodata & $[100.3\pm2.6],[56.6\pm5.5],{\bf25.2\pm4.7}$ \\
MWTD & $[131.7\pm7.0],[39.3\pm6.1],{\bf0.4\pm0.4}$ & $154.6\pm11.4,[47.0\pm4.4],{\bf2.5\pm1.2}$ & \nodata \\
GSE & $[46.9\pm2.7],[45.1\pm1.9],{\bf1.5\pm0.4}$ & $[35.6\pm2.0],[67.4\pm1.5],{\bf7.5\pm2.2}$ & $[28.7\pm5.9],[52.0\pm4.8],{\bf4.7\pm2.7}$ \\
RHS & $7.5\pm7.2,103.1\pm4.7,{\bf2.4\pm0.6}$ & $[-126.0\pm40.1],[62.1\pm16.0],{\bf0.4\pm1.5}$ & $[-74.8\pm32.4],[84.5\pm10.8],{\bf0.9\pm0.4}$ \\
\cline{1-4}
\multicolumn{4}{c}{$1 < |Z| (\rm kpc) \leq 2$} \\
\cline{1-4}
TD+thin disk & \nodata & \nodata & $[194.0\pm14.0],34.0\pm4.0,{\bf77.1\pm3.1}$ \\
SP & \nodata & \nodata & $[100.3\pm2.6],[56.6\pm5.5],{\bf17.3\pm3.7}$ \\
MWTD & $[131.7\pm7.0],[39.3\pm6.1],{\bf0.1\pm0.1}$ & $155.9\pm11.2,[47.0\pm4.4],{\bf1.2\pm0.5}$ & \nodata \\
GSE & $[46.9\pm2.7],[45.1\pm1.9],{\bf0.3\pm0.1}$ & $[35.6\pm2.0],[67.4\pm1.5],{\bf2.8\pm0.6}$ & $[28.7\pm5.9],[52.0\pm4.8],{\bf0.0\pm0.6}$ \\
RHS & $2.6\pm8.1,97.0\pm4.6,{\bf0.6\pm0.1}$ & $[-126.0\pm40.1],[62.1\pm16.0],{\bf0.2\pm0.5}$ & $[-74.8\pm32.4],[84.5\pm10.8],{\bf0.4\pm0.2}$ \\
\cline{1-4}
\multicolumn{4}{c}{$0.5 < |Z| (\rm kpc) \leq 1$} \\
\cline{1-4}
TD+thin disk & \nodata & \nodata & $[196.9\pm15.2],33.0\pm4.3,{\bf90.4\pm3.5}$ \\
SP & \nodata & \nodata & $[100.3\pm2.6],[56.6\pm5.5],{\bf7.0\pm2.9}$ \\
MWTD & $[131.7\pm7.0],[39.3\pm6.1],{\bf0.0\pm0.1}$ & $166.2\pm13.6,[47.0\pm4.4],{\bf0.6\pm0.2}$ & \nodata \\
GSE & $[46.9\pm2.7],[45.1\pm1.9],{\bf0.0\pm0.0}$ & $[35.6\pm2.0],[67.4\pm1.5],{\bf1.0\pm0.3}$ & $[28.7\pm5.9],[52.0\pm4.8],{\bf0.6\pm0.5}$ \\
RHS & $-17.1\pm36.9,94.8\pm15.4,{\bf0.1\pm0.1}$ & $[-126.0\pm40.1],[62.1\pm16.0],{\bf0.0\pm0.1}$ & $[-74.8\pm32.4],[84.5\pm10.8],{\bf0.2\pm0.1}$ \\
\enddata
\tablenotetext{a}{Values shown in brackets are those derived from specific $|Z|$ or $\rgc$ bins but held fixed in the final iteration (see text). Fractional contributions (boldface) are computed in each $|Z|$ bin.}
\end{deluxetable*}

\begin{deluxetable*}{lccc}
\tablecaption{Rotational Velocities of Individual Components in bins of $\rgc$
\label{tab:tab3}}
\tabletypesize{\scriptsize}
\tablehead{
   \colhead{} &
   \multicolumn{3}{c}{Mean $\vphi$ ($\kms$), Dispersion ($\kms$), Fraction (\%)\tablenotemark{\scriptsize a}} \\
   \cline{2-4}
   \colhead{Stellar} &
   \colhead{Very Metal-poor Group} &
   \colhead{Metal-poor Group} &
   \colhead{Metal-int + Metal-rich Groups} \\
   \colhead{Population} &
   \colhead{$\langle {\rm [Fe/H]} \rangle \approx-2.4$} &
   \colhead{$\langle {\rm [Fe/H]} \rangle \approx-1.5$} &
   \colhead{$-1.0\la \langle {\rm [Fe/H]} \rangle \la-0.2$}
}
\startdata
\cline{1-4}
\multicolumn{4}{c}{$11 < R_{\rm GC} (\rm kpc) \leq 13$} \\
\cline{1-4}
TD+thin disk & \nodata & \nodata & $157.6\pm1.4,44.7\pm1.0,{\bf 33.2\pm1.1}$ \\
SP & \nodata & \nodata & $[100.3\pm2.6],[56.6\pm5.5],{\bf1.2\pm1.0}$ \\
MWTD & $123.2\pm11.4,20.0\pm10.4,{\bf0.3\pm0.4}$ & $142.2\pm12.5,51.2\pm8.9,{\bf0.0\pm0.3}$ & \nodata \\
GSE & $[46.9\pm2.7],[45.1\pm1.9],{\bf16.7\pm0.9}$ & $[35.6\pm2.0],[67.4\pm1.5],{\bf25.3\pm1.5}$ & $[28.7\pm5.9],[52.0\pm4.8],{\bf14.4\pm1.9}$ \\
RHS & $[21.2\pm5.3],[130.1\pm6.1],{\bf4.9\pm0.5}$ & $[-126.0\pm40.1],[62.1\pm16.0],{\bf0.5\pm0.5}$ & $[-74.8\pm32.4],[84.5\pm10.8],{\bf3.5\pm1.7}$ \\
\cline{1-4}
\multicolumn{4}{c}{$9 < R_{\rm GC} (\rm kpc) \leq 11$} \\
\cline{1-4}
TD+thin disk & \nodata & \nodata & $162.7\pm2.5,44.7\pm1.0,{\bf 32.1\pm2.3}$ \\
SP & \nodata & \nodata & $[100.3\pm2.6],[56.6\pm5.5],{\bf11.2\pm4.4}$ \\
MWTD & $146.2\pm8.2,23.0\pm3.2,{\bf0.8\pm0.3}$ & $135.4\pm17.8,52.7\pm7.0,{\bf1.6\pm1.0}$ & \nodata \\
GSE & $[46.9\pm2.7],[45.1\pm1.9],{\bf12.8\pm0.6}$ & $[35.6\pm2.0],[67.4\pm1.5],{\bf17.1\pm2.5}$ & $[28.7\pm5.9],[52.0\pm4.8],{\bf16.6\pm2.5}$ \\
RHS & $[21.2\pm5.3],[130.1\pm6.1],{\bf4.5\pm0.5}$ & $[-126.0\pm40.1],[62.1\pm16.0],{\bf0.5\pm1.4}$ & $[-74.8\pm32.4],[84.5\pm10.8],{\bf2.9\pm1.2}$ \\
\cline{1-4}
\multicolumn{4}{c}{$7 < R_{\rm GC} (\rm kpc) \leq 9$} \\
\cline{1-4}
TD+thin disk & \nodata & \nodata & $167.9\pm3.5,44.7\pm1.0,{\bf 20.5\pm2.1}$ \\
SP & \nodata & \nodata & $[100.3\pm2.6],[56.6\pm5.5],{\bf35.6\pm5.2}$ \\
MWTD & $134.9\pm5.9,28.5\pm2.6,{\bf1.4\pm0.3}$ & $138.7\pm7.2,46.9\pm4.2,{\bf3.8\pm0.8}$ & \nodata \\
GSE & $[46.9\pm2.7],[45.1\pm1.9],{\bf7.6\pm0.6}$ & $[35.6\pm2.0],[67.4\pm1.5],{\bf12.7\pm2.0}$ & $[28.7\pm5.9],[52.0\pm4.8],{\bf9.7\pm3.6}$ \\
RHS & $[21.2\pm5.3],[130.1\pm6.1],{\bf6.0\pm0.6}$ & $[-126.0\pm40.1],[62.1\pm16.0],{\bf0.6\pm1.6}$ & $[-74.8\pm32.4],[84.5\pm10.8],{\bf2.0\pm1.3}$ \\
\cline{1-4}
\multicolumn{4}{c}{$5 < R_{\rm GC} (\rm kpc) \leq 7$} \\
\cline{1-4}
TD+thin disk & \nodata & \nodata & $159.8\pm5.2,44.7\pm1.0,{\bf 17.8\pm2.1}$ \\
SP & \nodata & \nodata & $[100.3\pm2.6],[56.6\pm5.5],{\bf37.0\pm5.2}$ \\
MWTD & $133.1\pm6.2,31.4\pm3.5,{\bf1.5\pm0.3}$ & $139.5\pm8.2,47.0\pm4.6,{\bf5.3\pm1.1}$ & \nodata \\
GSE & $[46.9\pm2.7],[45.1\pm1.9],{\bf4.0\pm0.6}$ & $[35.6\pm2.0],[67.4\pm1.5],{\bf11.4\pm2.6}$ & $[28.7\pm5.9],[52.0\pm4.8],{\bf11.9\pm3.8}$ \\
RHS & $[21.2\pm5.3],[130.1\pm6.1],{\bf8.1\pm0.8}$ & $[-126.0\pm40.1],[62.1\pm16.0],{\bf1.3\pm2.1}$ & $[-74.8\pm32.4],[84.5\pm10.8],{\bf1.7\pm1.5}$ \\
\cline{1-4}
\multicolumn{4}{c}{$3 < R_{\rm GC} (\rm kpc) \leq 5$} \\
\cline{1-4}
TD+thin disk & \nodata & \nodata & $182.6\pm22.4,44.7\pm1.0,{\bf 4.1\pm2.2}$ \\
SP & \nodata & \nodata & $[100.3\pm2.6],[56.6\pm5.5],{\bf43.5\pm6.4}$ \\
MWTD & $131.7\pm7.0,39.3\pm6.1,{\bf2.5\pm0.5}$ & $142.2\pm7.5,51.2\pm4.1,{\bf7.5\pm1.3}$ & \nodata \\
GSE & $[46.9\pm2.7],[45.1\pm1.9],{\bf2.7\pm0.6}$ & $[35.6\pm2.0],[67.4\pm1.5],{\bf14.7\pm3.4}$ & $[28.7\pm5.9],[52.0\pm4.8],{\bf10.6\pm4.3}$ \\
RHS & $[21.2\pm5.3],[130.1\pm6.1],{\bf10.1\pm0.7}$ & $[-126.0\pm40.1],[62.1\pm16.0],{\bf1.7\pm2.7}$ & $[-74.8\pm32.4],[84.5\pm10.8],{\bf2.4\pm2.1}$ \\
\enddata
\tablenotetext{a}{Values shown in brackets are those derived from specific $|Z|$ or $\rgc$ bins but held fixed in the final iteration (see text). Fractional contributions (boldface) are computed in each $\rgc$ bin.}
\end{deluxetable*}

Results from our modeling in all metallicity groups are summarized in Tables~\ref{tab:tab2} and \ref{tab:tab3}, in bins of $|Z|$ and $\rgc$, respectively. The uncertainties in $\meanvphi$ and $\sigmavphi$ are computed from the $10^3$ simulated data sets, as described above. The fractional contributions of individual populations in Tables~\ref{tab:tab2} and \ref{tab:tab3} represent those computed in each $|Z|$ or $\rgc$ slice.

\subsection{Spatial Distributions}\label{sec:spatial}

\begin{figure*}
\centering
\includegraphics[scale=0.42]{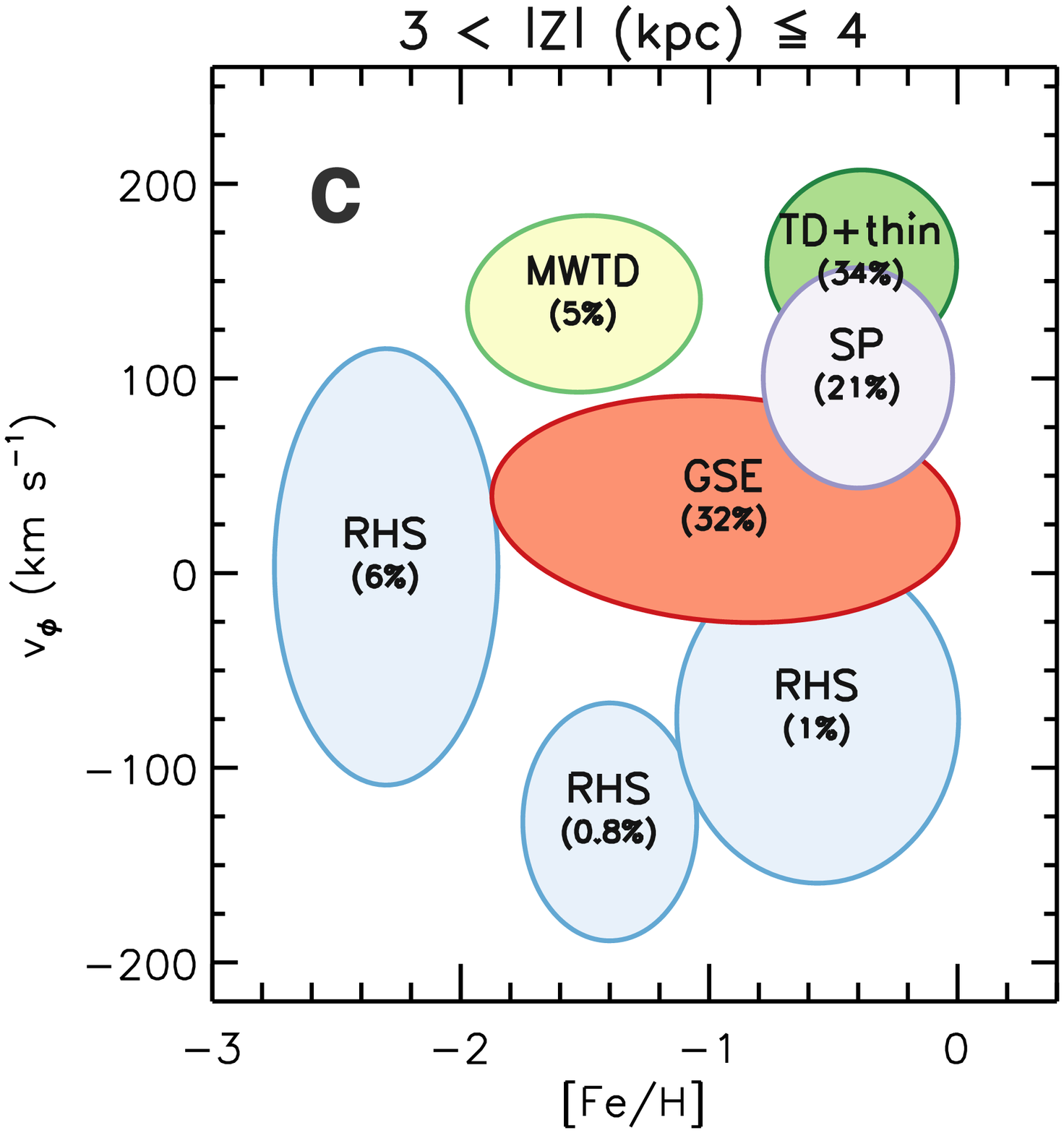}
\includegraphics[scale=0.42]{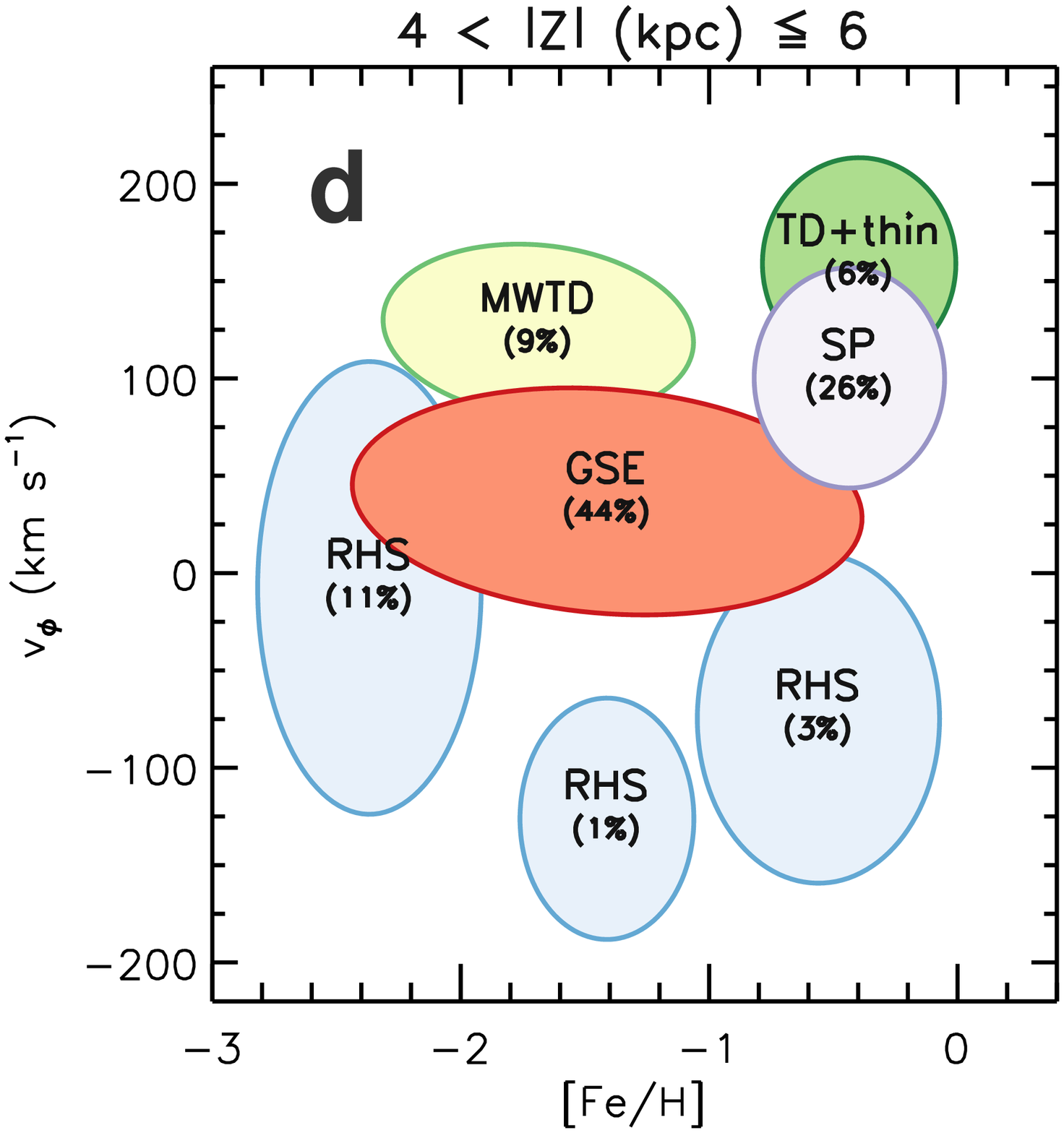}
\includegraphics[scale=0.42]{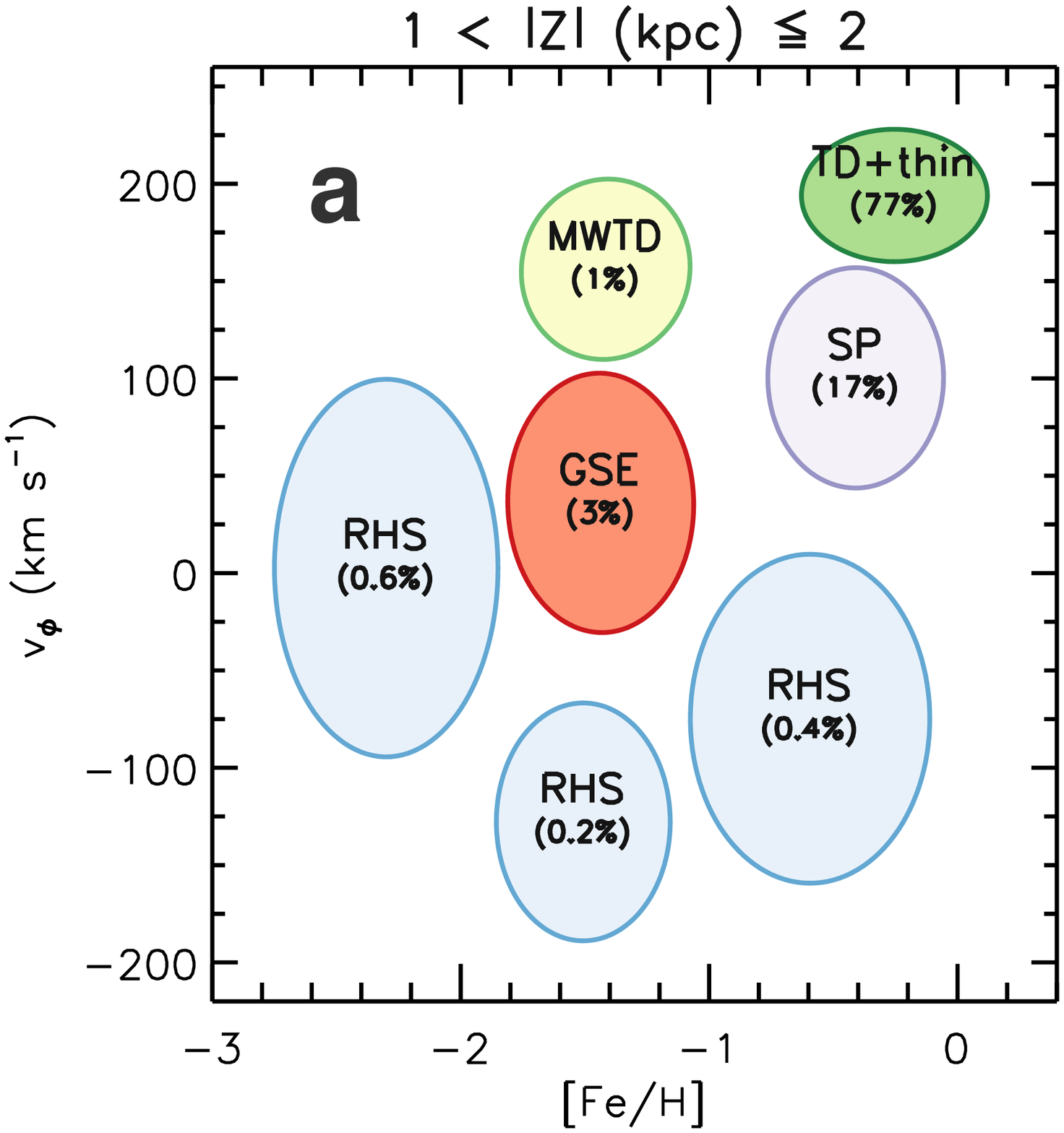}
\includegraphics[scale=0.42]{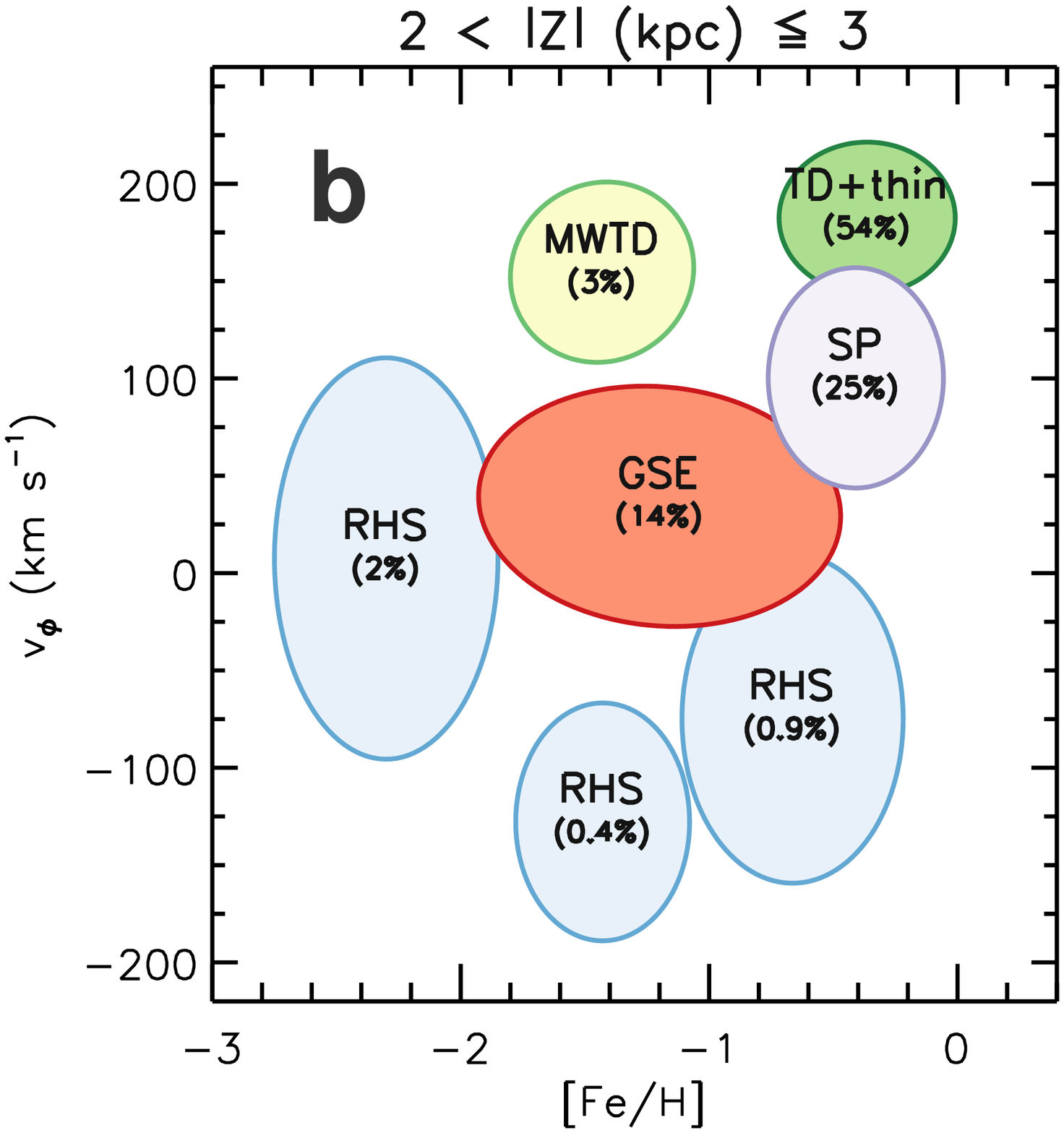}
\caption{Schematic diagrams showing the approximate mean positions and extents of the fitted components. The name of the stellar population that belongs to each component is indicated, along with the fractional contribution to the sample in each $|Z|$ bin.}
\label{fig:pops}
\end{figure*}

Figure~\ref{fig:pops} shows schematic diagrams of the mean positions and $1\sigma$ dispersions in the $\vphi$-[Fe/H] plane for the various stellar populations discussed in this study. Because GSE and the MWTD appear across different metallicity groups, we merge them into a single component, weighted by the fraction of each contributing subcomponent. Nonetheless, the mean location of the MWTD is more weighted by the metal-poor group, because the contribution from the very metal-poor group is significantly smaller. GSE represents a sum of the Gaussian components from all metallicity groups, which displays a mild anticorrelation between $\vphi$ and [Fe/H] (see below). The total fraction estimated in each $|Z|$ slice, based on our Gaussian decomposition, is also shown in parentheses. In comparison to our previous results in Papers~I and II, Figure~\ref{fig:pops} provides better identifications and puts the characterization of individual stellar populations on firmer ground.

\begin{figure*}
\centering
\fig{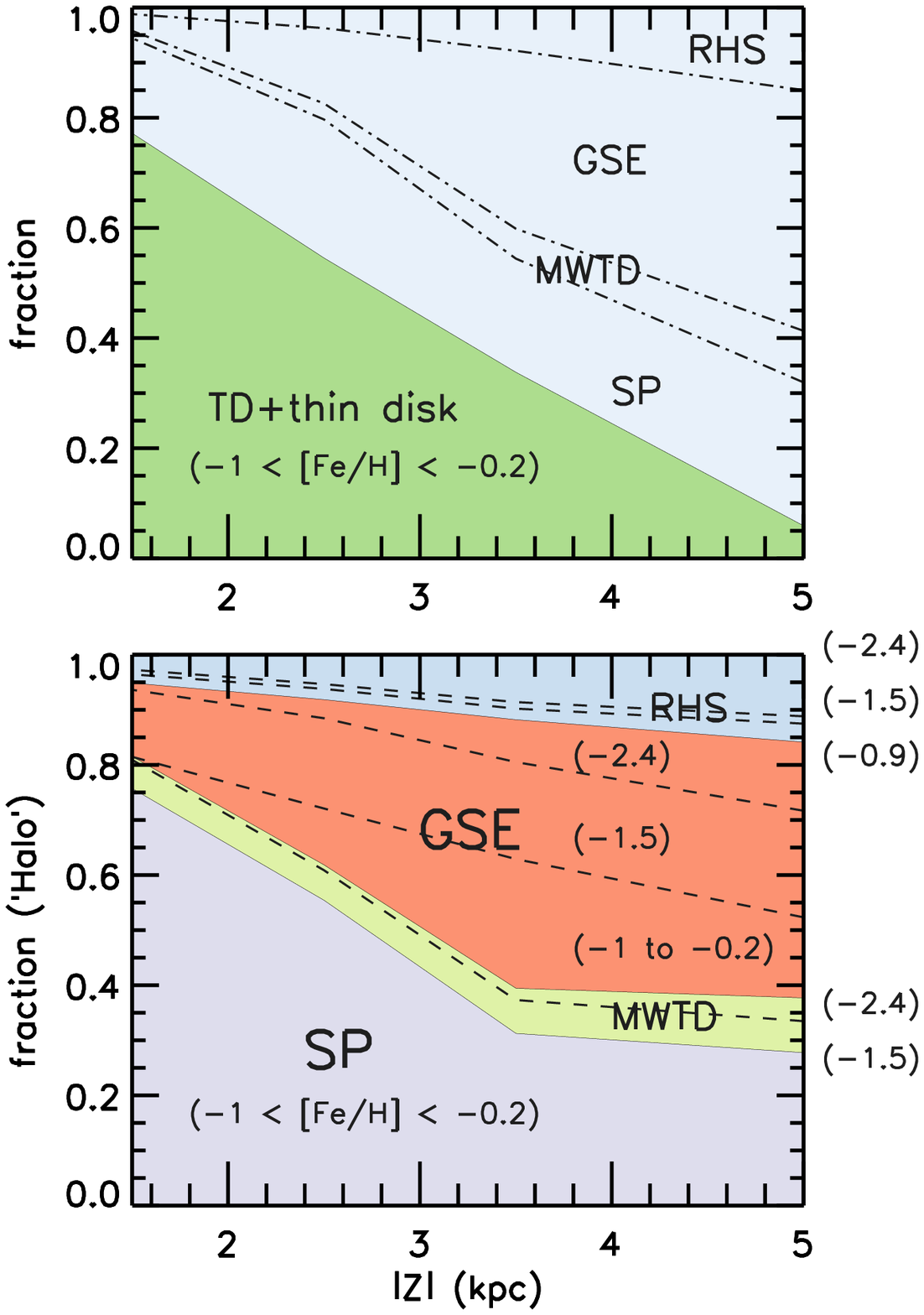}{0.45\textwidth}{\textbf{(a) Fractions as a function of $|Z|$}}
\fig{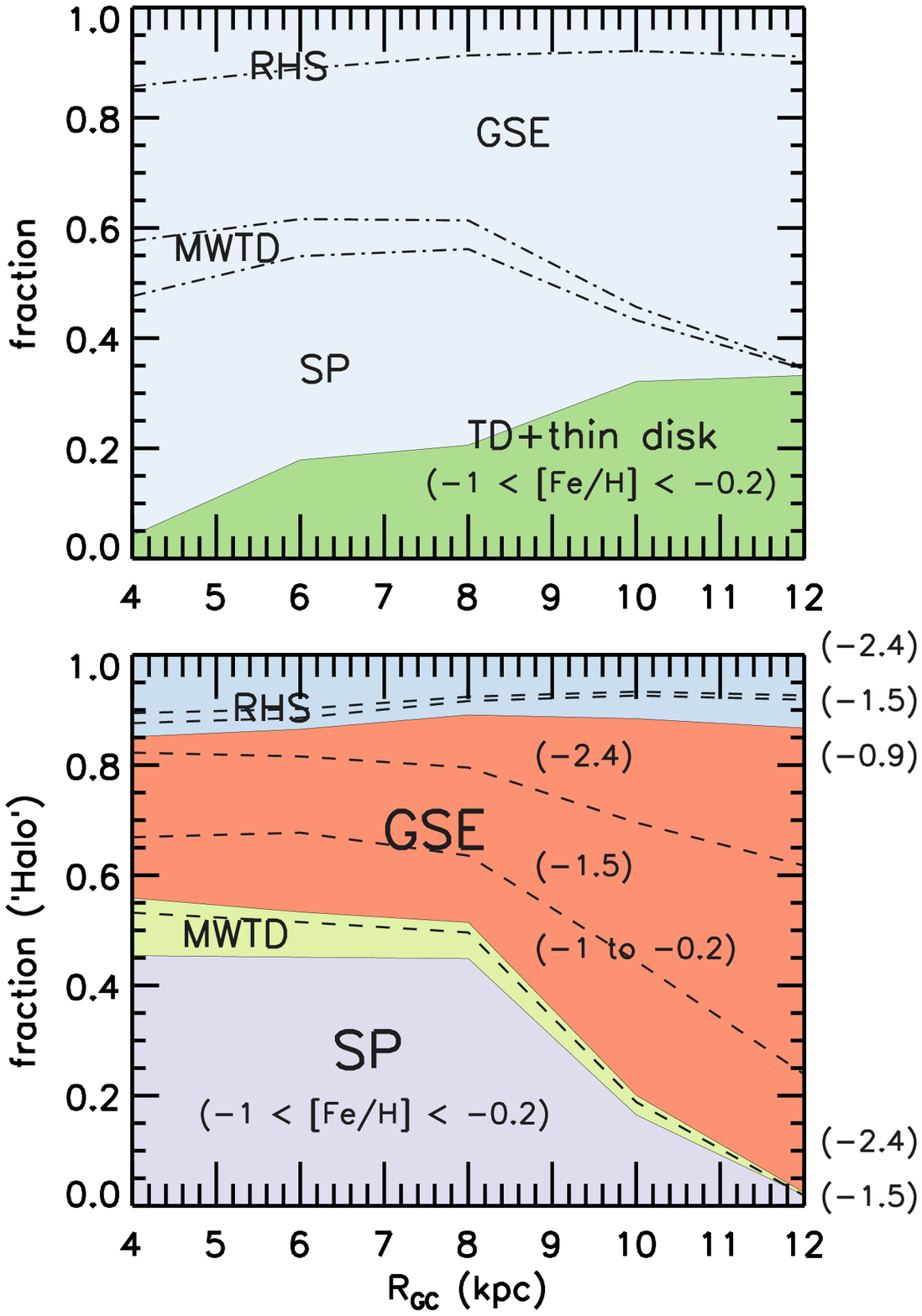}{0.45\textwidth}{\textbf{(b) Fractions as a function of $\rgc$}}
\caption{Fractions of individual stellar components, as a function of $|Z|$ (left panels) and $R_{\rm GC}$ (right panels; $|Z| > 3$~kpc), arranged in order from the lower-metallicity components at the top to the higher-metallicity components at the bottom of each panel. The bottom panels show the fractions of individual ``Halo'' components denoted in the top panels (blue shaded). The contribution from each population is further divided by dashed line(s) to indicate fractional contributions from individual metallicity groups (their approximate mean metallicities are shown in parentheses), in order from lower metallicity at the top to higher metallicity at the bottom.}
\label{fig:frac}
\end{figure*}

Figure~\ref{fig:frac} shows the fractional contributions of individual stellar populations based on our detailed decomposition in Figures~\ref{fig:vphi}--\ref{fig:vphi3} at $\langle |Z| \rangle =$ [$1.5$, $2.5$, $3.5$, $5.0$]~kpc (left panels) and $\langle \rgc \rangle =$ [$4$, $6$, $8$, $10$, $12$]~kpc (right panels). In the top panels, the sum of all halo components analyzed in this work (the SP, GSE, MWTD, and RHS) is displayed by a blue shaded region and is segregated into individual stellar populations by a dotted-dashed line. The fraction of the disk system (TD+thin disk) decreases monotonically with $|Z|$ from $\sim0.8$ at $1.5$~kpc to $\sim0.05$ at $5$~kpc. This can be compared to the disk fraction of [$0.69$, $0.66$, $0.60$, $0.43$] at $|Z|$=[$1.5$, $2.5$, $3.5$, $5.0$]~kpc from previous Milky Way models based on SDSS star counts \citep{juric:08}. Their larger disk fractions are likely caused by inclusion of SP stars in the TD component in the three-component (thin disk, TD, and halo) modeling of the Milky Way. The bottom panels show fractions from individual halo populations, which are further divided by dashed lines to indicate contributions from individual metallicity groups, in increasing order of metallicity (from lower metallicity at the top to higher metallicity at the bottom). As in Figures~\ref{fig:vphi}--\ref{fig:vphi3}, the $\rgc$ diagrams in the right panels are constructed from stars at $|Z| > 3$~kpc to minimize contributions from disk stars.

At $|Z| > 3$~kpc, where the halo components make up more than half of the sample in this study, the SP and GSE account for $\sim80\%$ of the local halo populations in approximately equal proportions. GSE exhibits an increasing contribution to the halo with $\rgc$, in contrast to a decreasing fraction of the SP. The rest of the populations (MWTD and RHS) constitute the remaining $\sim20\%$ of the local halo. In Paper~II, we found about equal proportions of the SP, MRH, and MPH at $4 < |Z| \leq 6$~kpc. The large differences from the current work are mainly rooted in the revised classification scheme in this work, where the MPH and the MRH are interpreted as the backbone of GSE (see \S~\ref{sec:blueprint}).

The range of $\rgc$ in this study is smaller than the ``break'' or ``apocenter pileup'' radius of $\rgc \sim 20$~kpc, inside of which stars in strong radial orbits (mostly GSE stars) are believed to make a significant contribution to accreted halo stars \citep{deason:18,lancaster:19,naidu:20}. Previously, \citet{mackereth:19} found that about $67\%$ of the APOGEE halo sample in a volume similar to that of this work has high orbital eccentricities and abundance patterns characteristic of a massive merger; given the chemo-kinematical properties, the majority of these stars likely belong to GSE. Although their estimated fraction is larger than our value ($\sim46\%$ at $4 < |Z| \leq 6$~kpc), the difference is only suggestive, because their sample covers a slightly larger volume (up to $|Z| \sim10$~kpc), where GSE likely dominates over the SP. Spectroscopic target selection biases could be the other potential cause of the discrepancy.

\citet{naidu:20} used detailed chemical information and full three-dimensional dynamics for a comparatively small sample of $5684$ giants to identify population structures and estimate their fractions over a much larger volume ($\rgc < 50$~kpc) than in this study. In the volume that overlaps with our study ($2 < |Z| \leq 6$~kpc), they found approximately equal proportions of GSE, the SP (their ``in situ'' halo and ``high-$\alpha$ disk''), and the rest of the populations (including their ``unclassified'' stars). At an intermediate distance $|Z| = 3.5$~kpc, their estimated fractions of GSE, the SP, and the MWTD are $\sim0.30:0.40:0.05$, respectively, while our estimates are GSE:SP:MWTD $=0.49:0.31:0.08$. In our decomposition, LMS-1 (Wukong) and the Helmi streams are likely absorbed into the MWTD and/or GSE components, but the estimated fractions of these populations are only at the few percent level \citep[see][]{naidu:20}. ``Unclassified'' stars in their sample may account for some of the difference of the GSE fraction. Several substructures in the RHS, including Sequoia, Arjuna, Thamnos, and I'itoi (see above), constitute $\la5\%$ of the local halo in their sample; the RHS fraction ($8\%$) in our decomposition is only marginally higher. 

\begin{figure*}
  \centering
  \gridline{\fig{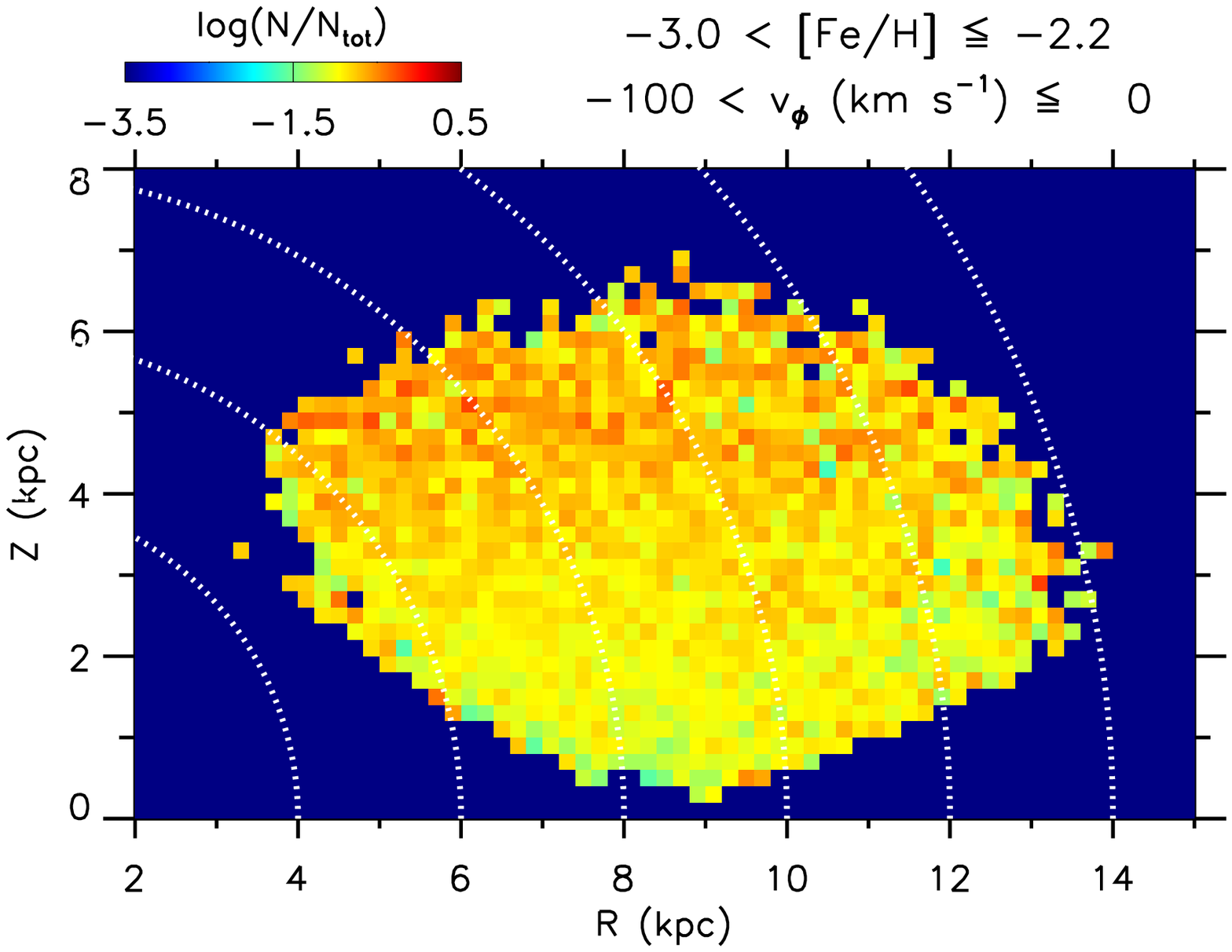}{0.45\textwidth}{}
  \fig{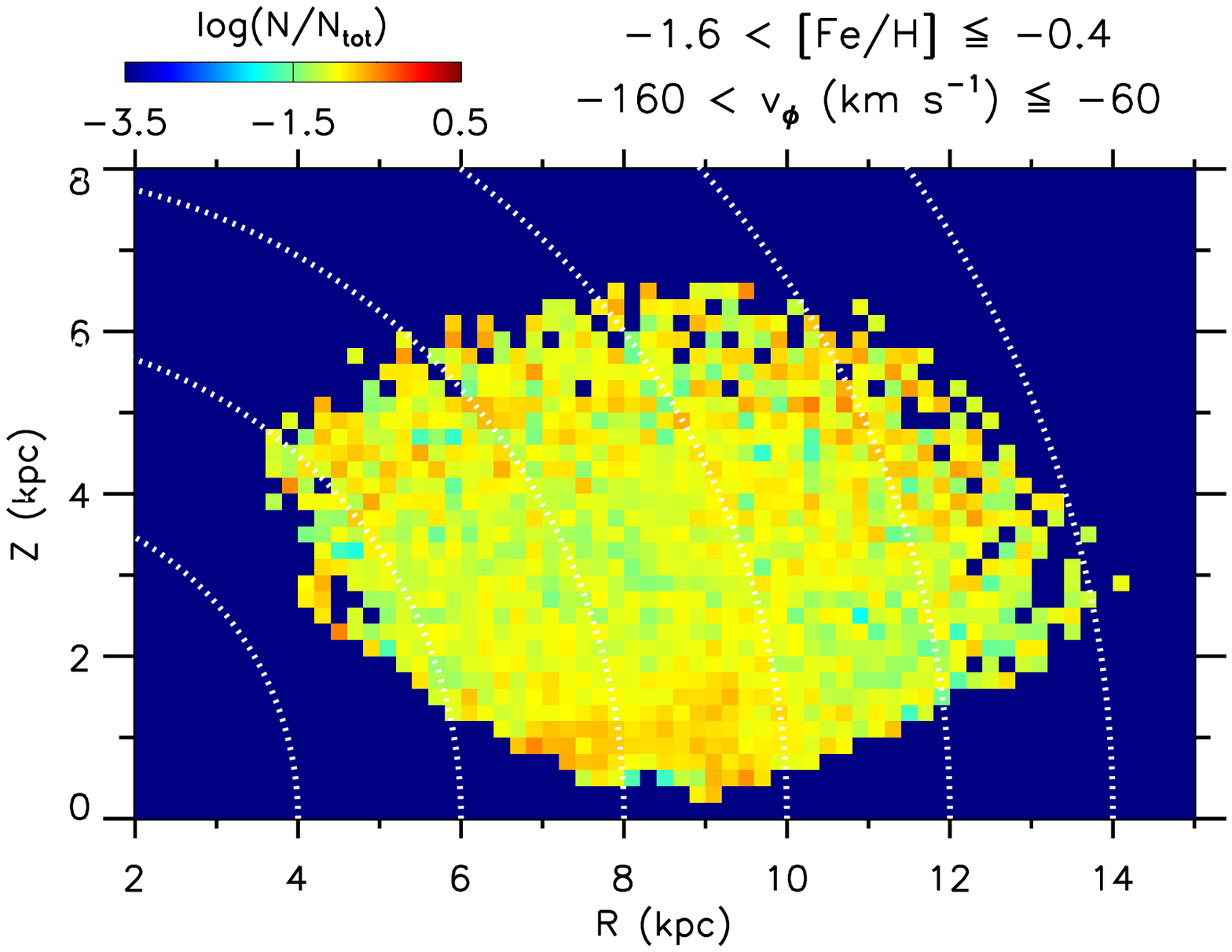}{0.45\textwidth}{}
  }{\textbf{(a) Retrograde Halo Structures (RHS)}}
  \gridline{\fig{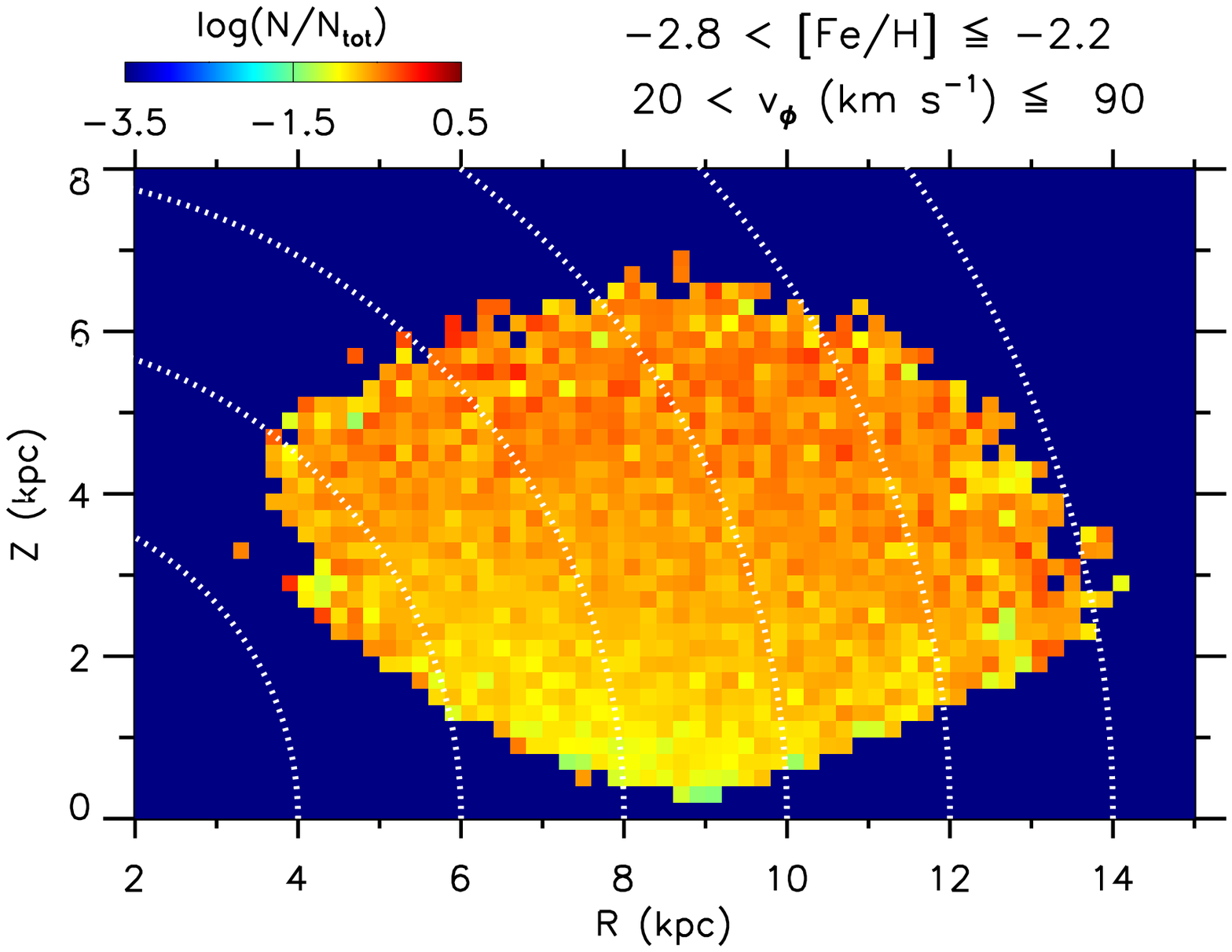}{0.45\textwidth}{}
  \fig{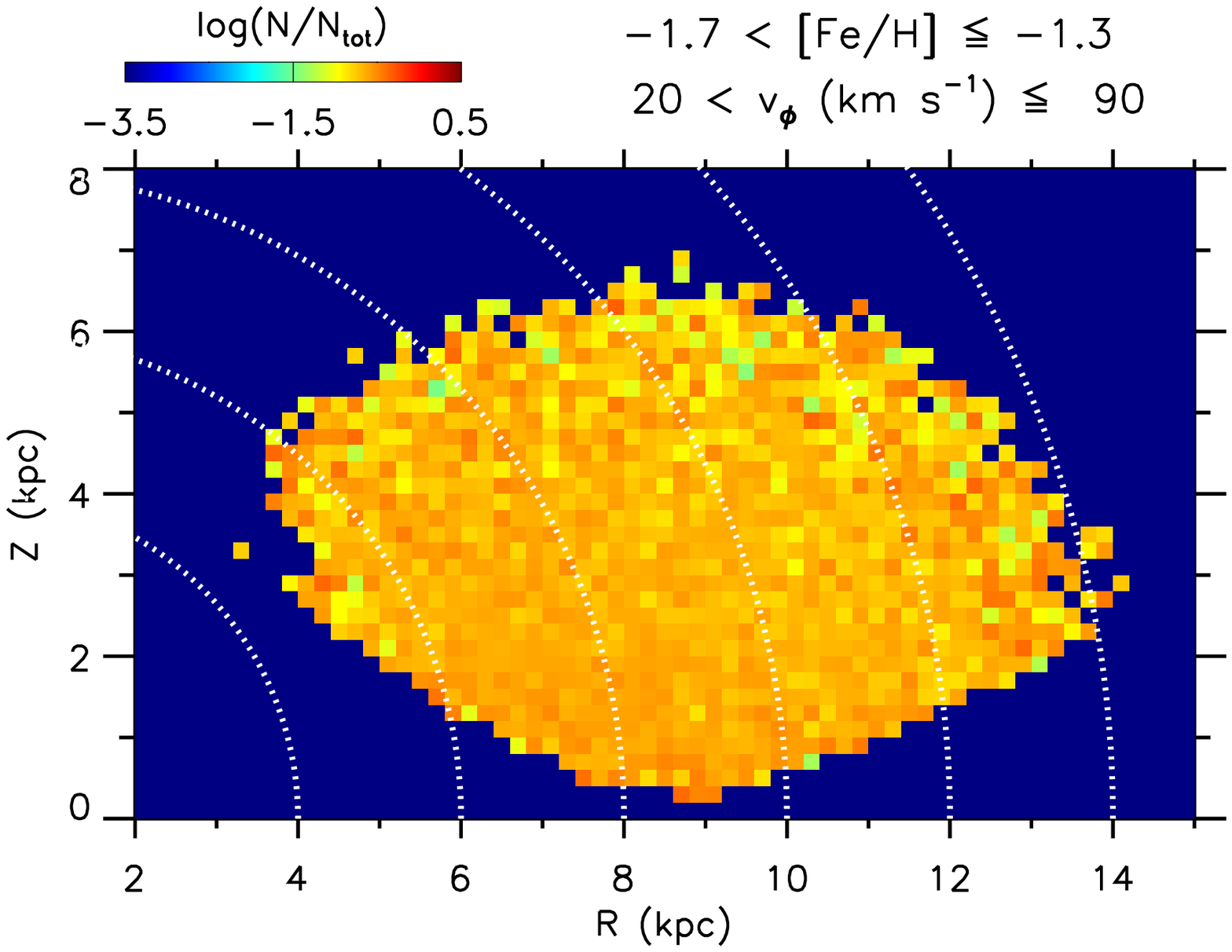}{0.45\textwidth}{}}
  \gridline{\fig{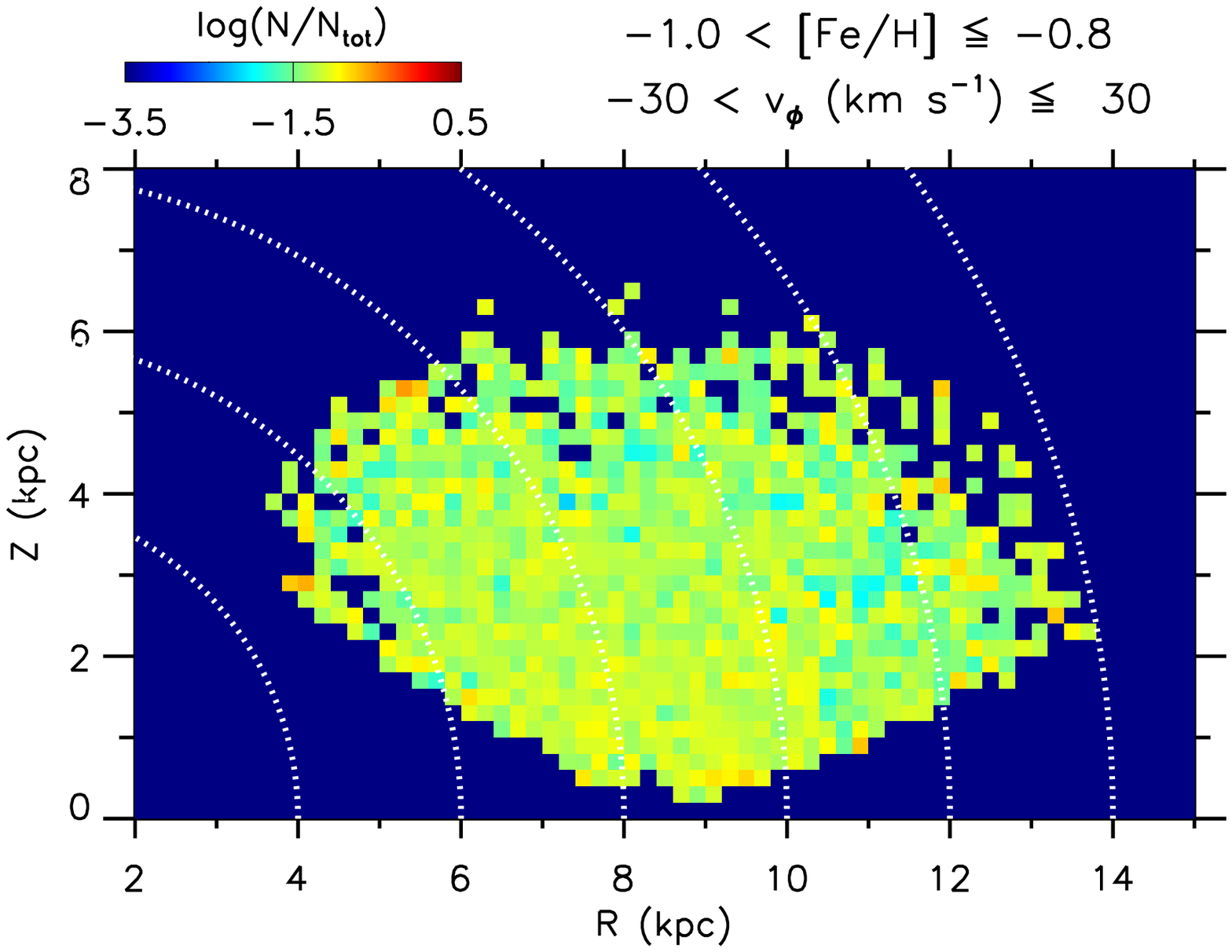}{0.45\textwidth}{}
  }{\textbf{(b) Gaia Sausage/Enceladus (GSE)}}
  \caption{Relative density distributions of stars in specific bins of $\vphi$ and [Fe/H] in the Galactocentric cylindrical coordinate system along the Galactic prime meridian. Each $\vphi$-[Fe/H] bin is chosen to represent individual components of stellar populations. Colors indicate the logarithmic fraction of stars with respect to the total number of stars with $-3 < {\rm [Fe/H]} < -1$ in each pixel. The thin dotted lines represent equal distances from the Galactic center for reference.  Note that the RHS and GSE components are divided into two and three panels, respectively, to depict spatial distributions from different metallicity groups.}
\label{fig:spatial}
\end{figure*}

\begin{figure*}
  \centering
  \figurenum{10}
  \gridline{
  \fig{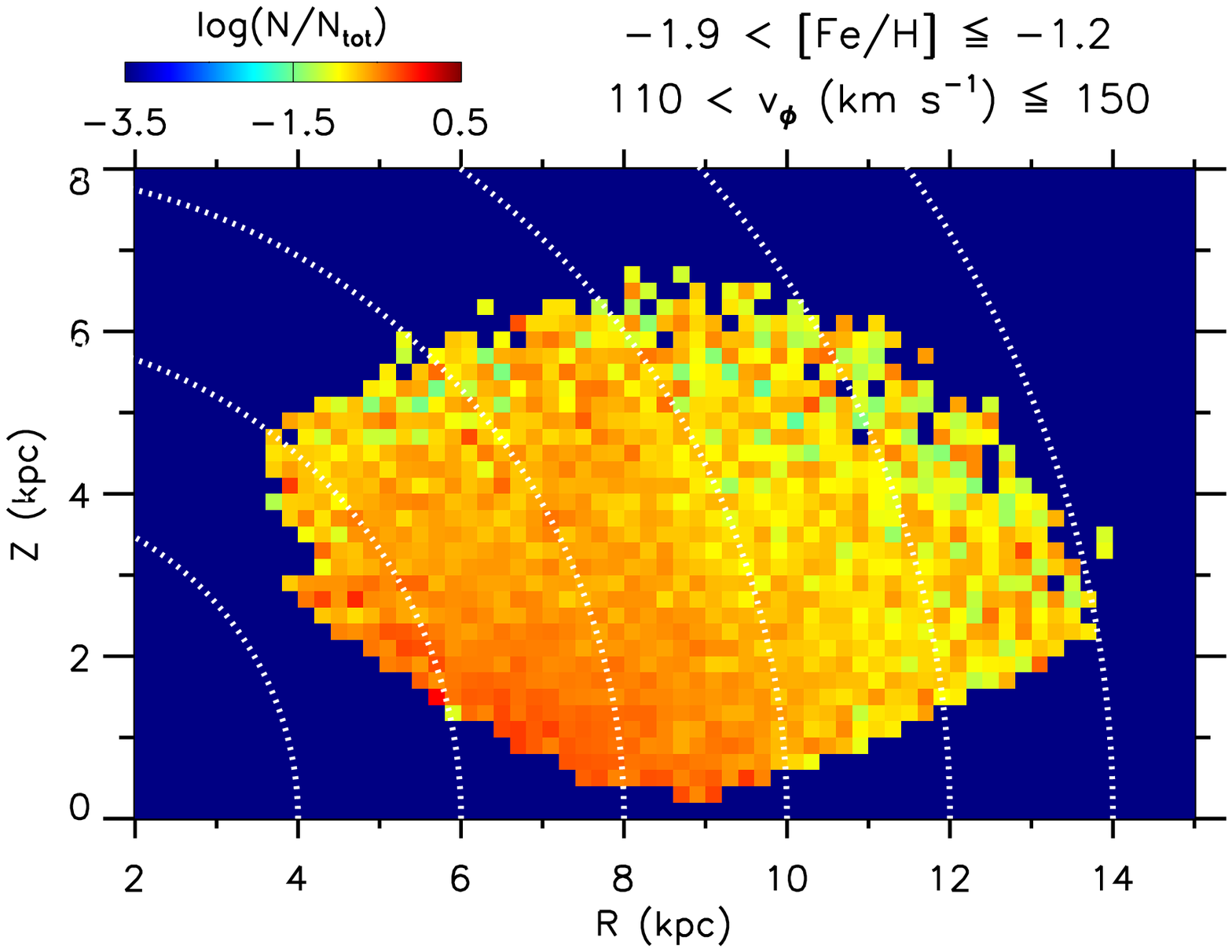}{0.45\textwidth}{\textbf{(c) Metal-Weak Thick Disk (MWTD)}}
  \fig{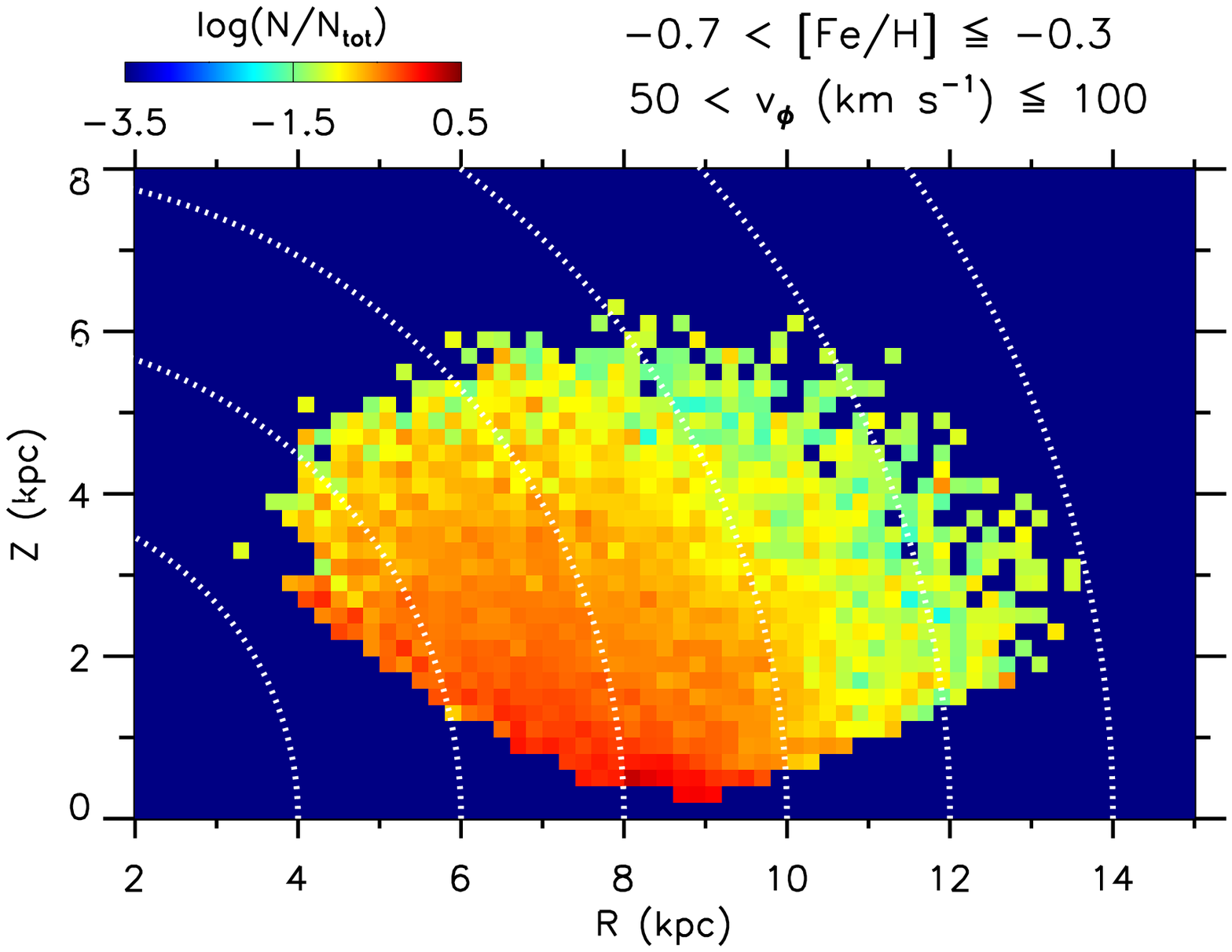}{0.45\textwidth}{\textbf{(d) Splash (SP)}}}
  \gridline{
  \fig{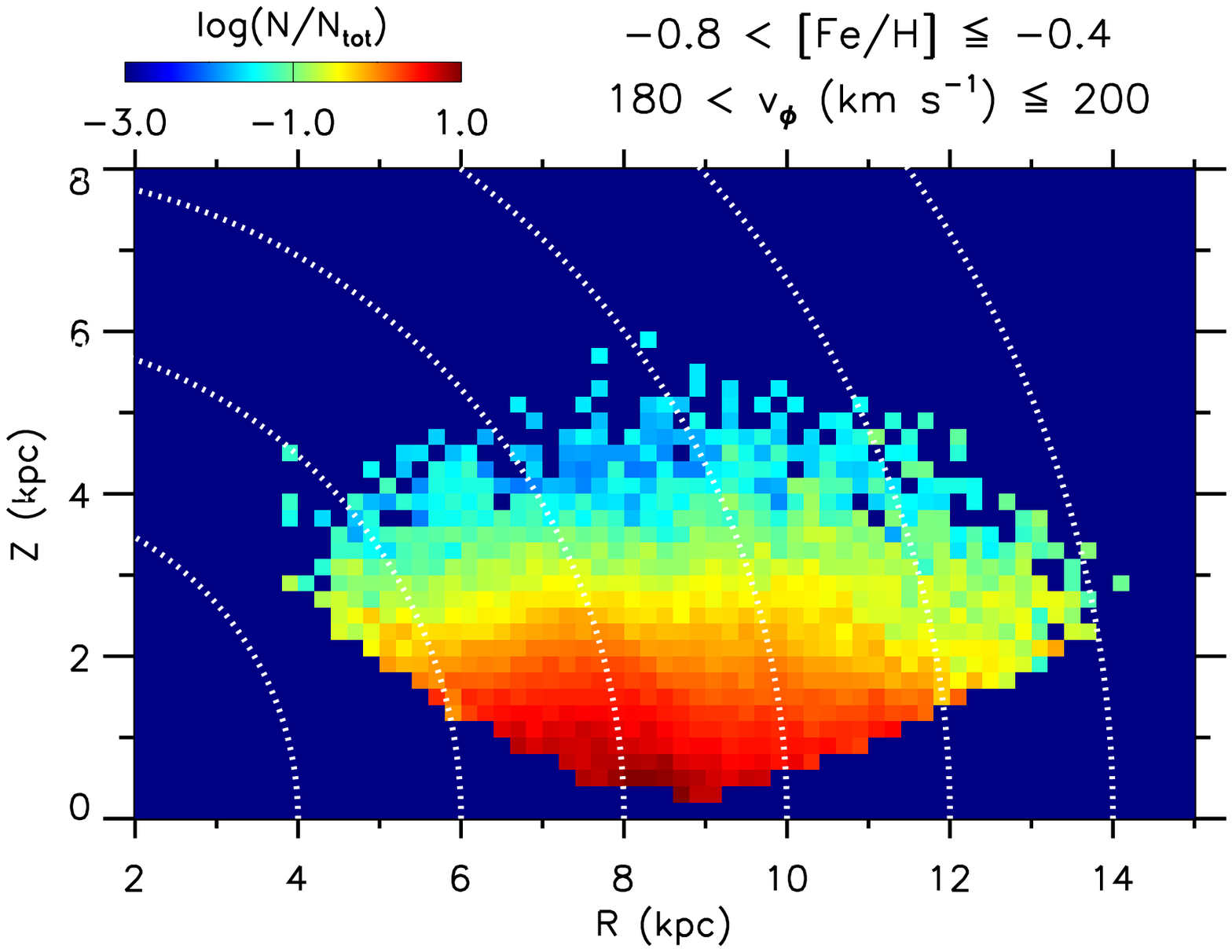}{0.45\textwidth}{\textbf{(e) Thick Disk (TD)}}
  \fig{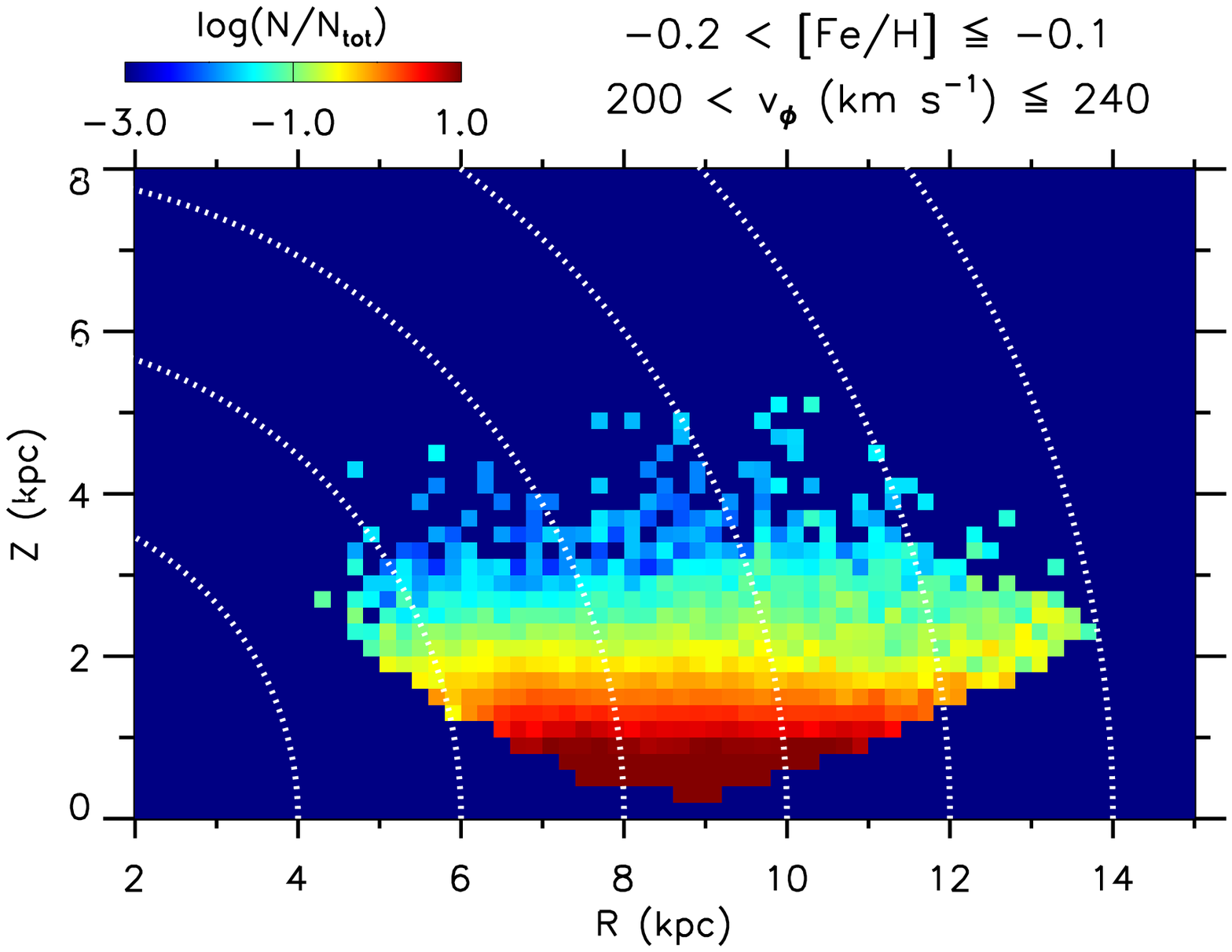}{0.45\textwidth}{\textbf{(f) Thin Disk}}}
  \caption{(Continued.)}
  \label{fig:spatial2}
\end{figure*}

To provide some insight on the spatial distributions of individual stellar populations, Figure~\ref{fig:spatial} displays a logarithmic number density plot in the plane of the Galactic prime meridian. Unlike in Figure~\ref{fig:frac}, a simple selection box is used in the $\vphi$-[Fe/H] space to depict the spatial extents of individual populations in the finer grid of $Z$--$R$ in the Galactocentric cylindrical coordinate system. The range of $\vphi$ and [Fe/H] is set to minimize the contamination from other populations and capture the relative extent of each population, but a significant overlap between various stellar populations is unavoidable. Furthermore, the distribution is normalized with respect to the total number of metal-poor stars with $-3 < {\rm [Fe/H]} < -1$ in each $Z$--$R$ pixel, in order to simplify the intercomparisons of various populations, without relying on specific bias corrections \citep[see, e.g.,][for the mass-based selection]{an:13}. Therefore, any inhomogeneity of the density plot indicates a deviation from the power-law density distribution of the traditional metal-poor ([Fe/H] $<-1$) halo stars. Because the number of stars in the southern Galactic hemisphere with available $u$-band photometry from SDSS is relatively small along the prime meridian, only stars in the northern Galactic hemisphere are displayed in Figure~\ref{fig:spatial}. The properties of each stellar population observed in Figures~\ref{fig:frac} and \ref{fig:spatial} are described below in detail.

\subsubsection{SP}\label{sec:splash}

In Figures~\ref{fig:frac} and \ref{fig:spatial}, the SP shows a strong concentration of stars toward the Galactic center. \citet{sahlholdt:19} obtained photometric metallicities from the SkyMapper Southern Sky Survey \citep[SMSS;][]{wolf:18}, and estimated relative fractions of the blue sequence versus red sequence stars in the Gaia color-magnitude diagrams. Although they found that the ratio depends on the adopted $\vphi$ cutoff in the sample, their result is consistent with a more centrally concentrated distribution of the red sequence stars, which mostly belong to the SP, rather than the blue sequence stars (mostly GSE). Similarly, \citet{belokurov:20} used spectroscopic samples to show that metal-rich ([Fe/H] $>-0.7$) stars (mostly SP) are more strongly populated toward the Galactic center than metal-poor ([Fe/H] $<-0.7$) samples (mostly GSE). As shown in Figure~\ref{fig:frac}, our decomposition reveals an almost flat fraction of the SP at $\rgc \la 8$~kpc and then a rapid decrease beyond the solar circle, indicating an oblate distribution of the SP. Interestingly, such a trend is also seen in the measurements of \citet{sahlholdt:19}, when minimum tangential velocity cuts of $250\ \kms$ or less are applied to their samples.

\subsubsection{MWTD}\label{sec:mwtd}

The MWTD also exhibits a centrally concentrated distribution, as is seen for the SP. At $|Z| > 3$~kpc, the fraction of the MWTD with respect to the total halo populations is $10\%$ in $3 < \rgc \leq 5$~kpc but decreases to $0.5\%$ in $11 < \rgc \leq 13$~kpc. In Figure~\ref{fig:frac}, its relative fraction is nearly constant over many $|Z|$ bins, but this indicates that its density drops rapidly with increasing $|Z|$. The $Z$--$R$ density plot in Figure~\ref{fig:spatial} also supports a large concentration of MWTD stars inside the solar circle.

For the MWTD, an approximate constant fraction from the two metallicity groups implies a weak or no metallicity gradient with $|Z|$ or $\rgc$. The fractional contributions of MWTD stars in the two metallicity groups are used as a crude representation of an MDF (i.e., measured at [Fe/H] $\sim-2.4$ and $-1.5$), from which we find an essentially zero dependence on $\rgc$ ($\Delta {\rm [Fe/H]} / \Delta\rgc = 0.00\pm0.07$~dex~kpc$^{-1}$; $|Z| > 3$~kpc) and on $|Z|$ ($\Delta {\rm [Fe/H]} / \Delta |Z| = -0.08\pm0.23$~dex~kpc$^{-1}$; $|Z| \geq 1$~kpc). These estimates are consistent with results in \citet{ruchti:11}, who found $\Delta {\rm [Fe/H]} / \Delta\rgc = -0.01\pm0.04$~dex~kpc$^{-1}$ and $\Delta {\rm [Fe/H]} / \Delta |Z| = -0.09\pm0.05$~dex~kpc$^{-1}$, respectively, based on spectroscopic analyses of metal-poor ([Fe/H] $<-1.2$) MWTD stars.

\subsubsection{GSE}\label{sec:gse}

Figure~\ref{fig:frac} shows evidence for the chemical differentiation of GSE with distance. As divided by dashed lines in the bottom panels, the {\it relative} fractions of each component of GSE, from the very metal-poor group, the metal-poor group and a combination of the metal-intermediate and metal-rich groups (from top to bottom), change as a function of $|Z|$ and $\rgc$. The ratios between these metallicity groups are $0.10:0.52:0.38$, respectively, at $3 < \rgc \leq 5$~kpc ($|Z| > 3$~kpc) but become $0.30:0.45:0.26$ at $11 < \rgc \leq 13$~kpc, demonstrating the greater presence of more metal-deficient stars at larger $\rgc$. This trend can also be seen as a function of $|Z|$, from $0.09:0.91:0.00$ at $1 < |Z| \leq 2$~kpc to $0.27:0.42:0.32$ at $4 < |Z| \leq 6$~kpc. A schematic representation of each metallicity group is shown in three different panels of Figure~\ref{fig:spatial}. Despite a full overlap of GSE with the RHS in the very metal-poor group (Figure~\ref{fig:vphi}), the simple box selection shows that the metal-poor GSE stars are more enhanced at larger $|Z|$, in contrast to nearly uniform relative density distributions from the metal-poor and metal-intermediate/rich groups.

\begin{figure}
\centering
\includegraphics[scale=0.58]{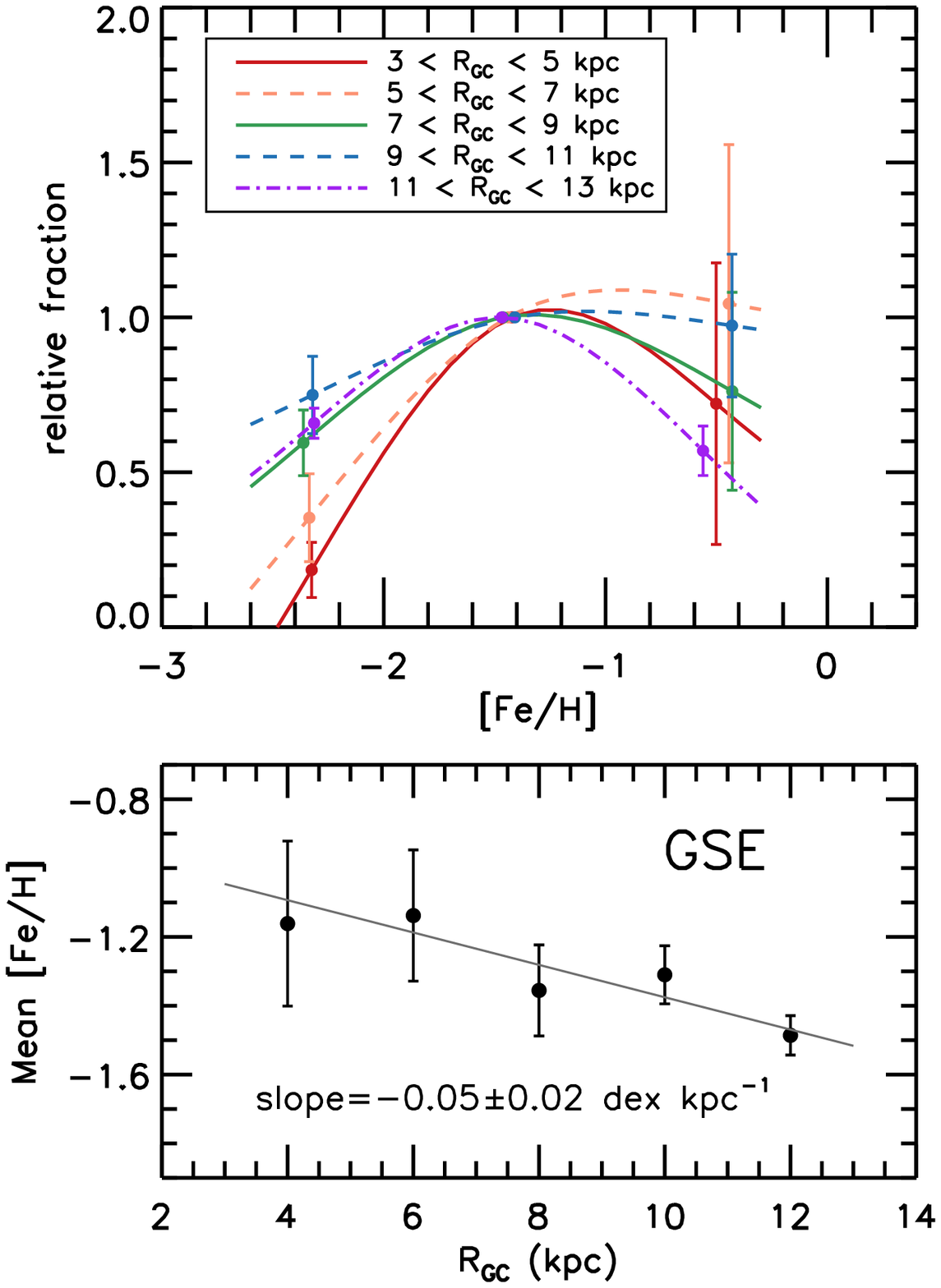}
\caption{Top: the MDFs of GSE as a function of $\rgc$. In each $\rgc$ bin, the data points show relative fractions of stars extracted from the very metal-poor group ($\langle {\rm [Fe/H]} \rangle \sim-2.4$), the metal-poor group ($\langle {\rm [Fe/H]} \rangle \sim-1.5$) and a combination of the metal-intermediate and metal-rich groups ($\langle {\rm [Fe/H]} \rangle \sim-0.5$), respectively. The solid line shows a cubic spline fit to the data points. The MDFs are normalized with respect to the fraction in the middle point ([Fe/H] $\sim-1.5$). Bottom: the mean [Fe/H] of GSE as a function of $\rgc$. The solid line is a weighted linear fit to the data, and its derived slope is indicated in the panel.}
\label{fig:gse}
\end{figure}

Figure~\ref{fig:gse} displays the MDFs and mean metallicities of GSE, as a function of $\rgc$, where stars at $|Z| > 3$~kpc are included. Because of our grouping of GSE stars into individual metallicity groups (Figures~\ref{fig:vphi}--\ref{fig:vphi3}), the relative fractions of stars are shown in three [Fe/H] bins only. Although the MDFs in the top panel are coarse representations of the underlying true distributions, it is readily apparent that GSE stars tend to be more metal-rich at smaller $\rgc$. This metallicity dependence is further examined in the bottom panel, where we compute the mean metallicity of GSE stars as a weighted sum of metallicity ($\langle {\rm [Fe/H]} \rangle \sim$[$-2.4$, $-1.5$, $-0.5$]) by relative number fractions. The error bars represent uncertainties propagated from the bootstrap sampling. As shown by the solid line, we obtain $-0.05\pm0.02$~dex~kpc$^{-1}$ for the $\rgc$ dependence of the mean metallicity of GSE from a weighted least-squares regression. The change with $|Z|$ is more subtle ($\Delta {\rm [Fe/H]} / \Delta |Z| = -0.01\pm0.09$~dex~kpc$^{-1}$), although the fraction of GSE in the very metal-poor group increases with $|Z|$.

\subsubsection{RHS}

In our present analysis, the RHS is extracted from all metallicity groups. Each of them may well be composed of multiple stellar streams and overdensities but the resolution of our map is too low to dissect the structures to identify them. In Figure~\ref{fig:frac}, the RHS is divided by dashed lines to indicate contributions from the three sets of metallicity groups, according to which the component from the very metal-poor group constitutes the majority of the RHS in the survey volume. The spatial distribution of the very metal-poor component also exhibits a mild dependence on $|Z|$ or $\rgc$, with relatively more stars in this group populating at greater distances. This trend can also be seen in Figure~\ref{fig:spatial}, where we display number density distributions in the $Z$--$R$ plane for the very metal-poor group and for more metal-rich RHS stars. Besides, there is a weak excess of stars at $Z \sim 1$~kpc in the latter panel, which tends to be more stronger at higher metallicity ([Fe/H] $> -1$) and larger velocities ($\vphi > -100\ \kms$). They may be distant background giants with large retrograde motions, or a group of stars in the foreground of the interstellar dust. Alternatively, they could be a low-$\vphi$ extension of the metal-rich GSE stars, but more detailed chemical and kinematical properties of these stars are required to confirm their likely association.

In this work, we do not include, nor do we detect, other major structures such as the ``inner Galaxy structure'' \citep{horta:21}. Its progenitor galaxy called ``Heracles'' may be closely related to ``Kraken'' or ``Koala,'' which was spotted through the analysis of globular clusters \citep{kruijssen:19,kruijssen:20,massari:19,forbes:20}. However, the inner Galaxy structure is located in the Galactic bulge and therefore is outside the SDSS imaging footprints. Some of the stars on low-energy orbits in \citet{carollo:21} could be part of these structures that may overlap with GSE at a nearly zero rotation in the $\vphi$-[Fe/H]-$|Z|$ space. However, they are mostly confined to small vertical heights, residing below the latitude limit of our sample ($|b| > 20\arcdeg$).

Our photometric metallicity estimator based on $ugriz$ photometry is only sensitive to the overall metal contents of stars and is unable to discriminate between stars with different elemental abundance ratios as long as their total metallicities are the same. As a consequence, stellar populations with systematically higher (or lower) $\alpha$-element abundances than the base [$\alpha$/Fe]-[Fe/H] relation of the isochrones adopted in this work\footnote{Linearly interpolated values from [$\alpha$/Fe] $=+0.4$ at [Fe/H] $=-3.0$, [$\alpha$/Fe] $=+0.3$ at $-2.0 \leq {\rm [Fe/H]} \leq -1.0$, [$\alpha$/Fe] $=+0.2$ at [Fe/H] $=-0.5$, and [$\alpha$/Fe] $=+0.0$ at [Fe/H] $\geq -0.3$ \citep{an:13}.} may appear more (or less) enhanced in metallicity (parameterized as [Fe/H] in this work) than their actual [Fe/H] (the iron abundance with respect to the solar value). In this regard, the recently claimed structure in the inner halo, called ``Jurassic,'' with a large number of silicon-enhanced stars \citep{fernandeztrincado:19,fernandeztrincado:20} is difficult to distinguish from other halo populations in our mapping.

\section{Discussion and Conclusions}\label{sec:discussion}

\subsection{The Dual-halo Paradigm}

In the seminal works of \citet{carollo:07,carollo:10} and \citet{beers:12}, a systematic change of spectroscopic MDFs as a function of $|Z|$ was interpreted as the presence of two spatially overlapping groups of stars -- the IH and the OH -- with distinct peak metallicities of [Fe/H] $\approx-1.6$ and $-2.2$, respectively. Their kinematic analyses suggested a coupling between metallicities and mean rotation speeds. They also demonstrated that the shape of the halo depends on the metallicities of stars, leading to a plausible hypothesis on the duality of the Galactic halo. See \citet{carollo:21} for a recent reinterpretation on the IH/OH dichotomy, based on a larger number of spectroscopic samples than in \citet{carollo:07}, including data from Gaia DR2. In light of the new perspective on stellar populations obtained in this series of papers, the proposed dichotomy of stellar populations into the IH and OH can be interpreted as influenced by contributions from the above populations as described below.

The metallicity range of the IH overlaps with the metal-poor group as defined in this work, while the OH matches up with the very metal-poor group. There is more than a single population associated with each of the metallicity groups, but the main body of GSE likely forms the majority of the IH sample; the MWTD and the relatively metal-rich ([Fe/H] $>-2$) RHS constitute only minor fractions. On the other hand, the original OH sample in \citet{carollo:07,carollo:10} is likely composed of a combination of the RHS in the very metal-poor group and the metal-poor tail of GSE. The MWTD is also assigned in the very metal-poor group, and can be integrated into the OH, although its fraction is negligible. The SP, which was identified later, is too metal-rich ([Fe/H] $>-1$) and was not included in \citet{carollo:07,carollo:10}.

Given the dichotomy of stellar populations as described above, the other properties of dual halos, such as the mean rotational velocities and spatial extensions, can be understood in terms of the sum of individual stellar populations. The main body of GSE at $-2 \la {\rm [Fe/H]} \la -1$ rotates with a net prograde orbit ($\meanvphi \approx 36\ \kms$), which agrees with a net prograde rotation of the IH. The metal-poor tail of GSE (previously identified as the MPH in Papers~I and II) also shows a net prograde orbit ($\meanvphi \approx 47\ \kms$). However, the RHS that belongs to the OH exhibits a slower net rotation $-20 \la \meanvphi \la 20\ \kms$ and includes a significant fraction of stars with highly retrograde orbits. Because their fractional contributions in the very metal-poor group are comparable to each other at all distances (except in the lowest $|Z|$ bin), their mean $\vphi$ becomes correspondingly smaller than that of the IH, which is indeed seen as an important difference between the IH and OH.

Our mapping also produces spatial extensions that are qualitatively similar to a global shape of the IH and OH. In Figure~\ref{fig:frac}, the summed fraction of the IH subcomponents (GSE, the MWTD, and the RHS in the metal-poor group) is nearly constant or slowly increasing with $|Z|$ or $\rgc$. On the other hand, the OH subcomponents (GSE, the MWTD, and the RHS in the very metal-poor group) make greater contributions to the halo at larger $|Z|$ and $\rgc$. This result is consistent with the picture drawn in \citet{carollo:07,carollo:10} and \citet{beers:12} that the OH has a more extended and rounder structure than the IH. Historically, it can now be recognized that the original identification of the the IH and OH by these authors was the crucial first step required to draw attention to the more refined picture of structure within the Galactic halo revealed by later studies, including our own.

\subsection{Our Previous MDF Decomposition}

Similarly, our previous results based on photometric metallicity estimates \citep{an:13,an:15} can be rephrased based on more accurate kinematic data than available before the Gaia data releases. The MDFs of the Galactic halo presented in these analyses were obtained from a relatively small area along the Stripe~82 region of the SDSS. These MDFs served as an independent constraint on the fractions of the IH/OH components, since their metallicities are based on observations of the least biased, in situ sample of halo stars ($5 \la |Z| \la 10$~kpc). These photometric samples supported the proposition by \citet{carollo:07,carollo:10}, in that a group of more metal-deficient stars tend to show a smaller $\meanvphi$. The two-component Gaussian fitting to the observed MDFs yielded that about $20\%$--$55\%$ of halo stars ([Fe/H] $\la -1$) belong to the OH.

However, the current work demonstrates that our earlier interpretation of the observed MDFs is an oversimplification of the underlying populations. As discussed above, the metal-rich component, which was interpreted as the IH in \citet{an:13,an:15}, represents almost the same entity as the metal-poor group -- a combination of the main body of GSE, the MWTD, and the metal-rich RHS. On the other hand, the metal-poor component (interpreted as the OH) represents the very metal-poor group -- mostly a sum of the metal-poor tail of GSE and the RHS. In the current data set, which is based on a significantly larger sample of stars than in Stripe~82, the ratio between the very metal-poor group and the metal-poor group is $\sim4:6$ at $3 < |Z| \leq 4$~kpc and $\sim5:5$ at $4 < |Z| \leq 6$~kpc. The $|Z|$ range covered by the samples in \citet{an:13,an:15} is $3 \la |Z| \la 8$~kpc, but the strict mass-metallicity-distance limit imposed on these samples favors closer stars. Given this, our estimated fractions of the OH ($43\%$--$48\%$) are slightly larger than the values of $20\%$--$35\%$ reported in \citet{an:13} but agree with the values of $35\%$--$55\%$ based on more accurate $u$-band photometry \citep{an:15}.

\subsection{An Inventory of Local Halo Populations}

\citet{nissen:10} demonstrated that local stars with halo-like kinematics fall into two chemical groups with distinct [$\alpha$/Fe]-[Fe/H] sequences. These groups of stars also exhibit different kinematic properties, where high-$\alpha$ stars tend to exhibit a faster net rotation in the direction of the disk rotation. The same behavior is seen in the larger sample of stars from APOGEE considered by \citet{hayes:18}. In addition, numerous studies have found that each $\alpha$-element group can be matched to the Gaia's double sequence from stars with large transverse motions, according to which stars in the blue (red) sequence mostly belong to the low (high) [$\alpha$/Fe] sequence \citep{haywood:18,gallart:19,sahlholdt:19}. As demonstrated in this series of papers, these sequences characterize two dominant stellar populations (GSE and the SP, respectively) in the local halo.

\citet{schuster:12} found that, based on orbital integration of the spectroscopic targets in \citet{nissen:10}, high-$\alpha$ stars (mostly SP) reach out to a maximum $\rgc\sim16$~kpc, with a maximum excursion out to $|Z| \sim 6$--$8$~kpc. This contrasts with the low-$\alpha$ stars on highly eccentric orbits (mostly GSE), having maximum $\rgc \sim 30$--$40$~kpc and $|Z| \sim 18$~kpc. The spatial distributions of these stars are consistent with our results, in which the SP shows a more compact morphology than GSE; the SP exhibits a sharp decline in the number density fraction at $\rgc \ga 8$~kpc, even though its stars are present out to $\rgc \sim 12$~kpc (see also \S~\ref{sec:splash}). The sample included in \citet{schuster:12} is more localized than in this work, and therefore none of these stars are present inside the solar circle, which makes a direct comparison difficult. Nonetheless, a similar maximum $\rgc$ of their high-$\alpha$ stars validates our result on the compact morphology of the SP.

The spatial extension of the SP can be best explained in the theoretical framework of \citet{zolotov:09,zolotov:10}, according to which mergers at high redshifts caused metal-rich stars formed in situ in the primeval bulge or disk to scatter into halo-like orbits in the inner few tens of kiloparsecs of the Galaxy through violent relaxation. Because of active star formation in the inner region of the primordial Milky Way, these stars can be discriminated from accreted stars with lower [$\alpha$/Fe] in the local halo. Indeed, the SP-like population induced by the merger-related heating is found to be a generic feature of recent numerical simulations in a cosmological context \citep{purcell:10,font:11,mccarthy:12,tissera:13,bonaca:17,bignone:19,park:19,belokurov:20,grand:20}. Some of these studies further demonstrated that the SP-like stars exhibit a nearly spherical and compact spatial distribution \citep[see][]{belokurov:20,grand:20}. Although the extent of the SP can be affected by a number of physical conditions, such as the mass and orbital properties of a merger, as well as the mass and the velocity dispersion of the primordial disk, these numerical results qualitatively agree with our mapping in Figures~\ref{fig:frac} and \ref{fig:spatial}.

\citet{grand:20} found that the fraction of in situ stars (mostly SP) on retrograde orbits generally increases with the stellar mass ratio of the GSE-like progenitor in their simulations (see their Figure~10). Our decomposition produces a significantly smaller fraction of such stars than in \citet{belokurov:20}, who found $\meanvphi = 25\ \kms$ with $\sigmavphi = 54\pm11\ \kms$ at $2 < |Z| \leq 3$~kpc, or $\sim30\%$ of SP stars in counterrotating orbits. This is in contradiction with our estimate of only $4\%$ (or $6\%$ if we adopt a scale error in $\vphi$ of $15\ \kms$; see \S~\ref{sec:blueprint}). The latter fraction results in a stellar mass ratio of less than $\sim0.1$ according to \citet{grand:20}, which implies the GSE progenitor's stellar mass of $\la10^9 M_\odot$ assuming a total stellar mass $\sim10^{10} M_\odot$ of the Milky Way at the time of the merger \citep[e.g.,][]{helmi:18}. Therefore, our lower fraction of counterrotating SP stars produces a progenitor stellar mass in better agreement with the range of stellar mass ($\sim0.3$--$5\times10^9\ M_\odot$) inferred from other independent studies \citep{helmi:18,mackereth:19,myeong:19,vincenzo:19,das:20} and the upper limit set by the current stellar mass of the halo \citep{deason:19,mackereth:20}.

In this work, we provide evidence for the metallicity differentiation of GSE stars with distance (\S~\ref{sec:gse}). More metal-poor stars are found at larger $\rgc$, giving more weight to the contribution from metal-poor stars in its MDF. The negative correlation found between metallicity and $\rgc$ imposes a constraint on the negative metallicity gradient of the progenitor galaxy, in which metal-rich stars are more tightly bound in deeper layers of the potential well, like most of the dwarf galaxies in the Local Group \citep{kirby:11}, including the Sagittarius dwarf galaxy \citep[][and references therein]{hayes:20}. In this regard, it is interesting to note that the range of $\langle {\rm [Fe/H]} \rangle$ in Figure~\ref{fig:gse} is approximately the same as the expected metallicity difference of $0.3$~dex between the central region and the outskirts of the GSE progenitor in the best-matching simulation of \citet{koppelman:20}.

In addition, if we apply a systematic zero-point shift in $\vphi$ on the order of $\sim15\ \kms$ (see \S~\ref{sec:blueprint}), the metal-rich ([Fe/H] $>-1$) stars of GSE exhibit $\meanvphi \approx 15\ \kms$, while its main body ([Fe/H] $<-1$) rotates at a significantly higher rate, $\meanvphi\sim30\ \kms$ and $\sim20\ \kms$ from the two metal-poor groups in increasing order of metallicity. As these metallicity groups are spaced by $\Delta {\rm [Fe/H]} \sim 1$~dex, the velocity gradient becomes $\Delta \vphi / \Delta \langle {\rm [Fe/H]} \rangle \sim -10\ \kms$~dex$^{-1}$. Taken at face value, our result suggests that the progenitor galaxy plunged on a slightly prograde trajectory, followed by a rapid decay into a radialized orbit, since a massive satellite system like the GSE progenitor rapidly loses its orbital energy by strong dynamical friction \citep{amorisco:17,naidu:21}. In other words, the merging galaxy would leave behind stars at smaller $\rgc$ through successive pericentric passages, gradually stripping away stars in its outer region, and eventually losing its most strongly bound, metal-rich stars deeper in the Milky Way's halo.

On the other hand, the above inference is in contradiction to other studies, which argue for a highly retrograde orbit of the GSE progenitor at infall \citep{belokurov:18,helmi:18,koppelman:20,naidu:21}. Their work is based on an observational notion that GSE stars exhibit a nearly zero net rotation \citep{belokurov:20,naidu:20} or a retrograde rotation, with a proposition that some of the structures found in highly retrograde orbits are the debris of the merger \citep{helmi:18,helmi:20,koppelman:20}. Most notably, \citet{naidu:21} proposed that the retrograde stars in Arjuna were part of the GSE progenitor and were stripped early by tidal interactions with the Milky Way, because they were at the outskirts of the progenitor galaxy. This serves as a strong constraint on the highly radialized retrograde orbit in their model and implies an almost flat metallicity gradient of the progenitor galaxy before the merger. Our mapping disfavors their suggested model, but it should be noted that the sample volume in the current study recorded events only from the last few pericentric passages of the merger and therefore may not provide necessary information for reconstructing a full history of the merger event.

Finally, our decomposition shows that the spatial distribution of MWTD stars is more or less similar to that of the SP, while being more centrally concentrated than GSE (Figure~\ref{fig:frac}). The nearly spherical density distribution is also seen in the finer $Z$--$R$ map (Figure~\ref{fig:spatial}), although this map is based on a simple box selection of stars. The observed trend precludes a previous claim on its possible connection to the Monoceros Ring at low latitudes toward the Galactic anticenter \citep[e.g.,][]{carollo:10}. A direct accretion of stars from dwarf galaxies was also considered unlikely in \citet{ruchti:11}, based on high $\alpha$-element abundances of MWTD stars \citep[see also][]{carollo:19}. Moreover, the mean rotation of the MWTD ($\sim140\ \kms$) is significantly larger than $\meanvphi$ of more metal-rich SP stars ($\sim100\ \kms$), which seems to make a direct connection between the two components less likely.

Instead, we speculate that the MWTD stars originated from the dynamical heating of the primordial disk by minor mergers at high redshifts ($z > 2$), while the disk was still growing in size and mass through gas accretion. At later times ($z\sim2$) when the Milky Way encountered GSE \citep[e.g.,][]{bonaca:20}, the primordial disk became sufficiently large and massive enough to sustain the formation of metal-rich ([Fe/H] $> -1$) stars, which have subsequently been displaced to halo-like orbits by the GSE merger, eventually forming the SP population. Because of this catastrophic merger event, these stars would have a larger velocity lag than the MWTD. Active star formation in the inner region of the primordial disk at early epochs would have ensured that MWTD stars possess high $\alpha$-element abundances and old ages. Their almost flat metallicity gradients \citep[][see also \S~\ref{sec:mwtd}]{ruchti:11} may also imply spatially well-mixed stars from the chaotic nature of such process. However, our results do not preclude other possible formation channels, such as a gas-rich merger \citep{ruchti:11}. Additional constraints from both observational and numerical simulations would help us understand the origin of these stars.

\subsection{Future Prospects}

Our ``blueprint procedure,'' as described in this series of papers, is readily extensible to other data sets. These include large-scale ongoing and planned narrow/medium-bandwidth photometric surveys \citep[see Table~1 in][]{huang:21}, such as the SMSS, the Pristine survey \citep{youakim:20}, the Stellar Abundance and Galactic Evolution survey \citep[SAGE;][]{zheng:18}, the Javalambre-Physics of the Accelerated Universe Astrophysical Survey \citep[J-PAS;][]{benitez:14}, the Javalambre/Southern Photometric Local Universe Survey \citep[J/S-PLUS;][]{cenarro:19,splus}, and the Multi-channel Photometric Survey Telescope (Mephisto; X.\ Z.\ Er et al. 2021, in preparation).

The narrow/medium bands adopted by those surveys provide the possibility to perform precise estimates  of stellar atmospheric parameters (e.g., effective temperature, $T_{\rm eff}$, surface gravity, log\,$g$, and metallicity, [Fe/H]; see \citealt{casagrande:19,huang:19,whitten:19,whitten:21,youakim:20}).  In the case of J-PAS, J-PLUS, and S-PLUS, narrowband filters placed on individual stellar absorption features will even allow photometric estimates of a number of elemental abundances, such as C, Na, Mg, and Ca, once they are fully calibrated.  Such elemental abundance measurements could, in turn, be used to study the expected differences in the chemical evolution histories of the different stellar populations identified in the survey blueprints.  

Ultimately, the exquisite $ugriz$ photometry to be obtained by the Legacy Survey of Space and Time \citep[LSST;][]{ivezic:19} on the Vera Rubin Observatory (expected to be better than 1\% precision) will enable the production of much higher resolution blueprints for stellar populations in the Galaxy, owing to its depth of coverage. There should be several magnitudes of overlap between the LSST bright limit and the faint limits of several surveys currently employing narrow/medium-bandwidth filters, opening the possibility of studying multiple elemental abundances for very large numbers of stars in common.

\acknowledgements

We thank the anonymous referee for detailed comments that helped us to improve readability of the paper. D.A.\ acknowledges support provided by the National Research Foundation of Korea grant funded by the Korea government (MSIT) (No.\ 2021R1A2C1004117, No.\ 2018R1D1A1A02085433). T.C.B.\ acknowledges partial support from grant PHY 14-30152 (Physics Frontier Center/JINA-CEE), awarded by the U.S. National Science Foundation.

Funding for the Sloan Digital Sky Survey IV has been provided by the Alfred P.\ Sloan Foundation, the U.S.\ Department of Energy Office of Science, and the Participating Institutions. SDSS-IV acknowledges support and resources from the Center for High Performance Computing  at the University of Utah. The SDSS website is www.sdss.org.

SDSS-IV is managed by the Astrophysical Research Consortium for the Participating Institutions of the SDSS Collaboration including the Brazilian Participation Group, the Carnegie Institution for Science, Carnegie Mellon University, the Chilean Participation Group, the French Participation Group, Harvard-Smithsonian Center for Astrophysics, Instituto de Astrof\'isica de Canarias, The Johns Hopkins University, Kavli Institute for the Physics and Mathematics of the Universe (IPMU)/University of Tokyo, the Korean Participation Group, Lawrence Berkeley National Laboratory, Leibniz Institut f\"ur Astrophysik Potsdam (AIP),  Max-Planck-Institut f\"ur Astronomie (MPIA Heidelberg), Max-Planck-Institut f\"ur Astrophysik (MPA Garching), Max-Planck-Institut f\"ur Extraterrestrische Physik (MPE), National Astronomical Observatories of China, New Mexico State University, New York University, University of Notre Dame, Observat\'ario Nacional / MCTI, The Ohio State University, Pennsylvania State University, Shanghai Astronomical Observatory, United Kingdom Participation Group, Universidad Nacional Aut\'onoma de M\'exico, University of Arizona, University of Colorado Boulder, University of Oxford, University of Portsmouth, University of Utah, University of Virginia, University of Washington, University of Wisconsin, Vanderbilt University, and Yale University.

This work presents results from the European Space Agency (ESA) space mission Gaia. Gaia data are being processed by the Gaia Data Processing and Analysis Consortium (DPAC). Funding for the DPAC is provided by national institutions, in particular the institutions participating in the Gaia MultiLateral Agreement (MLA). The Gaia mission website is https://www.cosmos.esa.int/gaia. The Gaia archive website is https://archives.esac.esa.int/gaia.

\appendix

\section{Test of Parameter Estimates}\label{sec:scale}

\begin{figure}
\centering
\includegraphics[scale=0.5]{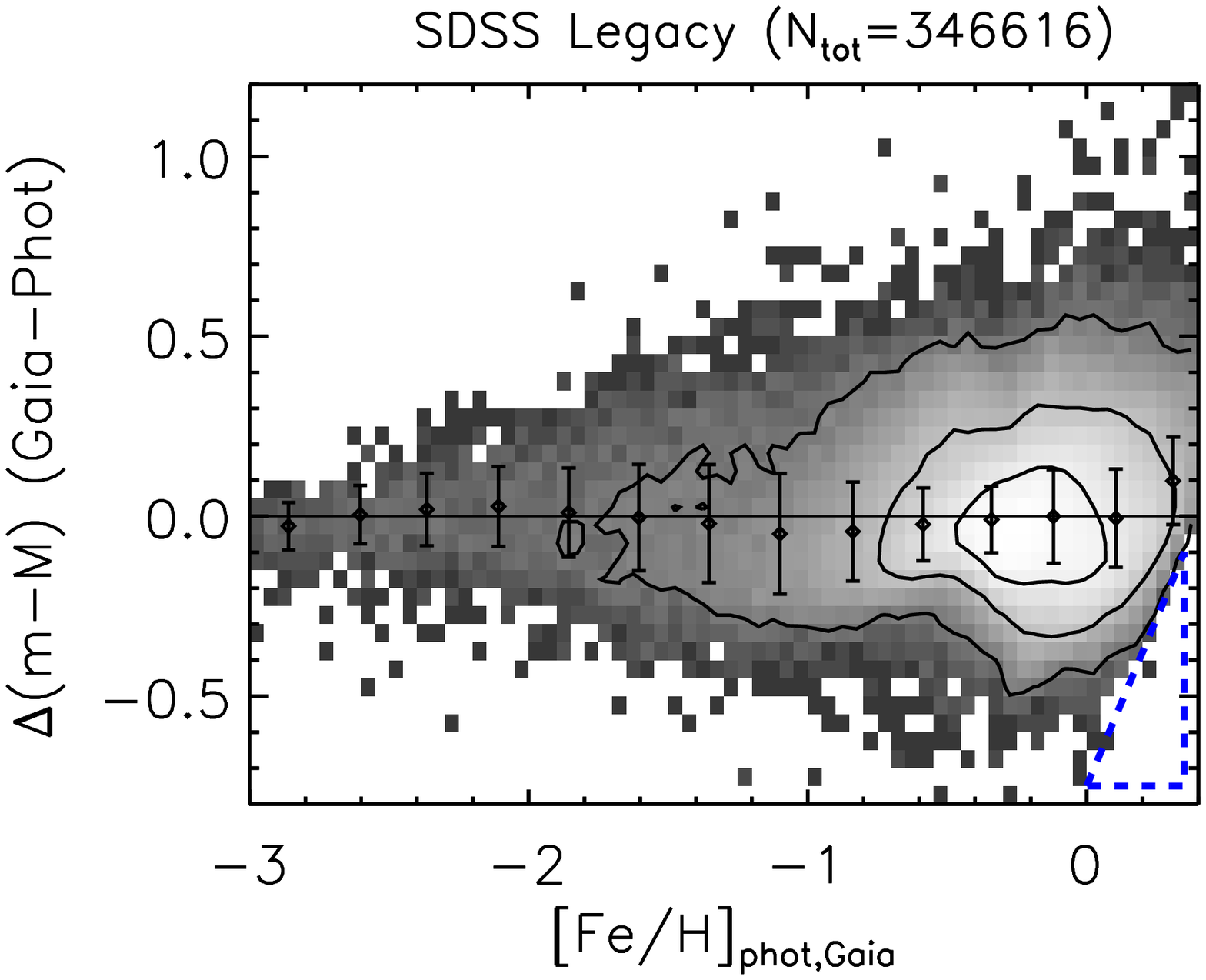}
\caption{Comparison of the distance moduli between Gaia EDR3 parallaxes ($\pi > 1$~mas and $\sigma_\pi/\pi \leq 0.1$) and photometric estimates of the main sample in this work. The differences are shown as a function of photometric metallicity, which is derived using Gaia parallaxes as a prior. The gray scale shows the logarithmic number density of the stars, with the contours indicating $50$th, $90$th, and $98$th percentiles. Moving averages and standard deviations are shown by the points with error bars. The blue dashed triangle in the lower right corner delineates an unoccupied region due to a positive correlation between photometric metallicity and distance estimates.}
\label{fig:deltadmn}
\end{figure}

Figure~\ref{fig:deltadmn} compares the distance moduli used in this work with Gaia parallaxes from EDR3. Comparisons are restricted to those with accurate parallaxes ($\sigma_\pi/\pi \leq 0.1$) among those that satisfy our selection criteria for the main photometric sample (see \S~\ref{sec:sample}). The overall agreement is satisfactory and does not show any strong systematic deviation as a function of photometric metallicity (derived using Gaia's parallaxes as a prior), with a maximum deviation of $\sim5\%$ at $-1.2 \la {\rm [Fe/H]} \la -0.8$. At supersolar metallicity ([Fe/H] $>0$), Figure~\ref{fig:deltadmn} is devoid of stars with negative differences (delineated by a blue dashed triangle), which is simply caused by a positive correlation of photometric distance with metallicity; the large mean offset in the most metal-rich bin ($+0.2 < {\rm [Fe/H]} < +0.4$) reflects this artificial cutoff.

\begin{figure}
\centering
\includegraphics[scale=0.44]{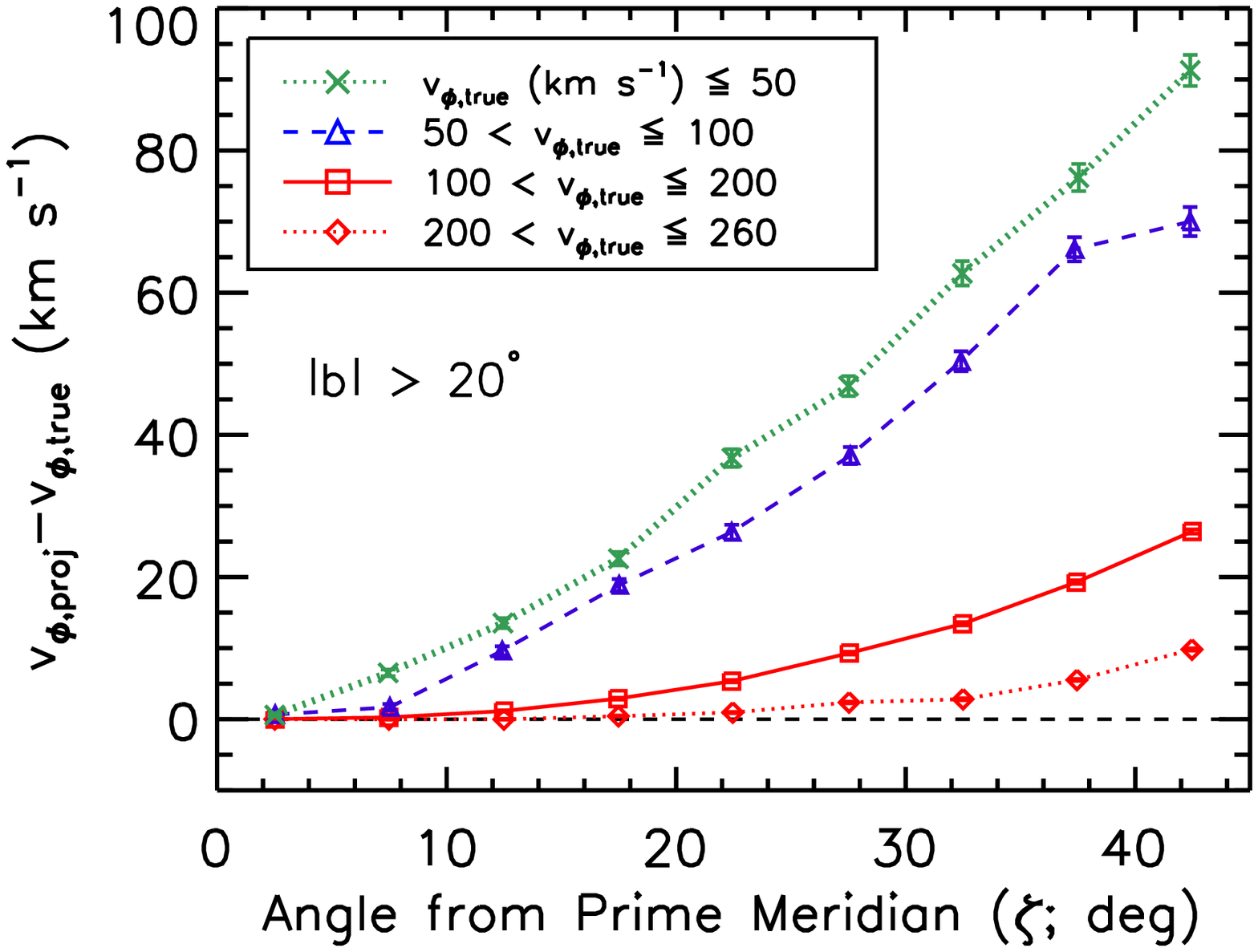}
\includegraphics[scale=0.44]{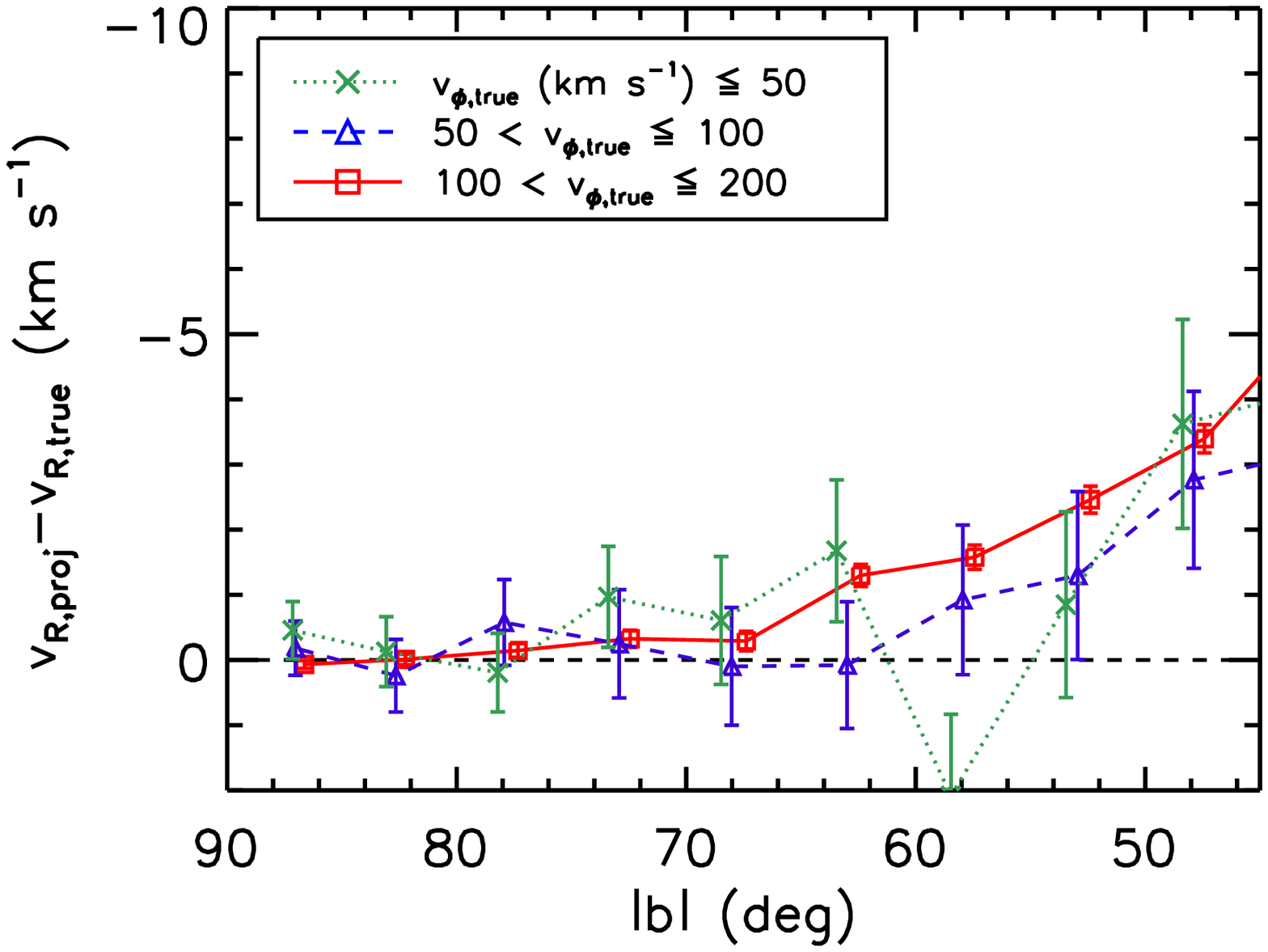}
\caption{Mean velocity differences of proper-motion-based estimates in this work ($v_{\rm proj}$) from the true values obtained using full three-dimensional motions with Gaia's proper motions and radial velocities ($v_{\rm true}$). The $v_{\phi,\rm proj}$ (left) and $v_{R,\rm proj}$ (right) represent azimuthal and radial velocity components in the Galactocentric cylindrical coordinate system, respectively, as derived from proper-motion measurements and distance estimates alone. Average differences and their errors are shown by different symbols for a number of $\vphi$ bins, from stars with similar $\vphi$ to the LSR (red diamonds) to those with the largest velocity lags in the sample (green crosses). Only stars with $\sigma_\pi/\pi < 0.2$ and proper-motion errors of $<30\%$ are included in the above comparisons.}
\label{fig:deltavphi}
\end{figure}

We test the accuracy of our $\vphi$ measurements using a sample of stars having both proper motions and radial velocities in Gaia EDR3, from which we estimate full three-dimensional velocity vectors. These stars are typically brighter than our main sample (main-sequence stars), so there are only a few objects with our photometric estimates. We use Gaia's parallaxes in the velocity estimate, by restricting the sample to those with $\sigma_\pi/\pi < 0.2$ and proper-motion uncertainties of $<30\%$. We also limit the sample to $0.5 < |Z| \leq 6$~kpc, although most stars are confined to $|Z| \la 3$~kpc owing to parallax uncertainties. A total of $23,802$ stars are used in the following test.

The left panel of Figure~\ref{fig:deltavphi} shows differences between the true velocities ($v_{\phi,\rm true}$) and those computed using proper motions alone, in the same manner as for the main photometric sample in this work ($v_{\phi,\rm proj}$). As described in Paper~I, the rotational velocities derived from proper motions ($v_{\phi,\rm ppm}$) are corrected for the inclination effect, using a first-order correction, $v_{\phi,\rm proj} = v_{\phi,\rm ppm} \sec \zeta$, where $\zeta$ is the angle from the Galactic prime meridian (here, $v_{\phi,\rm proj}$ has the same meaning as $\vphi$ in the main text). As shown in Figure~\ref{fig:deltavphi}, the velocity difference is a sensitive function of $\zeta$, but it also increases rapidly with increasing velocity lags. If we restrict stars to $|\zeta| < 30\arcdeg$, as adopted in this work, and also to $v_{\phi,\rm true} < 100\ \kms$ to select those with halo-like kinematics, the proper-motion-based velocities become systematically larger by $19.5\pm0.3\ \kms$ (a median of $8\ \kms$) than the true values based on full three-dimensional components. The $\vphi$ difference only weakly depends on $|Z|$, and differences in the northern and southern Galactic hemispheres are almost identical.

The right panel of Figure~\ref{fig:deltavphi} displays a similar comparison as the left panel, but for radial velocities in the Galactocentric cylindrical coordinate system ($v_R$). As expected from the zero net motion of stars in the radial direction, Figure~\ref{fig:deltavphi} shows negligible systematic differences; the mean $v_R$ difference is less than $1\ \kms$ for stars at $|b| > 60\arcdeg$, as employed in this work (see also Figure~\ref{fig:vrad}).

\section{$\vphi$-[Fe/H] Distributions Based on Gaia Parallaxes}\label{sec:disk}

\begin{figure*}
\centering
\includegraphics[scale=0.36]{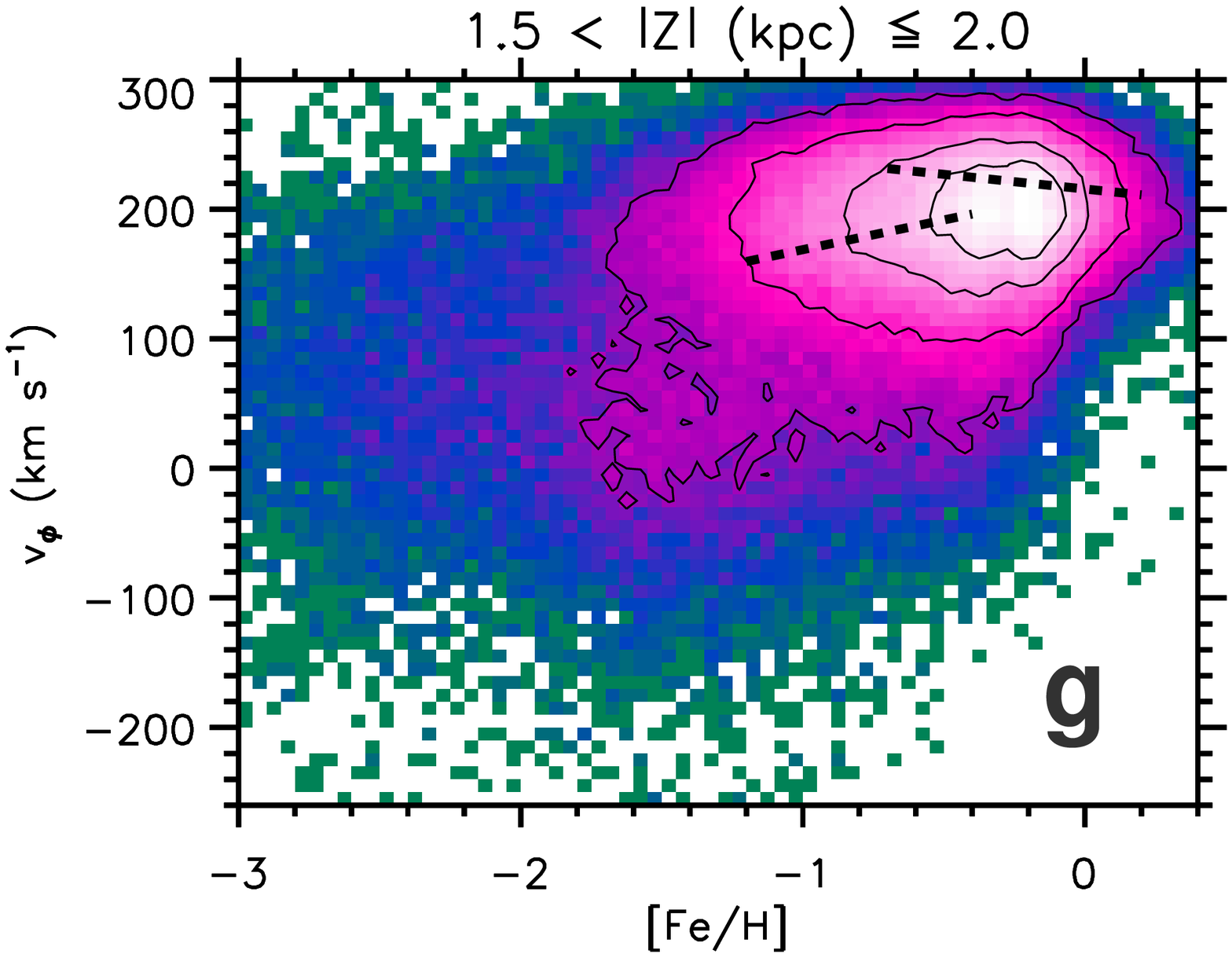}
\includegraphics[scale=0.36]{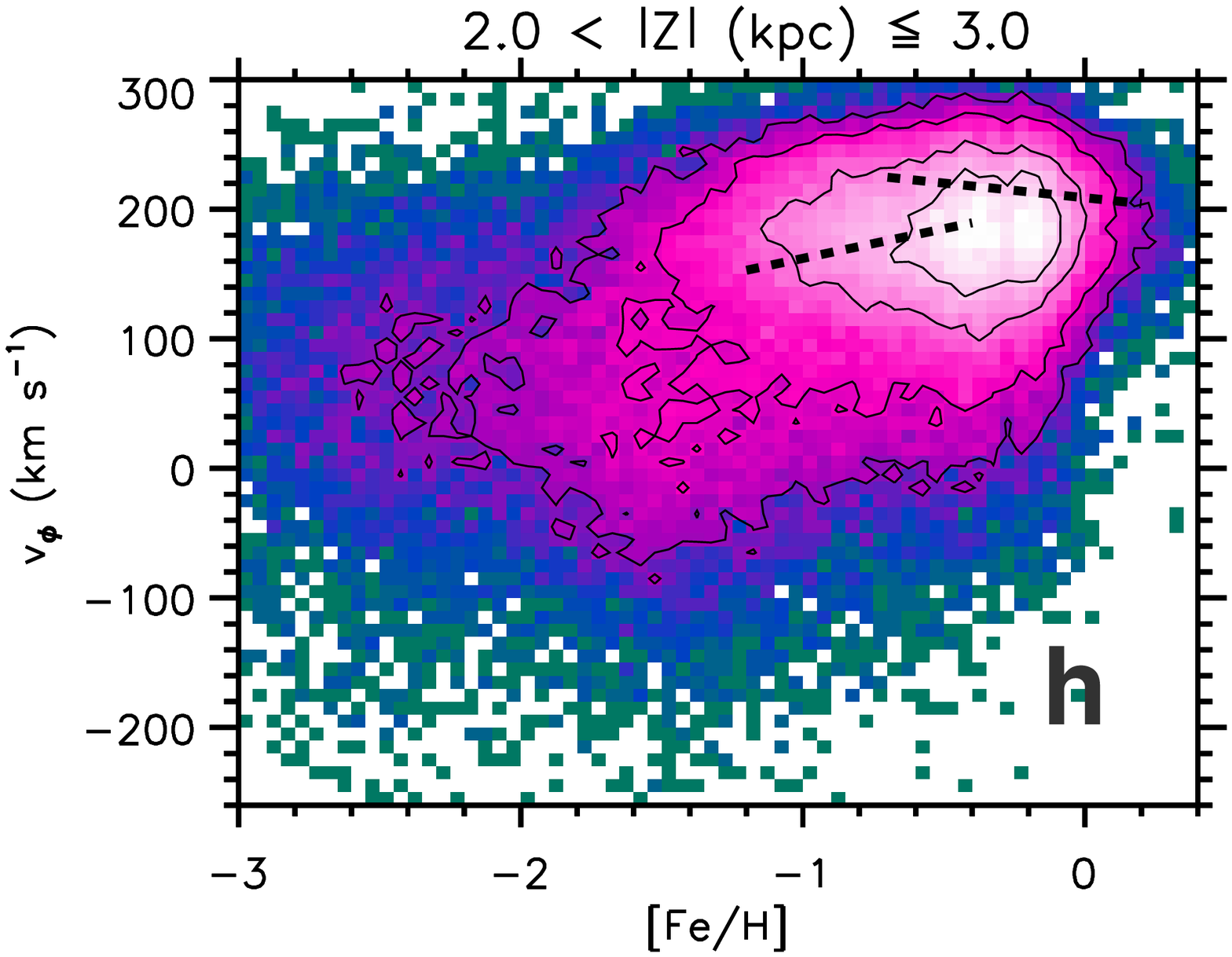}
\includegraphics[scale=0.36]{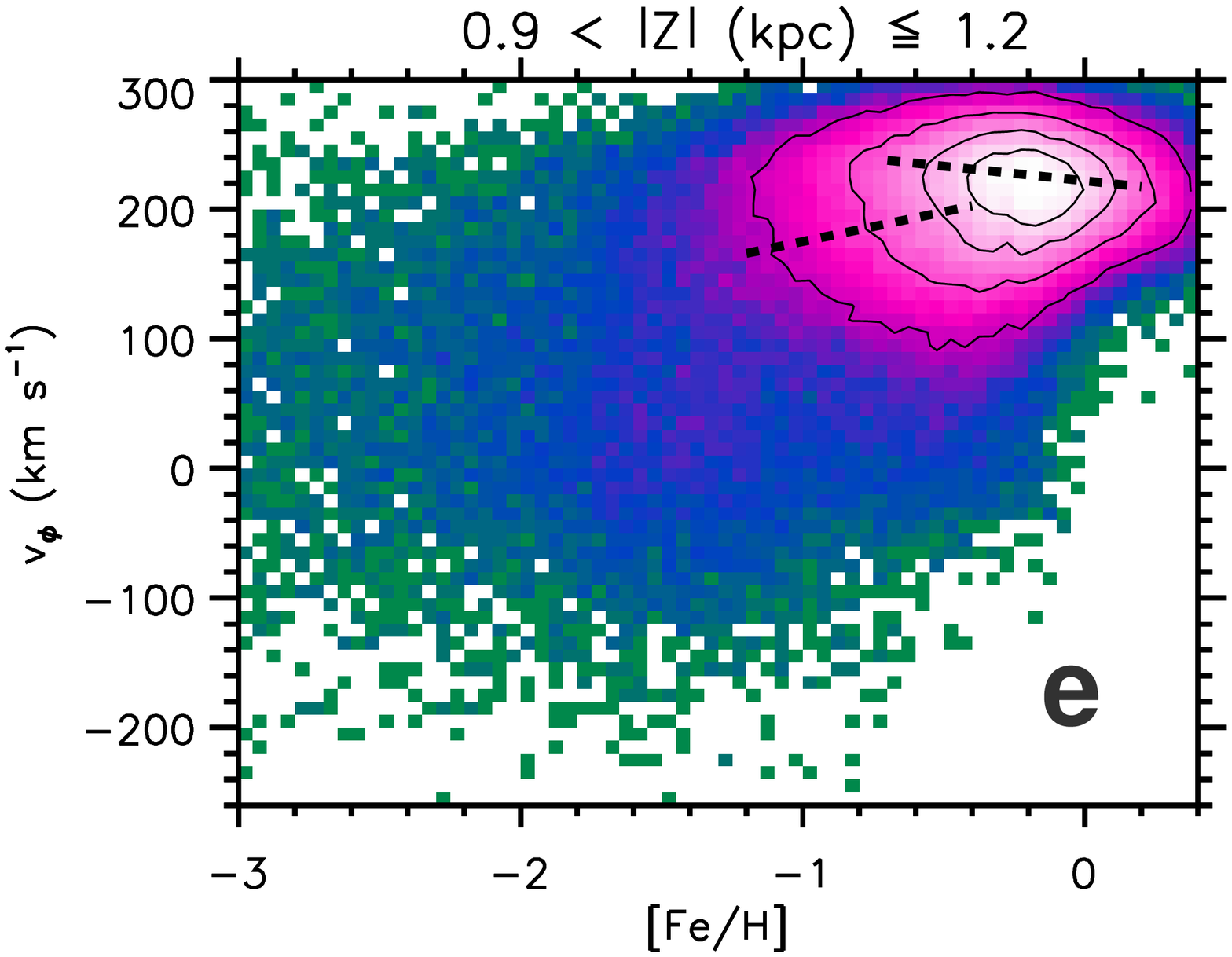}
\includegraphics[scale=0.36]{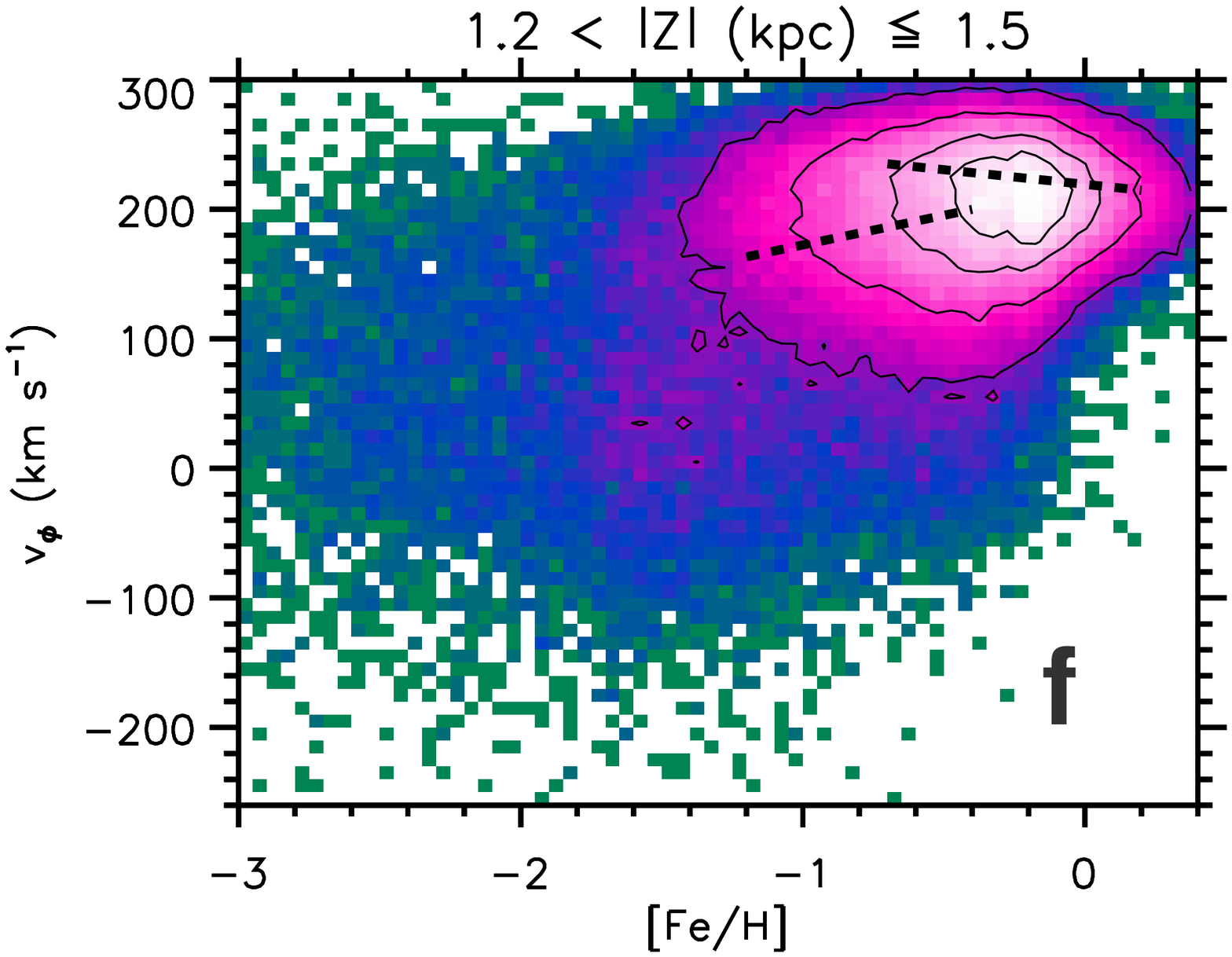}
\includegraphics[scale=0.36]{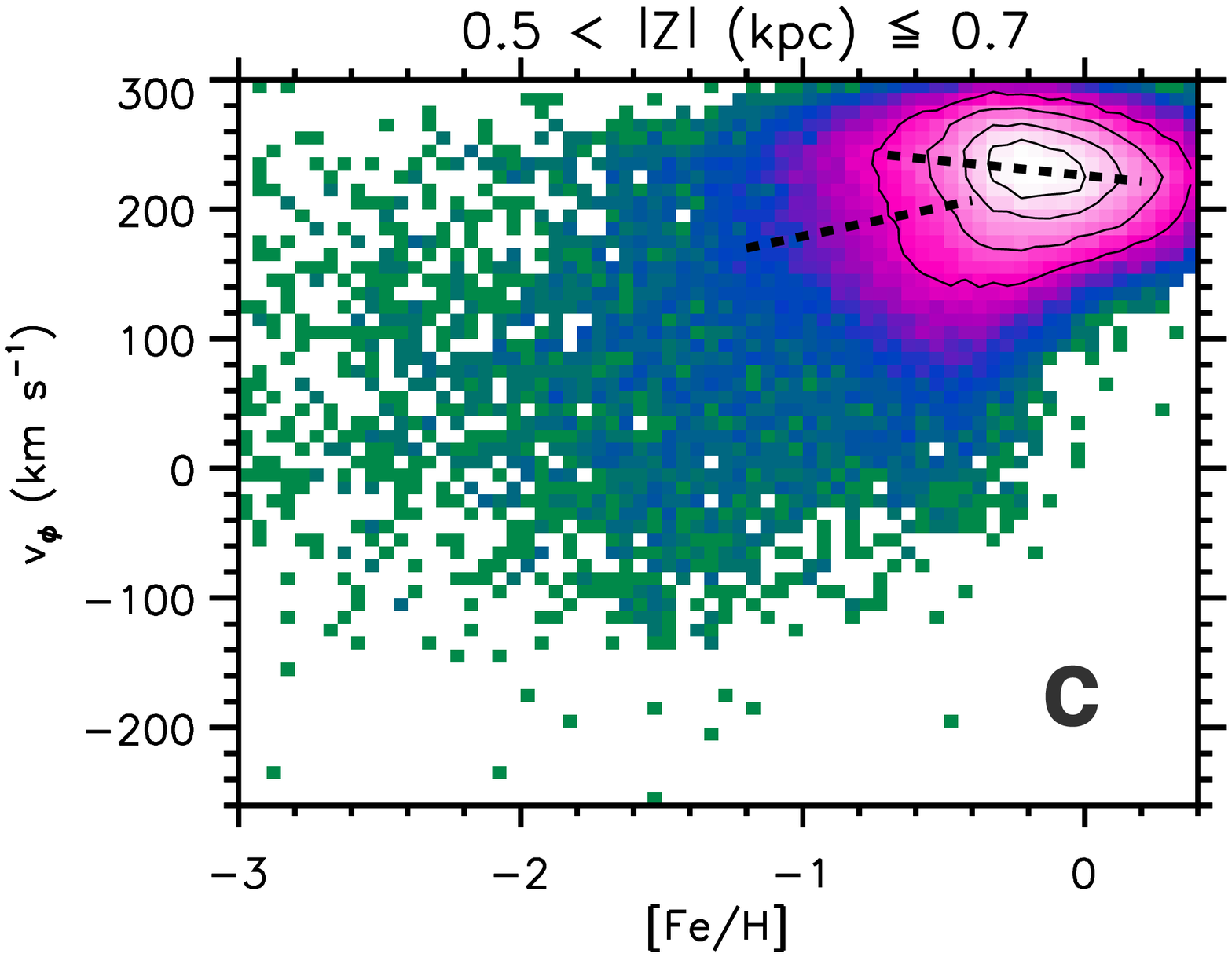}
\includegraphics[scale=0.36]{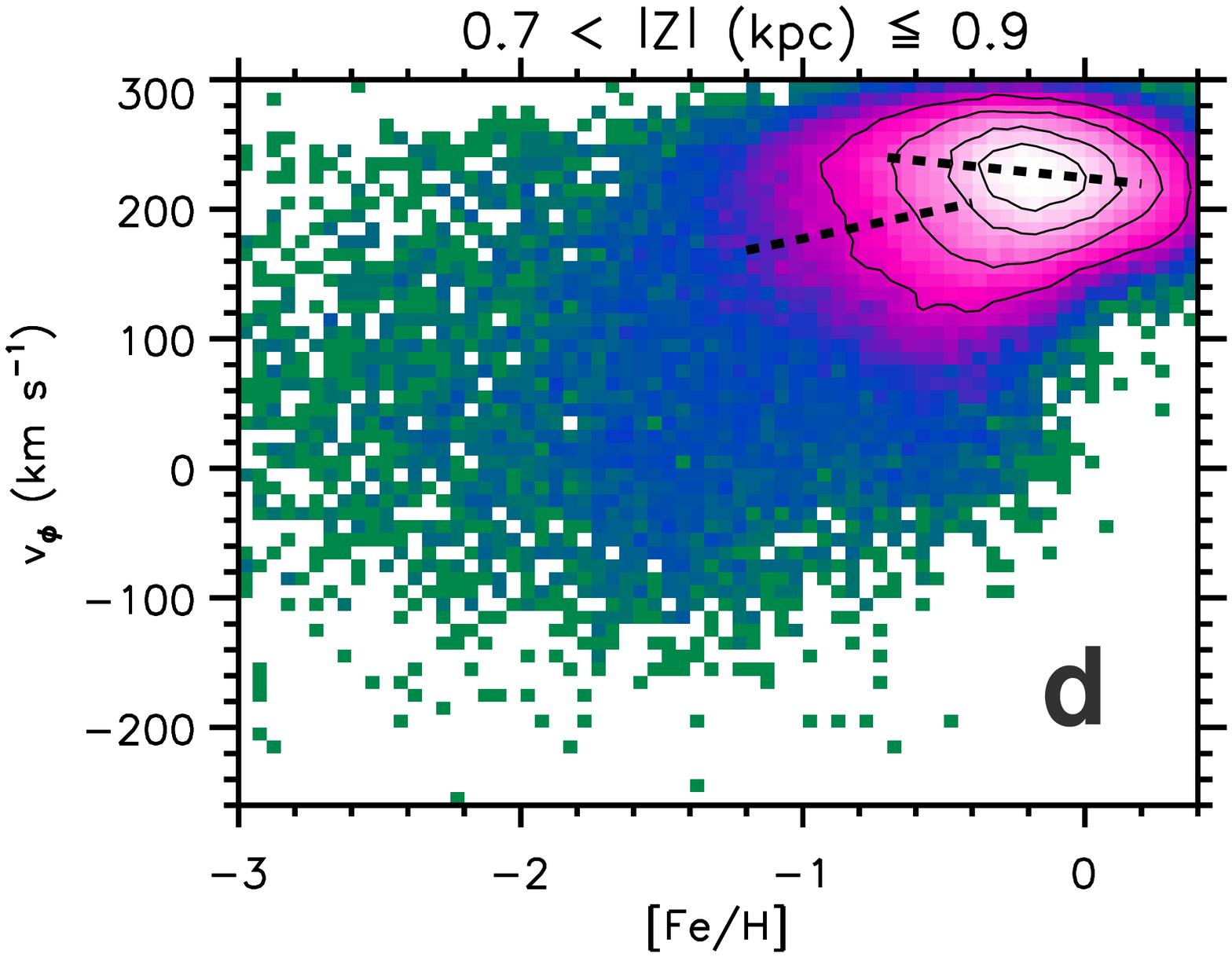}
\includegraphics[scale=0.36]{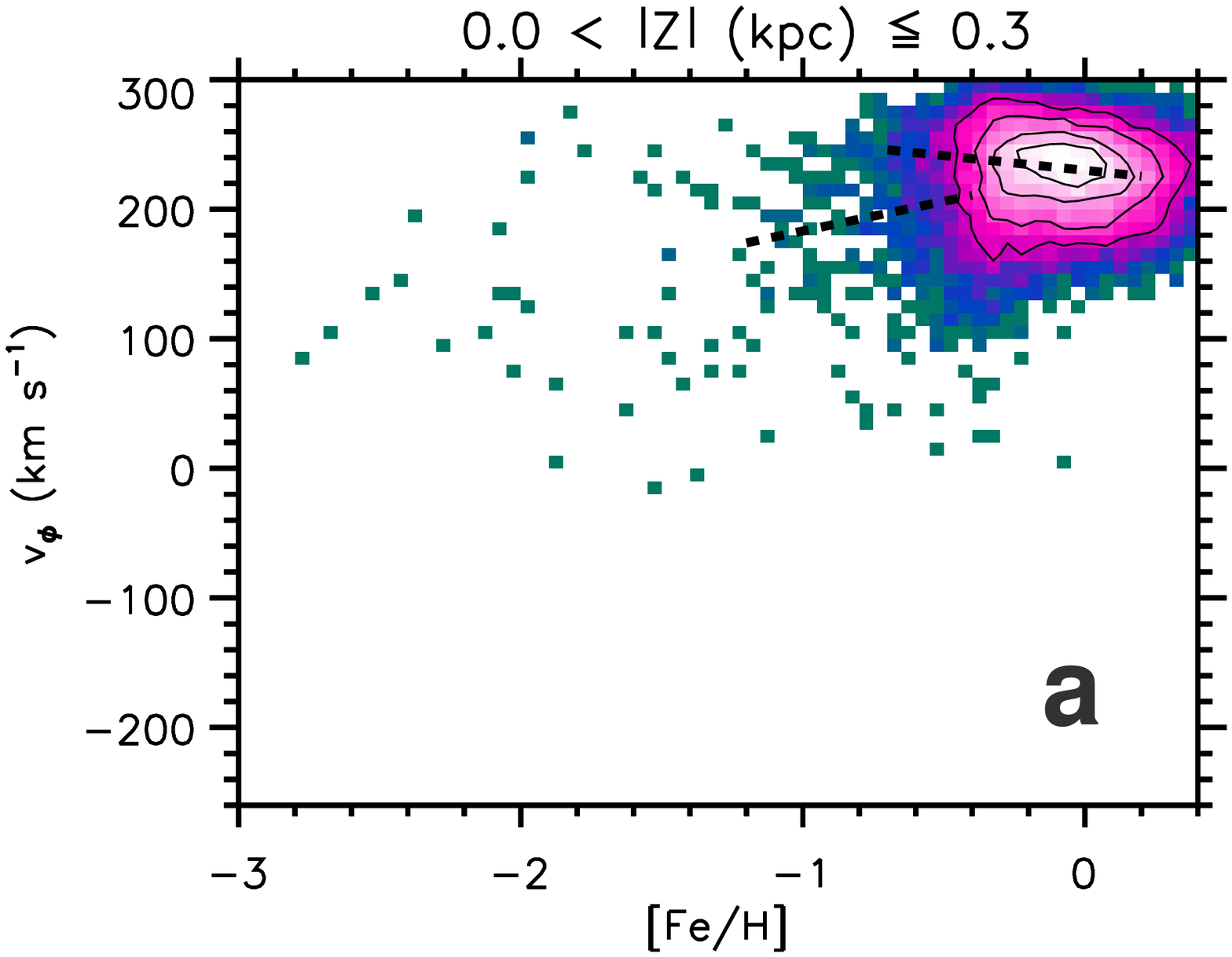}
\includegraphics[scale=0.36]{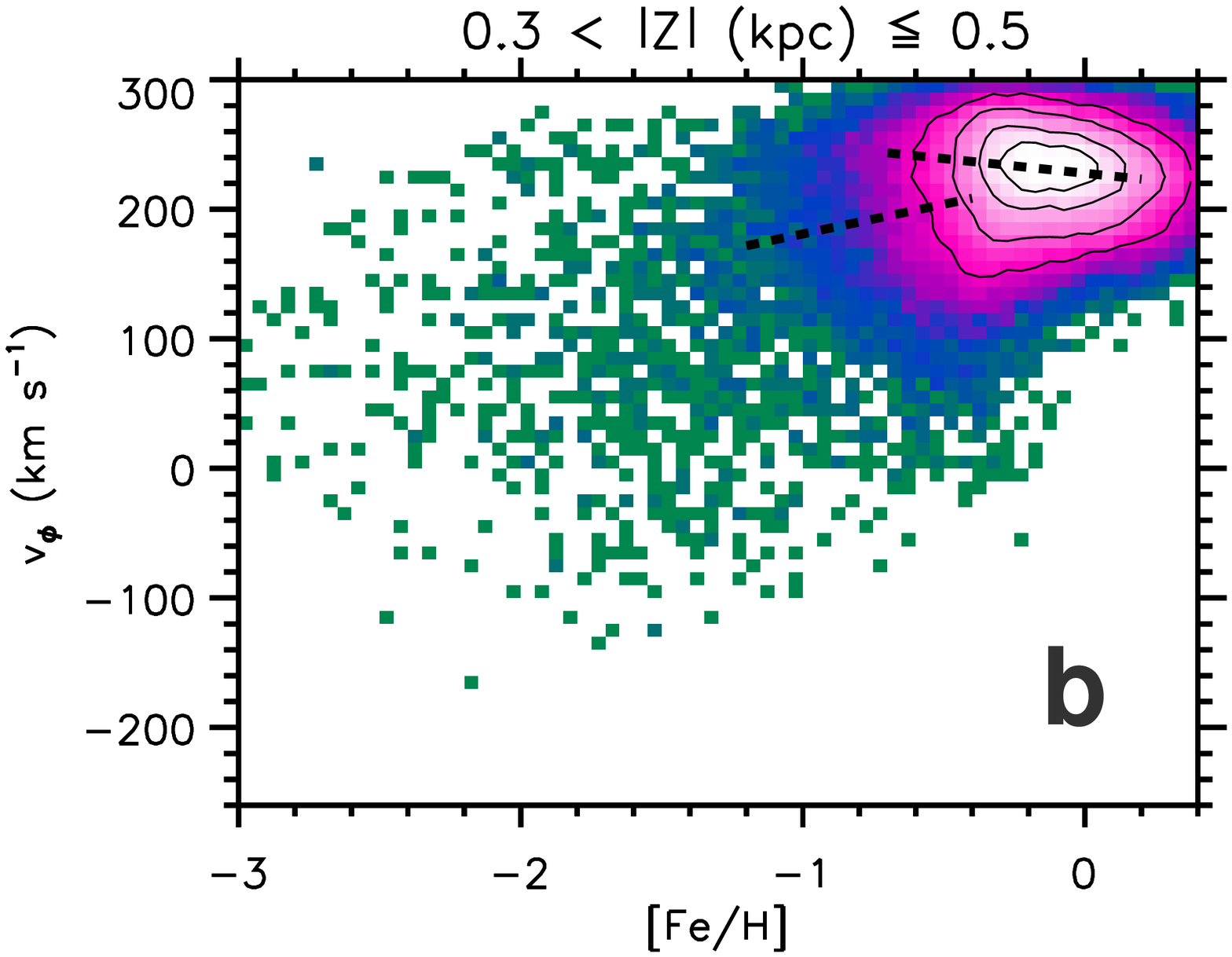}
\caption{Logarithmic number density distributions in the $\vphi$ vs.\ [Fe/H] plane, as a function of $|Z|$. Parallaxes ($\sigma_\pi/\pi < 0.3$) and proper-motion data from Gaia EDR3 are used to compute $\vphi$. The dotted lines indicate mean $\vphi$-[Fe/H] relations for the chemically defined thin-disk and TD samples \citep{lee:11}, which exhibit negative and positive correlations between $\vphi$ and [Fe/H], respectively.}
\label{fig:disk}
\end{figure*}

At $|Z|\sim1$~kpc, the number of thin-disk stars is approximately equal to the number of TD stars \citep[e.g.,][]{juric:08}. Therefore, the metal-rich group in our main sample should contain a sizable fraction of thin-disk stars at large $\vphi$. In particular, both the thin disk and the TD are known to exhibit a strong correlation between $\vphi$ and [Fe/H], as well as their dependence on $|Z|$. However, since our grouping of stars into four metallicity bins is coarse, the projected distribution of $\vphi$ loses a significant amount of information on the properties of individual components. Below we inspect three-dimensional distributions of metal-rich stars to gain insight on their relative strengths in the local volume.

Figure~\ref{fig:disk} shows a logarithmic number density distribution of stars in the $\vphi$-[Fe/H] plane in bins of different $|Z|$, which provides a closer look into the $\vphi$-[Fe/H] correlations of the thin disk and TD. The Gaia parallaxes and proper motions are used to compute $\vphi$, while metallicities from the fully photometric solutions are employed. Only stars with good parallaxes ($\sigma_\pi/\pi < 0.3$) are shown; other constraints are the same as in the main sample of this work (see \S~\ref{sec:sample}).

The observed $\vphi$-[Fe/H] distributions as a function of $|Z|$ are consistent with our previous results in Papers~I and II -- a gradual shift of the main stellar populations from the metal-rich, high-$\vphi$ disk stars to the metal-poor, low-$\vphi$ stars. However, Figure~\ref{fig:disk} clearly exhibits a change of the $\vphi$-[Fe/H] slope for disk stars. At small $|Z|$, the stars show a negative correlation, but the sign of the slope is reversed at $|Z| \ga 1.5$~kpc. The negative $\vphi$-[Fe/H] correlation of the thin disk is often interpreted as a negative metallicity gradient of the disk with respect to $\rgc$, along with epicyclic motions \citep[e.g.,][]{allendeprieto:16}, while the opposite trend for the TD can be understood by an inside-out formation \citep[e.g.,][]{schonrich:17} or by successive dynamical heating of the proto-disk \citep{grand:20}. Besides, an extended distribution toward lower $\vphi$ from the main sample of disk stars is visible in all distance bins, which is attributed to the SP. Interestingly, the dispersion drops abruptly at [Fe/H]$\ \ga-0.2$, which may set the end of the period of an early disruption of the primordial disk.

To compare with previous findings, the dotted lines in Figure~\ref{fig:disk} show the mean $\vphi$-[Fe/H] relations of the thin-disk and TD samples in \citet{lee:11}, which were selected based on distinct [$\alpha$/Fe]-[Fe/H] distributions of their stars \citep[see also][for a summary of the estimates in the literature]{yan:19}. A $|Z|$-dependence of their relations is also taken into account, with respect to a fiducial distance ($|Z|=0.6$~kpc), where we find a satisfactory eyeball match of our sample to their relations. The observed trends from our sample are strikingly similar to the spectroscopic results; their relation for the thin disk matches our map at $|Z| \leq 1$~kpc, while their TD relation is consistent with the observed trend at $1 < |Z| \leq 3$~kpc. At intermediate distance ($|Z| \sim1$~kpc), the ridgeline is almost horizontal, indicating  approximately equal contributions from the two disk components \citep[see the discussion in][for an early dispute on this issue]{ivezic:08}.

The bottom two panels of Figure~\ref{fig:spatial} show the spatial distributions ($Z$ vs.\ $R$, in the Galactocentric cylindrical coordinate system) of the canonical TD and the thin disk, respectively. Even though the division into the so-called chemically defined TD and thin-disk samples is difficult without information on the [$\alpha$/Fe] abundances of these stars, we use photometrically derived quantities ([Fe/H] and $\vphi$) to narrow down each sample based on the above $\vphi$-[Fe/H] correlations. As expected, our map captures the flattened structures of both components and the larger scale height of the TD \citep[e.g.,][]{hayden:15}, which validate the relative accuracy of the metallicity and distance determinations adopted in this work. Interestingly, the density distribution of the TD also exhibits small-scale structures, which may have been induced by vertical and horizontal oscillations in the Galactic disk \citep[e.g.,][]{antoja:18,an:19}.

{}

\end{document}